\documentclass[twocolumn]{article}
\usepackage{epubtk}    
\usepackage{natbib}
\usepackage[font=small,labelfont=bf]{caption}

\usepackage{graphicx}
\usepackage{bm}
\usepackage{amssymb}
\usepackage{aas_macros}
\usepackage{color}

\showlistoftablesfalse

\newcommand{\rd}{{\rm d}}
\newcommand{\del}{{\bm\nabla}}

\newcommand{\curl}{\del\times}
\newcommand{\beq}{\begin{equation}}
\newcommand{\eeq}{\end{equation}}
\newcommand{\beqa}{\begin{eqnarray}}
\newcommand{\eeqa}{\end{eqnarray}}
\newcommand{\bcom}{}
\newcommand{\blh}{}
\newcommand{\bkh}{}

\usepackage{polski}
\usepackage[english]{babel}
\usepackage{hyperref}

\begin{document}

\title{Magnetic fields in non-convective regions of stars}

\author{\epubtkAuthorData{Jonathan Braithwaite}{
Argelander Institut f\"ur Astronomie \\
Universit\"at Bonn, Auf dem H\"ugel 71, 53121 Bonn}{
jonathan.braithwaite@gmail.com}{
http://www.astro.uni-bonn.de/~jonathan}
\and
\epubtkAuthorData{Henk C. Spruit}{
Max Planck Institut f\"ur Astrophysik \\
Karl-Schwarzschild-Str. 1, 85741 Garching}{
henk@mpa-garching.mpg.de}{
http://www.mpa-garching.mpg.de/~henk}
}

\date{}
\maketitle

\vspace{ -1\baselineskip}
\par \noindent 
{\rm (arXiv version 7 April 2017)}\\ 
\bf
We review the current state of knowledge of magnetic fields inside stars, concentrating on recent developments concerning magnetic  fields in stably stratified (zones of) stars, leaving out convective dynamo theories and observations of convective envelopes. 
We include the observational properties of A, B and O-type  main-sequence stars, which have radiative envelopes, and the fossil field model which is normally invoked to explain the strong fields sometimes seen in these stars. Observations seem to show that Ap-type stable fields are excluded in stars with convective envelopes. Most stars contain both radiative and convective zones, and there are potentially important effects arising from the interaction of magnetic fields at the boundaries between them. Related to this, we discuss whether the Sun could harbour a magnetic field in its core.  Recent developments regarding the various convective and radiative layers near the surfaces of early-type stars and their observational effects are examined. We look at possible dynamo mechanisms that run on differential rotation rather than convection.  Finally we turn to neutron stars with a discussion of the possible origins for their magnetic fields.
\rm 

\section{\large Introduction}
Interest in magnetic fields in the interiors of stars, in spite of a lack of immediate observability, is rapidly increasing. It is sparked by progress in spectropolarimetric observations of surface magnetic fields as well as by asteroseismology and numerical magnetohydrodynamic simulations.

An important incentive also comes from developments in stellar evolution theory. Discrepancies between results and steadily improving observations has led to a newly  perceived need for evolution models `with magnetic fields'. At the same time  the demand for  results from stellar evolution have increased for application outside stellar physics itself. An example is the need for predictable colors of stellar populations in calculations of galaxy evolution.

Key questions concern the rate of mixing of the products of nuclear burning, since stellar evolution is sensitive to the distribution of these products inside the star. The heavier nuclei are normally produced later and deeper inside the star. Outside of convective zones, these nuclei reside stably in the gravitational potential.  Even weak mixing mechanisms in radiative zones, operating on long evolutionary time scales, can nevertheless change the distribution enough to affect critical stages in the evolution of stars. Possible mechanisms include hydrodynamic processes like extension of convective regions into nominally stable zones (`overshooting'), shear instabilities due to differential rotation, and large-scale circulations. Assumptions about the effectiveness of such processes are made  and tuned   to minimise discrepancies between  computed evolution tracks and observations. The presence of magnetic fields adds new mechanisms, some of which could compete with or suppress purely hydrodynamic processes.

 As well as mixing chemical elements\footnote{  The term `chemical elements' is used anomalously in astrophysics \blh (including this review) \bkh to mean atomic species.  }, magnetohydrodynamic processes in radiative zones should damp the differential rotation that is produced by the evolution of a star. 
 An indirect observational clue 
  has been the rotation rate of the end products of stellar evolution (Section~\ref{clurot}).  As the core of an evolving star contracts it tends to spin up. However, it is evident from the slow rotation of stellar remnants that angular momentum is transferred outwards to the envelope, and in many cases stellar remnants rotate much slower  than can be explained  even  with the known hydrodynamic processes. Maxwell stresses are more effective in transporting angular momentum than hydrodynamic processes (they can even transport angular momentum across a vacuum). This leads to the study of dynamo processes driven by differential rotation in stably stratified environments (Section~\ref{varint}).

 Note that there is a difference regarding the mixing of chemical elements. In purely hydrodynamic processes, the transport of angular momentum and chemical elements are directly related to each other, but this is not the case for magnetohydrodynamic processes. For a given rate of angular momentum transport, mixing by magnetohydrodynamic processes is less effective than in the case of hydrodynamic processes.

  
Until recently, the Sun was the only star for which direct measurements of  \emph{internal}  rotation were available, made possible by helioseismology.  For all other stars  the only source of information on angular momentum transport inside stars was the rotation of their end products. This has changed dramatically with the asteroseismic detection of rotation-sensitive oscillation modes in giants and subgiant stars by the Kepler and CoRoT satellites. These data now provide stringent tests for theories of angular momentum transport in stars (Sections~\ref{clurot}, \ref{varint}).

Possible internal magnetic fields come in two distinct kinds.  One kind is  time-dependent magnetic fields created and maintained by some kind of dynamo process,  running from some source of free energy.  Dynamos in convective zones have been studied and reviewed extensively before;  they are not covered  in this review except for a discussion of subsurface convection in O stars (Section~\ref{subsurf}). Another obvious source of free energy is differential rotation, and this could produce a  self-sustained small-scale magnetic field  in a radiative zone. This would be the candidate for transport of angular momentum and chemical elements described above.

The other kind of internal magnetic field is \emph{fossil} fields, remnants of the star formation process that have somehow survived in a stable configuration. The theory of such fields is discussed in Section~\ref{sec:theory}. 

A subset of intermediate-mass stars display strong magnetic fields, the chemically pecular Ap and Bp stars; and some more massive stars display similar fields. These are thought to be such fossil fields. They used to be interpreted in terms of configurations resembling simple dipoles. With the improved observations of the past decade a much larger range of configurations is found; this is reviewed in Section \ref{obsAp}. Theory indicates that only a small fraction of all imaginable magnetic equilibria in a star can be stable as observed.  Comparing these theoretically allowed configurations with the surface fields actually observed in individual stars gives clues about the internal structure of the fields. Together with statistical information on observed field strengths and configurations, this holds the promise of telling us something about the conditions under which the magnetic fields formed.

Though stably stratified throughout most of their interior, Ap stars still contain a small convective core. This raises the question to what extent convection interacts with the fossil field, or whether a fossil field is compatible at all with the presence of a convective zone somewhere in the star. A related question is whether stars with convective envelopes like the Sun might have fossil fields hidden in their stably stratified interiors. These questions are addressed in Section~\ref{conv}.

 Also thought to be of fossil nature are the magnetic fields in neutron stars. As in upper-main-sequence stars, there is a puzzlingly enormous range in field strengths, spanning five orders of magnitude. There are two obvious ways to explain this range: either it is inherited from the progenitor stars, in which case one still needs to explain the range in birth magnetic properties of main-sequence stars, or it is produced during the birth of the neutron star. It is possible that it is produced from the conversion of energy from differential rotation into magnetic, and that the same physics is at work during the birth of main-sequence stars. These issues are addressed in Section \ref{NSgen}. 

This review is organised as follows. In the next section, we look at observations of magnetic stars, {\blh with some focus on peculiarities that may hold clues on the origin of their fields.  Among the main-sequence star these are the Ap/Bp stars and the apparently non-magnetic intermediate-mass stars, next are the massive stars, and the magnetic white dwarfs then are discussed very briefly. \bkh} Section~\ref{sec:theory} is a review of the theory of static `fossil' magnetic fields in radiative zones, and in Section~\ref{sec:variations} we examine various scenarios which could explain where these fossil fields come from. In Sections~\ref{diffrot} and \ref{conv} respectively we look at the interaction of magnetic fields with differential rotation and with convection. In Section~\ref{neutron} we move onto neutron stars: their observational properties as well as likely theoretical explanations in terms of internal magnetic field.  Finally we summarise in Section~\ref{summary}. 

 This review goes into some depth in the magnetohydrodynamics of stars. The interested reader may wish to look at some literature on MHD, including the astrophysical context. \blh The classic book by \citet{Roberts67}  covers basic MHD in general contexts but is out of print. \bkh More recent is the monograph by \citet{Spruit13}, an introduction tailored specifically to astrophysicists and with an emphasis on physical intuition and visualisation rather than mathematics. The books by \citet{2004prma.book.....G} and \citet{2010adma.book.....G}  offer a more detailed look at various astrophysical contexts. Also worth a look are the books by \citet{1998pfp..book.....C} and \citet{Kulsrud05}, which have a greater emphasis on plasma effects, i.e. not using the single fluid approximation. 

\begin{figure}[htb]
\centerline{\includegraphics[angle=0,width=0.5\textwidth]{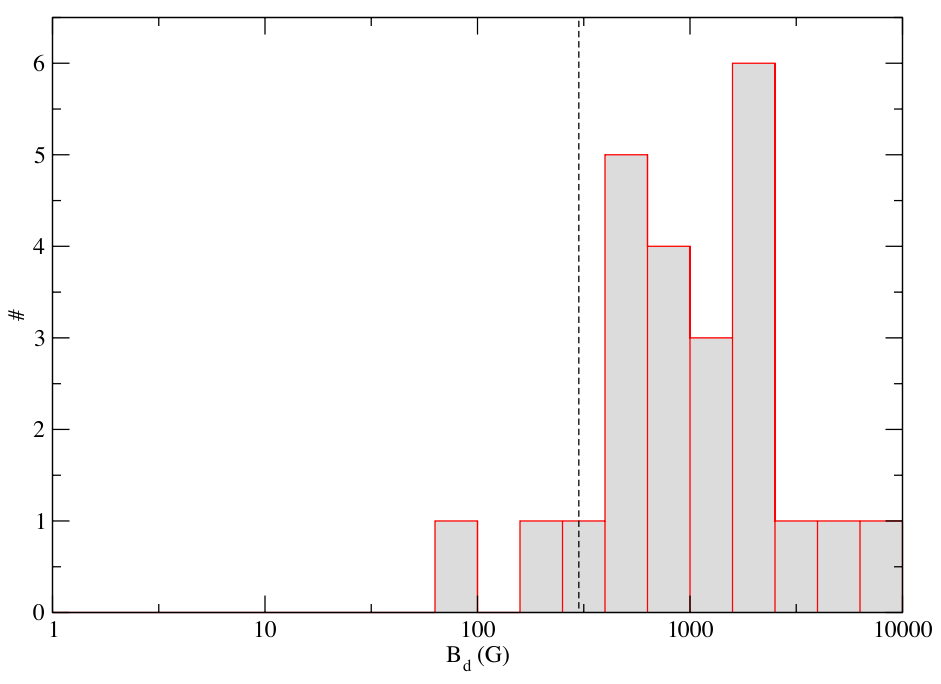}}
\caption{Measured magnetic fields in a sample of Ap stars in which either no magnetic field had yet been detected or in which there had been an ambiguous or borderline detection. In this study the detection limit is only a few gauss; every star in the sample is found to have a magnetic field much stronger than this.  The dashed line represents the cutoff at 300 G.  This result confirms convincingly that all Ap-type chemically peculiar stars have strong magnetic fields. In contrast, other A stars have never been found to have any magnetic field above a few gauss; there is a clear bimodality. From \citet{Auriere_07}.}
\label{bimod}
\end{figure}

\bigskip
\section{\large Observed Properties of Magnetic Stars}\label{obsAp}

The observational techniques used for measuring magnetic fields on the surface of stars have been reviewed by \citet{DonatiL09} and \citet{Mathys12}. In almost all types of magnetic star the Zeeman effect is used to detect the magnetic field. (For an excellent introduction of the Zeeman effect and its detection in stars see \citet{Landstreet11}.)

The most reliable observations of magnetic stars use `full Stokes' spectropolarimeters that record the complete information contained in linear and circular polarization.  The full set of polarization components of a single suitable spectral line is sufficient in principle to determine the strength of the magnetic field and its orientation with respect to the line of sight. The V-signal, which gives information about the line-of-sight component of the magnetic field, is easier to detect than Q and U because it is antisymmetric with respect to the center of the line and less sensitive to instrumental polarization.

To increase the signal-to-noise ratio of these measurements, many metal lines can be combined together to give a weighted mean of the Stokes I and V line profiles in a procedure known as Least Squares Deconvolution (LSD; \citealp{Donati_97}). Since this technique was introduced, the detection limit has dropped significantly. What this gives us is a disc-averaged line-of sight component of the magnetic field, the so-called \emph{mean longitudinal field} $B_z$; for many stars this is the only quantity that can be reliably measured. In other stars though it has also been possible to get extra information from Stokes Q and U. If the star is observed at several rotational epochs one can then construct a simple model of the magnetic field on the surface, e.g., dipole + quadrupole, and work backwards to find the best-fitting parameters of the model. 

\subsection{Ap stars}\label{apstars}

In this section we review the Ap stars, the spectroscopically 'peculiar' intermediate-mass main-sequence stars between about B8 and F0, their spectra showing very unusual abundances of the elements.

It has gradually become clear that there is a bimodality in the population of intermediate-mass stars ($1.5$ to $6\,M_{\odot}$), namely that all stars classified as Ap/Bp (with exception of the so-called mercury-manganese (HgMn) stars) host large-scale magnetic fields with mean longitudinal fields between around 200~G and 30~kG, and that the rest of the population lack magnetic fields above the detection limit of a few gauss (\citealp{Auriere_07}; Figure~\ref{bimod}). Ap stars account for a few percent of the A star population.

\begin{figure}[t]
\centerline{\includegraphics[width=0.5\textwidth]{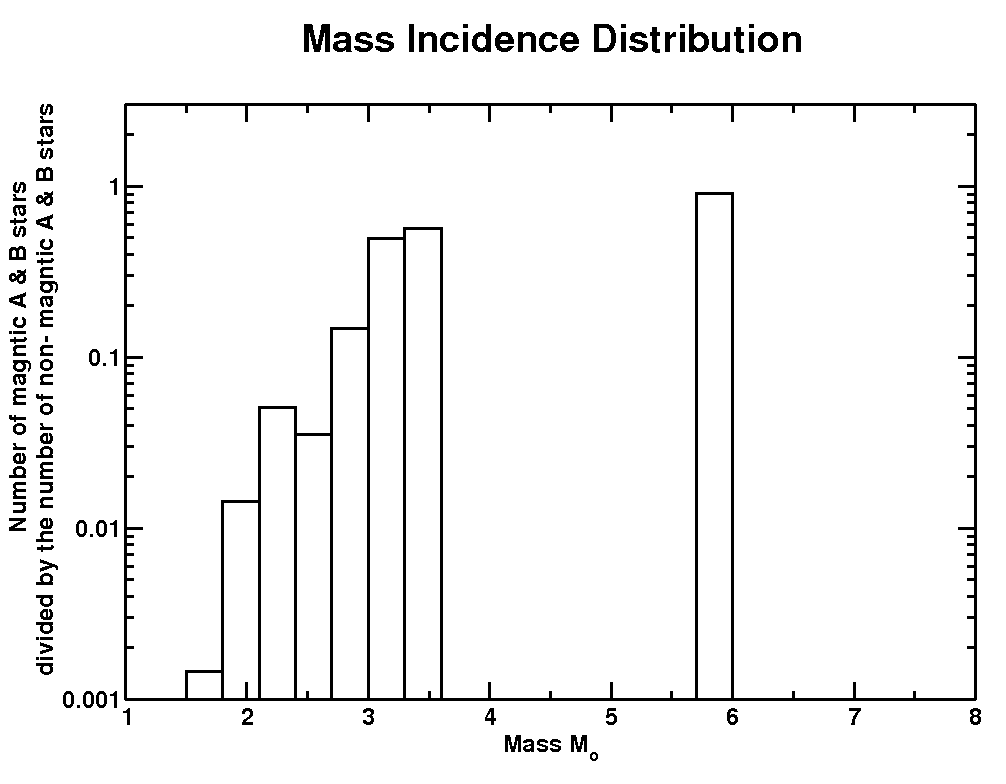}}
\caption{The fraction of intermediate-mass stars within 100pc in which a magnetic field has been detected; the sample is relatively complete. From \citet{Power_07}.}
\label{Power}
\end{figure}

Still, this leaves a factor of 100 in field strength to be explained by models for the origin of Ap star fields. A similar problem exists in the magnetic white dwarfs, where field strengths range from $<10^4$ to almost $10^9$~G, and in pulsars ($\sim10^{10}-10^{15}$~G). This problem may well reflect a basic property of the formation process of fossil fields (see also Section~\ref{sec:variations}). There are various observational clues to the origin of this bimodality in magnetic properties amongst A and late B stars. For instance, there is a strong correlation between mass and the magnetic fraction of the population. \citet{Power_07} examined a volume-limited sample of intermediate-mass stars -- all stars within 100 pc of the Sun -- finding that the magnetic fraction of the population increases from less than 1\% at $1.5\,M_{\odot}$ to $\gtrsim20\%$ at $3.5\,M_{\odot}$; see Figure~\ref{Power}.  The total magnetic fraction in the sample is only 1.7\%. The Ap phenomenon disappears completely at masses below $1.5\,M_{\odot}$ (around F0), which coincidences with the onset of efficient convection in the envelope \citep[e.g.,][]{Landstreet91}. Other clues come from the rotation and binarity -- see below.

In some strongly-magnetised, slowly-rotating Ap stars and most magnetic white dwarfs the Zeeman splitting is greater than the line width, in which case it is possible to measure the Zeeman splitting directly in Stokes I (intensity), without having to use polarimetry. This gives an average of the field strength over the visible disc, the so-called \emph{mean field modulus} $B_s$. Now, if a star's magnetic field is dominated by small-scale structure we will clearly expect that $B_s\gg |B_z|$, since the line-of-sight component from various patches on the surface will cancel each other in some kind of statistical $\sqrt{N}$ manner. For the Ap stars in which $B_s$ is measurable, we do not find this -- an important result showing that small-scale structure, if present, is not dominant.


Ap stars display a large variety of field geometries. In many stars, a good fit to the data can be achieved by assuming the simplest geometry of all, i.e., a dipole field, at the surface, which is inclined to the rotation axis at some angle. In other stars this produces poor results and a more complicated geometry produces better results, for instance dipole + quadrupole.

\begin{figure}[htbp]
\vspace{ 1\baselineskip}
\centerline{\includegraphics[width=0.5\textwidth]{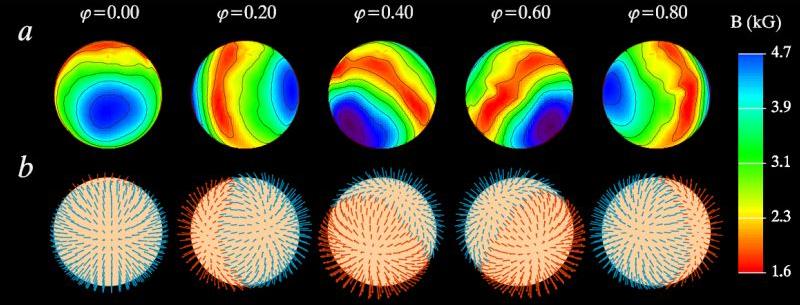}}
\vspace{2mm}
\centerline{\includegraphics[width=0.5\textwidth]{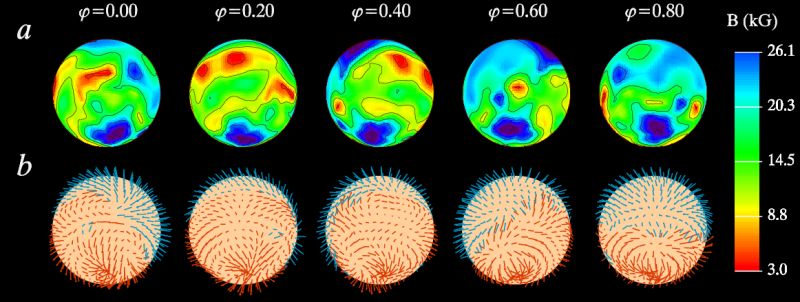}}
\caption{The magnetic fields of $\alpha^2$ CVn (upper panel) and 53~Cam (lower panel), viewed at five rotational phases. The upper rows in each panel show field strength and the lower rows the direction. Clearly, whilst $\alpha^2$ CVn has an almost perfect dipole field, the magnetic field of 53 Cam has a much more complex geometry. From \citet{Kochukhov_02, Kochukhov_04}.}
\label{a2cvn}
\end{figure}

Improved observations have made it possible to combine the Zeeman effect with the Doppler effect from the rotation of the star to get, in effect, some spatial resolution on the surface of the star, without having to make prior assumptions of this kind. \citet{PiskunovK03} have developed a technique called Magnetic-Doppler imaging and have used it to make some impressive maps of the magnetic field on a number of stars, such as 53~Cam \citep{Kochukhov_04}, $\alpha^2$CVn \citep{Kochukhov_02, KochukhovW10} and HD~37776 \citep{Kochukhov11a}. Two examples are shown in Figure~\ref{a2cvn}. A similar technique called Zeeman--Doppler imaging, developed by Donati \& Petit \citep[see, e.g.,][]{Donati01} has been used to make magnetic images of cool stars, e.g., \citet{Petit_04}, as well as some hot stars, e.g., \citet{Donati_06} and Figure~\ref{tauSco}. Using these techniques, some rather complicated geometries\footnote{The common useage of term `topology' in this context is sloppy. \blh Meant is distribution on the star's surface.  Topology is by definition \bkh a global property of the entire field configuration; nothing can be inferred about it from observations of the stellar surface alone.} have been found which appear   to indicate the presence of  meandering flux tubes just below the stellar surface.

\subsubsection{Chemical peculiarities}

Interesting and unique to intermediate-mass stars are processes near the surface: gravitational settling and radiative levitation, which cause separation of chemical elements in the atmospheres of the stars and result in a variety of observed chemical abundance phenomena \citep{Michaud70}. Ap/Bp stars are defined as a class showing peculiar (hence the `p' in `Ap star') abundances of rare earths and some lighter elements such as silicon, as well as inhomogeneities of these elements on the surface which show correlations with the magnetic field structure, albeit not the same kind of correlation in all stars.
 There is a strong  correlation between the Ap/Bp phenomenon and strong magnetic fields,  with the apparent exception of the subclass of the HgMn stars \citep{Kochukhov11b}. 

The origin of this phenomenon is inextricably linked to the presence and location of surface- and subsurface convection layers resulting from opacity bumps associated with helium and hydrogen ionization (see also Section~\ref{subsurf}). Convection obviously washes out the effects of any gentle separation processes, and a sufficiently strong magnetic field is expected to disrupt convection. A magnetic field  $\gtrsim$~200~G is above equipartition with the thermal pressure at the photosphere, and should thus inhibit convection \citep[][see also Section~\ref{subsurffossil}]{GoughT66, MossT69, Mestel70}. Indeed there is observational evidence for this in the form of a reduction in microturbulence velocities in Ap stars (D.\ Shulyak, 2013, priv.\ comm.). In spectral type the Ap/Bp phenomenon disappears around F0, corresponding to the onset of efficient convection at the surface, and at B8, corresponding to the appearance of stronger subsurface convection. 

The chemical peculiarities apparently develop after the magnetic field is already in place, appearing at some stage during the pre-main-sequence \citep{Folsom_13a}. Note though that chemical peculiarities are not restricted to the magnetic stars; amongst the other A stars, various other types of chemical peculiarity are seen, for instance in the slowly-rotating Am stars (of which Sirius is the best-known specimen), mercury-manganese stars and $\lambda$ Bootis stars. See \citet{Turcotte03} for a review of these `skin diseases'.

\subsubsection{Rotation}

It has long been known that most magnetic A stars rotate slowly compared to the non-magnetic stars \citep[see e.g.][]{AbtM95}. Whilst the non-magnetic A stars are generally fast rotators, with rotation periods of a few hours to a day, most Ap stars have periods between one and ten days, and some have periods much greater, with around 10\% of Ap stars having periods above 100 days \citep{Mathys08}. The slowest rotation periods are of order decades and in several cases only a lower limit can be stated. Note that whilst the rotation periods of the non-magnetic stars are estimated statistically from $v \sin i$, those of magnetic stars can be measured directly from the periodicity in the Zeeman signal, since the magnetic field is never perfectly symmetrical about the rotation axis.

\begin{figure}[t]
\centerline{\includegraphics[width=0.5\textwidth]{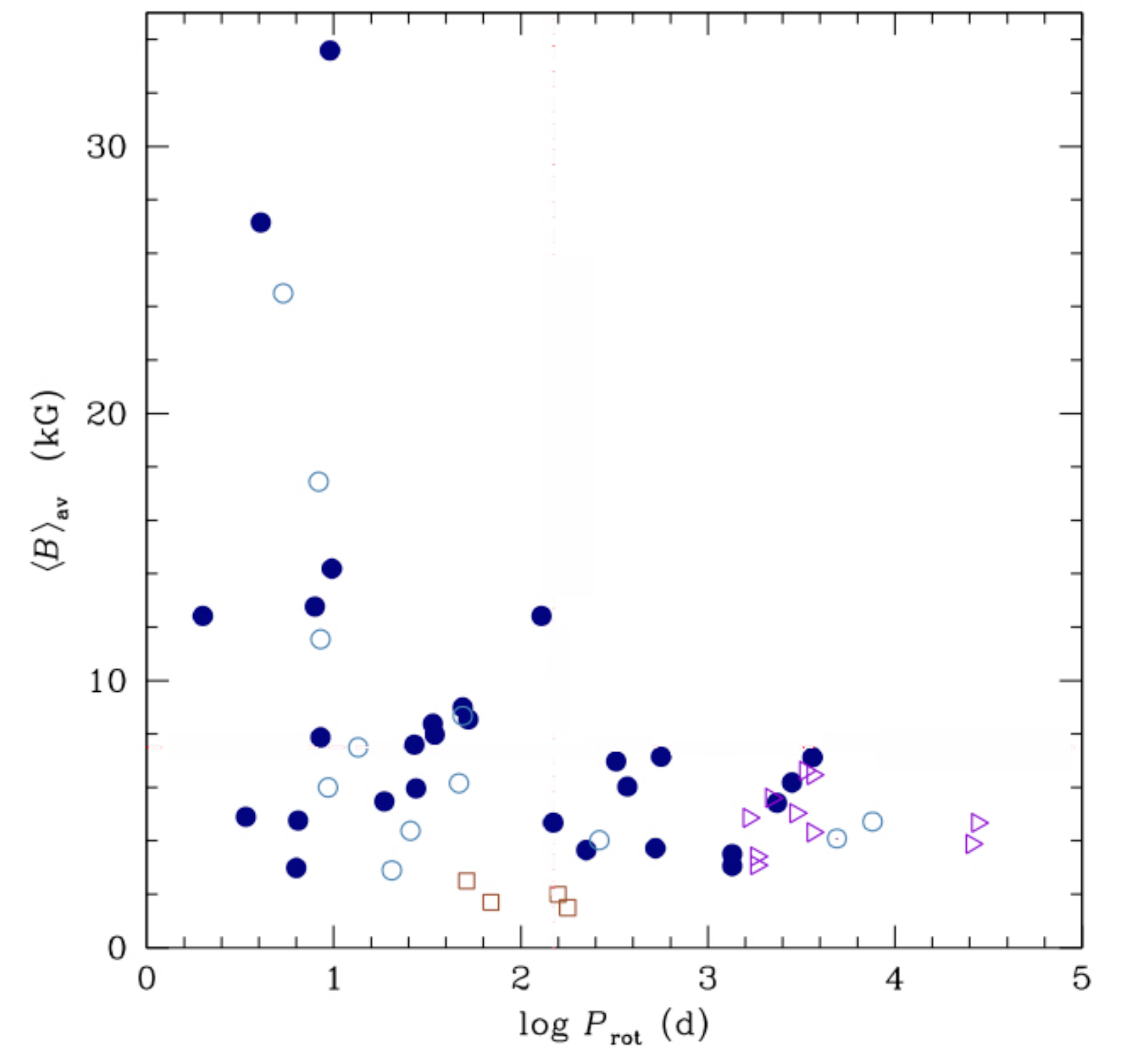}}
\caption{\blh Observed average of the mean magnetic field modulus against rotation period. Dots: stars with known Prot; triangles: only a lower limit of Prot is known. Open symbols : stars for which existing measurements do not cover the whole rotation cycle. From  \citet{Mathys16} \bkh}
\label{B-P}
\end{figure}

Some intriguing correlations between the rotation period and magnetic properties of Ap stars have been found. For instance, \citet{Mathys08} finds in a sample of slowly rotating stars that those with $P>100$ days lack fields in excess of 7.5~kG. \blh A recent compilation is shown in Fig.\ \ref{B-P}. \bkh  In addition, \citet{LandstreetM00} find  that the slower rotators ($P>25$ days) are more likely to have closely aligned magnetic and rotation axes. 

The slow rotation of Ap stars holds only in a very broad sense; rapidly rotating examples like CU Virginis (0.5~day) exist as well. Any given explanation for the slow average rotation may well miss the most important clue: the astonishingly large range in rotation periods, of at least four orders of magnitude. 

\subsubsection{Binarity}

The binary fraction amongst the magnetic stars is lower than in the non-magnetic stars \citep{AbtS73, Gerbaldi_85, Carrier_02, Folsom_13b}. There is apparently a complete lack of Ap stars in binaries with periods of less than about 3 days, except for one known example (HD~200405) with a period of 1.6 days. This speaks in favour of more than one of the formation scenarios (see Section~\ref{sec:variations}).

\subsubsection{Ages}
\label{herbig}
Pre-MS stars are known as T Tauri stars if they are later than spectral type F5 ($\log T\approx 3.8$) and Herbig Ae-Be (HAeBe) if they are earlier. We now know of over 100 HAeBe stars \citep[e.g.,][]{HerbigB88, Vieira_03}. Stars between about $1.5$ and $4\,M_{\odot}$ leave the birth line as  fully-convective  T Tauri stars.  Eventually they develop a radiative core, at which time they stop moving downwards on the HR diagram and move instead to the left on what is known as the Henyey track; the convective envelope shrinks and they become HAeBe stars.  More massive stars leave the birthline as HAeBe stars. Depending on the local and/or accretion conditions  as well as on the mass, Ê these stars may or may not become visible before they reach the MS; the most massive HAeBe stars observed  are around $20\,M_{\odot}$.

\begin{figure}[htb]
\vspace{1\baselineskip}
\centerline{\includegraphics[width=0.47\textwidth]{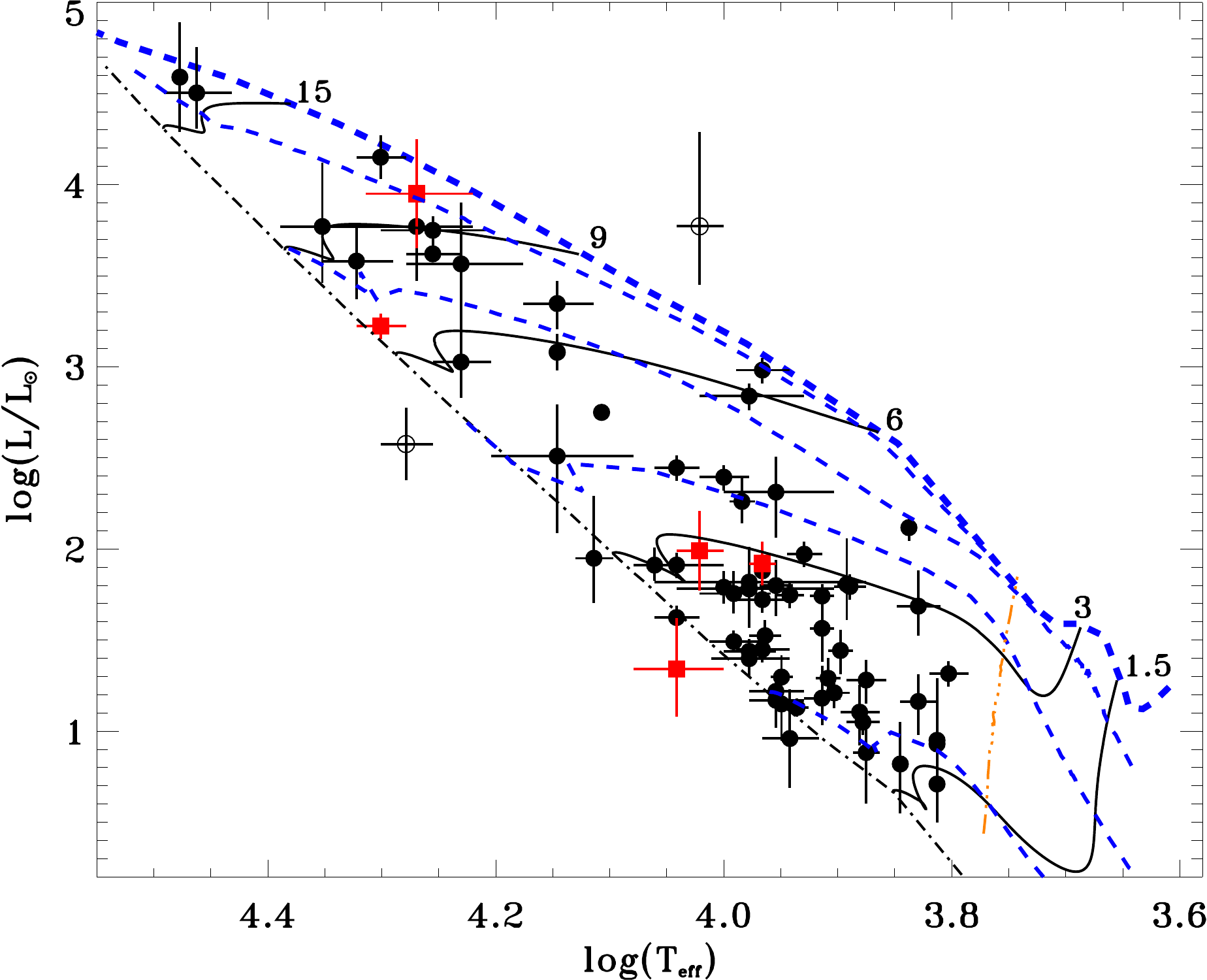}}
\caption{Magnetic (red squares) and non-magnetic (black circles) HAeBe stars on an H-R diagram. The thick blue dashed line is the birth line, the thin blue dashed lines are isochrones, the dot-dashed line is the ZAMS and the solid lines are  theoretical  evolutionary tracks from the birth line to the ZAMS for the masses indicated. The orange dot-dot-dot-dashed line represents a radiative/convective transition: to the left of this line, the convective envelope accounts for less than 1\% of the star's mass. A convective core appears towards the end of the PMS when the star moves downwards on the HR diagram.  The open circles correspond to HD 98922 (above the birthline) and IL Cep (below the ZAMS), whose positions cannot be reproduced with the theoretical evolutionary tracks considered.  From Figure~4 in \citet{Alecian_13b}.}
\label{HAeBe-mag}
\end{figure}

It is clear now that some fraction -- comparable to the fraction amongst main-sequence A and B stars -- of HAeBe stars are magnetic. \citet{Wade_05} presented the first detections of magnetic fields in HAeBe stars, and \citet{Alecian_13b, Alecian_13c} present the results from a survey of 70 Herbig Ae-Be stars: see Figure~\ref{HAeBe-mag}.

There was a claim \citep{Hubrig_00} that all Ap stars have passed through at least 30\% of their main-sequence lifetime. This result depended on determining the ages of stars by placing them on the HR diagram, which is very challenging since stars move very slowly across the HR diagram during the first part of the main-sequence,  and because of the distortion of apparent surface temperatures by the atmospheric abundance anomalies.  In light of this, and of the recent results on pre-MS stars, it looks unlikely that this result is correct. \citet{Landstreet_09a} looked instead at Ap stars in clusters -- where the ages can be determined much more accurately -- and found the opposite: that there is a negative correlation between the field strength in Ap stars and their age, greater than one would expect from flux conservation as the star expands along the main sequence. Possible explanations for this field decay are discussed in Section \ref{decay}.

\begin{figure}[h]
\centerline{\includegraphics[width=0.53\textwidth]{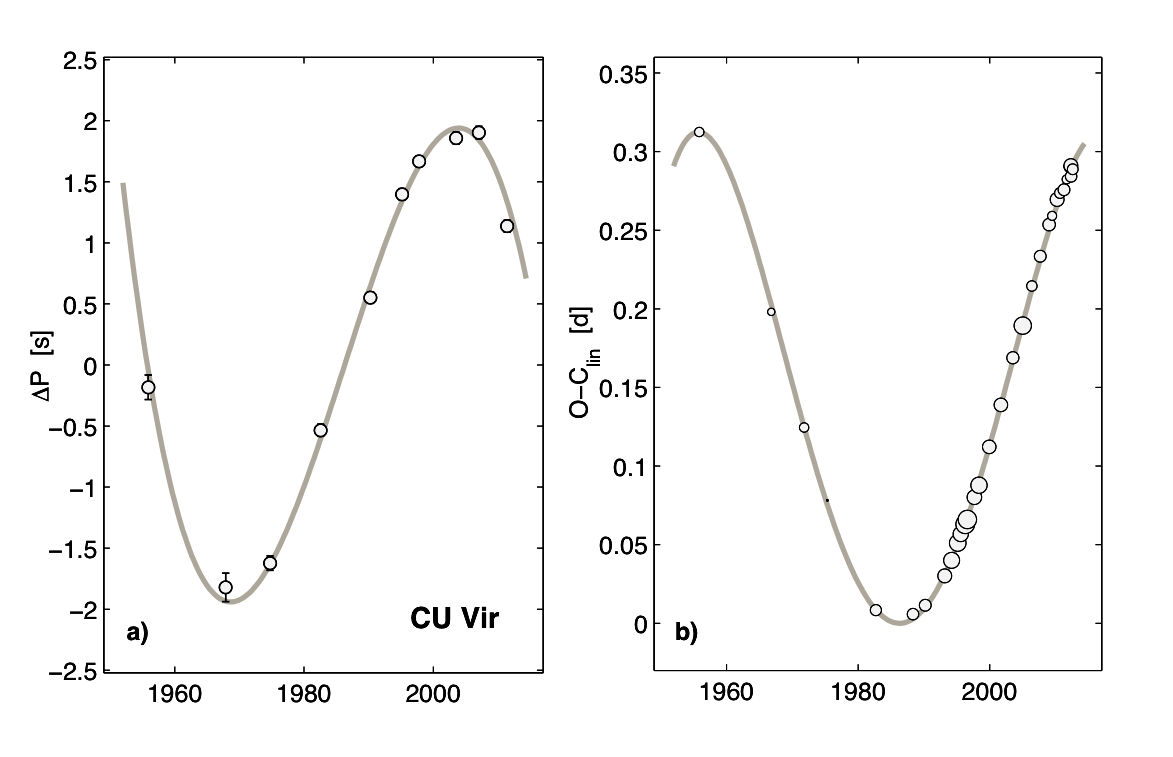}}
\caption{Variation in the rotation period of CU Virginis. The left panel shows the variation in the rotation period, and the right panel is the phase residue, assuming a constant period. From \citet{Mikulasek_11a}.}
\label{CUVir}
\end{figure}

\subsubsection{Variability in Ap stars}

Though Ap stars are characterised by stability of their magnetic and spectral signatures on all observable time scales (apart from rotational modulation), there exist a few curious cases where changes have been observed. The rapidly rotating star CU~Virginis (P~=~0.52~days, with observations stretching back to 1949 \citep{Deutsch52}), has increased its rotation period by about 50~ppm (Figure~\ref{CUVir}). The O-C data (the phase drift relative to a fixed period) can be fit by an increase in rotation period within a few years around 1984 (\citealp{Pyper_98, Pyper_13}\epubtkFootnote{The title of the paper incorrectly says `decrease'.}). This would be reminiscent of the `glitches' seen in pulsars. However, a more gradual change in period also seems compatible with the observations. \citet{Stepien98} suggested that the changes in CU Vir may not actually be monotonic, but could reflect some form of magnetic oscillation in the star, with typical time scales of several decades. This would imply that, at a given point in time, one would find period decreases about as frequently as increases. Period increases have been seen in other stars: spindown rates of order $P/\dot P\sim 10^{5-6}$ yr$^{-1}$ have also been reported in V901 Ori, $\sigma$ Ori e, HR 7355, SX Ari, and EE Dra \citep{Mikulasek_11b}. No clear case of a period decrease case has so far been seen in the sample of stars with changing periods, but there are hints that both CU Virginis and V901 Orionis may now be spinning up \citep{Mikulasek_11a}. The current limited sample looks therefore still to be statistically compatible with the idea; finding alternative explanations would be challenging. Speaking strongly in favour of the oscillation idea is that the spindown timescales measured are of order $10^{5-6}$ yr, which would otherwise be hard to reconcile with the ages inferred for these stars, of the order of $10^8$ yr.

\citet{Mikulasek_13} offer two alternative interpretations of the data on CU Virginis, namely that the rotation period is undergoing either some kind of Gaussian variation and will return to its original value, or a periodic variation with a minimum rotation period in 1968 and a maximum in 2005. In the oscillation scenario, the amplitude (at the surface) of the torsional oscillation  phase residue  in CU Vir would be approximately $2.3\pi$ and in V901 Ori at least $0.5\pi$ \citep{Mikulasek_13}. A challenge for the oscillation idea might be that the Alfv\'en timescale in CU Vir, with a surface field strength of around 3~kG, would be only about 3 years, assuming that field strength also in the star's interior. One expects the fundamental magnetic oscillation mode to have this period, and higher harmonics even shorter periods.

A remarkably rapid change in the field configuration, on a time scale of years, was reported in the pre-MS star HD190073 \citep{Alecian_13a}. It now appears that this was a spurious result \citep{Hubrig_13}.

\subsection{ Other stably stratified stars }

 Having looked in some detail at the strongly magnetic subset of intermediate-mass stars, which have a long history in the literature, we now turn to other stars which are stably stratified, at least on the outside. First of all, the rest of the intermediate-mass population and the more massive stars. We look then briefly at white dwarfs. Neutron stars are discussed separately in Sect.~\ref{neutron}.

\subsubsection{Vega \& Sirius}
\label{vega}
As far as the `non-magnetic' part of the population of intermediate-mass stars is concerned, an exciting  discovery has been the detection of magnetic fields in Vega and Sirius, the two brightest A stars in the sky. Zeeman polarimetric observations of these two stars have revealed weak magnetic fields: in Vega a field of $0.6 \pm 0.3$~G \citep{Lignieres_09, Petit_10} and in Sirius $0.2 \pm 0.1$~G \citep{Petit_11}. The field geometry is poorly constrained, except that the field should be structured on reasonably large length scales, as cancellation effects would prevent detection of a very small-scale field. Vega seems to have a strong ($\sim$~3~G) magnetic feature at its rotational pole. The existence or otherwise of time variability is unknown -- Petit et al.\ simply note that in Vega `no significant variability in the field structure is observed over a time span of one year'.  These two stars have similar mass but have significant differences: Vega is a rapidly rotating single star, and Sirius is a slowly rotating Am star which may well have accreted material from its companion \citep{deValBorro09}.  Given that we so far have detections in both stars observed, it seems very likely that the rest of the `non-magnetic' population also have magnetic fields of this kind. There are also theoretical grounds to expect this (see Section~\ref{failed}).

\subsubsection{Massive main sequence stars}\label{OBstars}

Direct detection via the Zeeman effect of magnetic fields in stars above around $6$ or $8\,M_{\odot}$ is significantly more challenging than in intermediate-mass stars. Firstly, because there are fewer lines in the spectrum, and since the signals from each line are normally added together with the LSD technique (see above) this leads to a smaller signal. Secondly, because there are various line-broadening mechanisms. Whilst the atmospheres of A and late B stars are very quiet, O and early B stars display a number of observational phenomena such as discrete absorption components, line profile variability, wind clumping, solar-like oscillations, red noise, photometric variability and X-ray emission. Much of this has to do with winds and wind variability, and much is not yet understood \citep[see e.g.][and refs therein]{Michaud13}. This complicates the life of the Zeeman observer, with the result that the detection limit on magnetic fields is perhaps 30 or 100 gauss, much higher than in A stars. See for instance \citet{Henrichs12} for a review of magnetism in massive stars.

\begin{figure}[htb]
\centerline{
\includegraphics[width=0.28\textwidth]{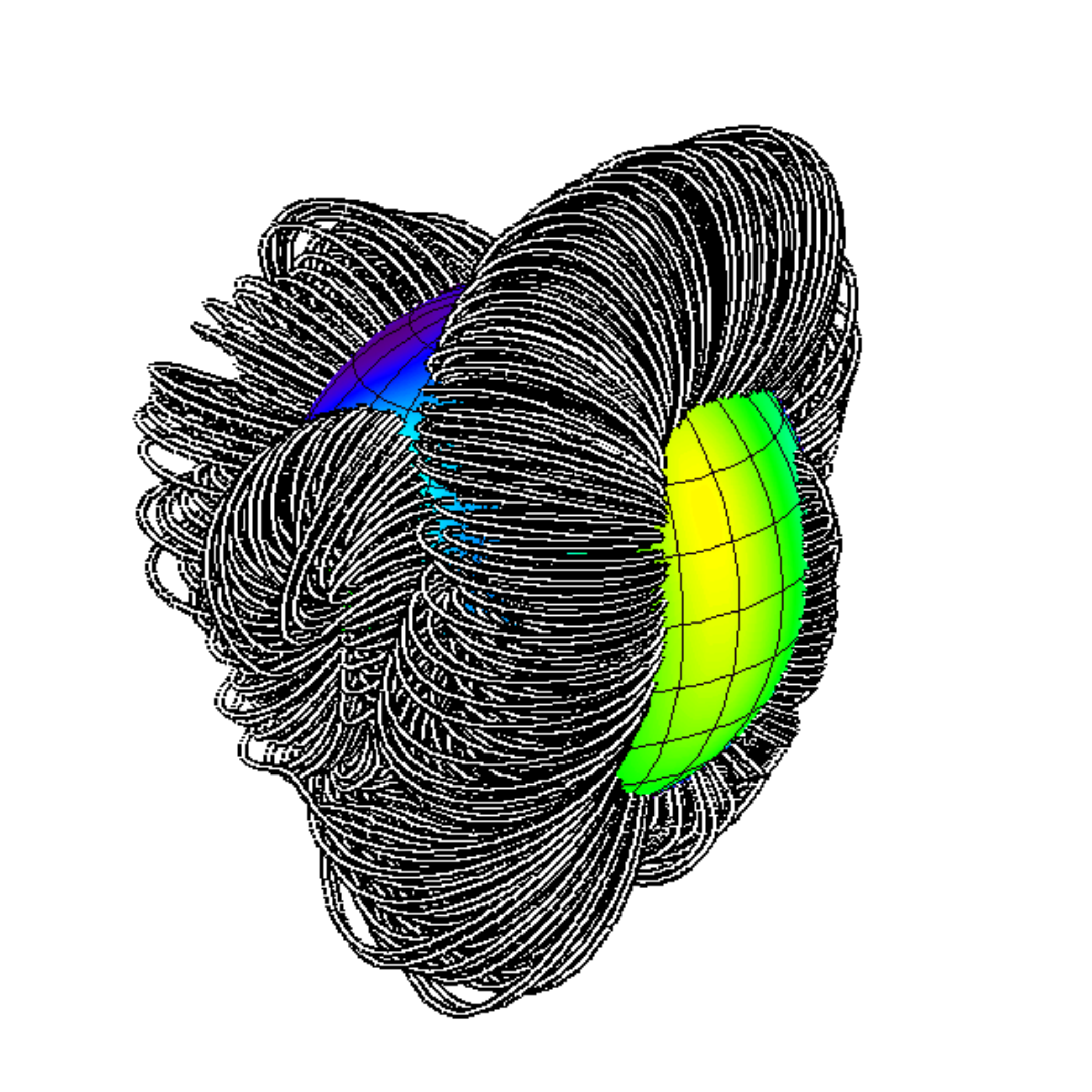}\hspace{-1cm}
\includegraphics[width=0.28\textwidth]{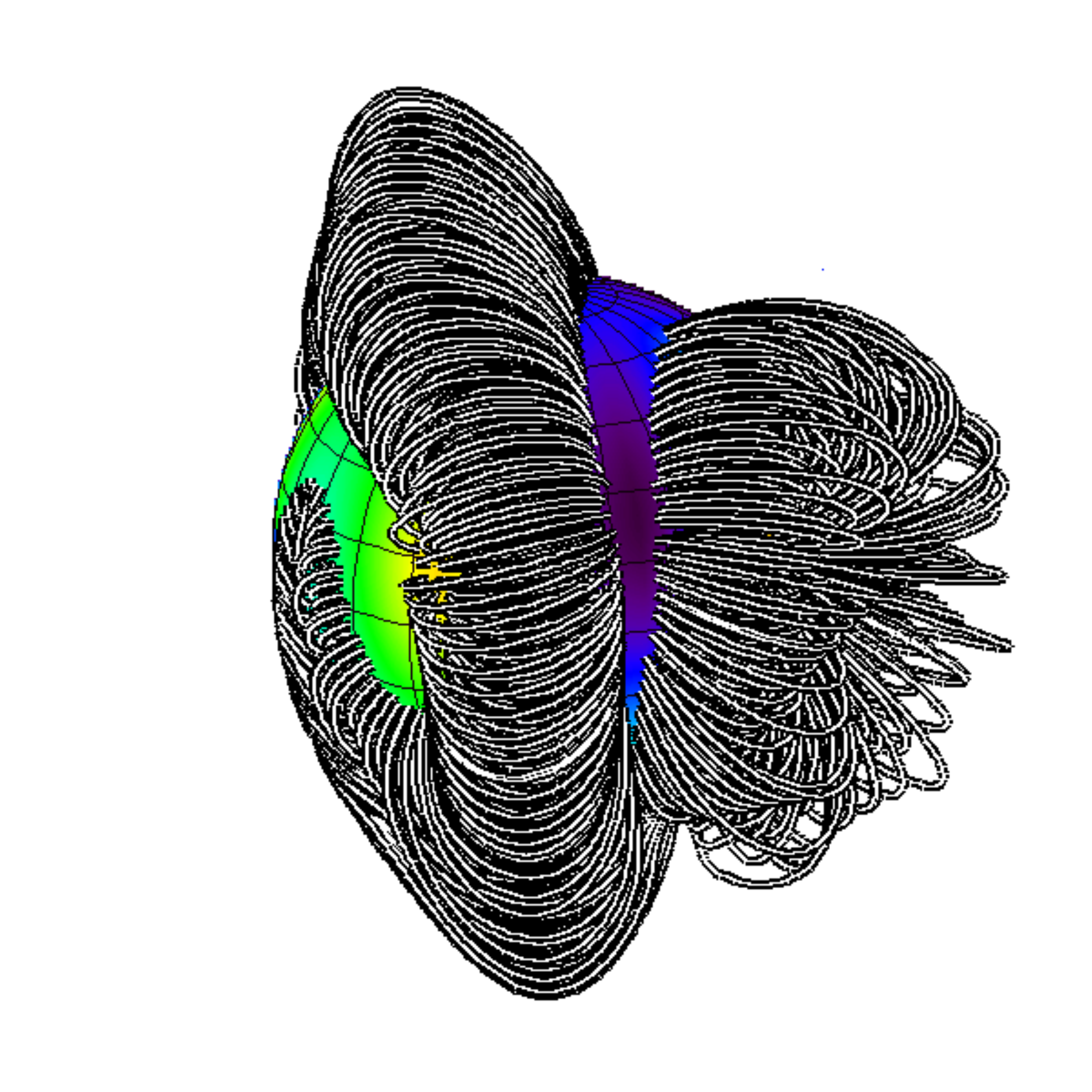}
}
\caption{The observed field geometry on the main-sequence B0 star $\tau$ Sco, at two rotational phases, using Zeeman-Doppler imaging. The paths of arched magnetic field lines are clearly visible; it seems likely they are associated with twisted flux tubes buried just below the surface -- see Section~\ref{nonax}. From \citet{Donati_06}. Observations of this star taken several years apart show the same topology, confirming a lack of variability also in stars with complex magnetic topologies \citep{DonatiL09}.}
\label{tauSco}
\end{figure}

The result of this is that magnetic fields were not detected in hot stars until relatively recently \citep{Henrichs_00}, but with the completion of recent surveys \citep[e.g., the MiMeS survey,][]{Wade_13} we have a much better picture of the incidence of large-scale magnetic fields in massive stars. It seems that around 7\% of the population host large-scale fields \citep{Petit_13}. The magnetic stars have fields of 300~G\,--\,10~kG, and a variety of geometries, as is found in the A stars. Whilst some of the magnetic stars have an approximately dipolar field, others are found to have a more complicated geometry -- see Figure~\ref{tauSco} for an example. In several other stars  similarly complex  magnetic fields have been found, dubbed the `$\tau$ Sco clones'.  Recently, it has been found that the magnetic flux in magnetic OB stars decays during the main-sequence (\citep{Fossati16}), the same as is found in Ap stars (Section \ref{herbig}). Possible explanations are discussed in Section \ref{decay}. 

It is tempting to conclude, therefore, that the phenomenon is simply a continuation of that seen in intermediate-mass stars, and indeed there are no particular theoretical reasons to think otherwise. The historical division between magnetism in intermediate- and high-mass stars is probably due only to the observational difficulty of observing the Zeeman effect in the hotter stars, and perhaps more importantly the fact that magnetic fields in hotter stars do not give rise to chemical peculiarities as seen in intermediate-mass stars, making them harder to identify. The reason for the lack of such peculiar abundances is presumably that the wind removes the outer layers before chemical-separating mechanisms have time to work. 

 The `non-magnetic' majority of the population may well have weaker and/or smaller-scale magnetic fields of some kind, and the various observational phenomena mentioned above could plausibly be at least partly the result of magnetic activity, not unlike that in the Sun -- see Section \ref{subsurf}. 

\subsubsection{White dwarfs}

The  stable fields observed in the subclass of magnetic white dwarfs (mWD) show parallels with those of Ap stars,  though it is not clear if this is more than a coincidence. The subclass is similarly small, the range in field  strength again large, ranging from a few times $10^3$ to $5 \times 10^8$~G.  As in the case of Ap stars, the large range in rotation periods of mWD is puzzling. Most rotate at periods from a fraction of a day to several weeks, but some of them have inferred periods as long as decades. For more on their properties we refer to reviews by \citet{Putney99} and \citet{Schmidt01}.

 White dwarfs have been important for the theory of fossil fields because, unlike early-type main-sequence stars, they do not contain any significant convective zones. Consequently, as far as the nature of their magnetic fields is concerned, there is essentially no alternative to a magnetic equilibrium. They are the same in this respect to the neutron stars, see Sect.~\ref{neutron}; indeed much of the physical processes discussed in that section also applies to white dwarfs. In Sect.~\ref{freeze} we discuss magnetic white dwarfs in the context of the origin of fossil fields in general. They also offer clues regarding the rotation of the cores of giant stars and angular momentum transport within stars in general (see Sect.~\ref{clurot}).

\bigskip
\section{\large Theory of Fossil Fields}
\label{sec:theory}

In radiative stars, the main challenges are to explain not just how these stars come to host magnetic fields at all, but to explain the large range in the magnetic and rotational properties of otherwise similar stars.  There are similarities in these respects between Ap stars, magnetic massive stars, magnetic white dwarfs (mWD), and to some extent also the neutron stars.

 An early idea triggered by these similarities, still widely referred to, is that of `flux freezing'. It does not attempt to explain the fields of Ap stars themselves, which is just taken as an observational given. Instead, it interprets the magnetic flux of white dwarfs and neutron stars as `inherited' from magnetic main sequence progenitors, the Ap, Bp and magnetic O stars. Ideas for the origin of these progenitors themselves are even less well developed. An obvious connection that must play a role is that with the magnetic fields observed in protostellar clouds, but this in itself does not tell us what to make of the puzzling range of field strengths in Ap stars, their low flux-to-mass ratio, and overall low frequency. This is discussed some more in Sect.~\ref{sec:variations}. 

\subsection{Flux freezing}
\label{freeze}

The large field strengths of mWD can be understood as a consequence of the greater compactness of WD compared to main sequence stars. An mWD may have a radius of $\sim5 \times 10^8$ cm, a factor $\sim100$ smaller than the core of a mid-A main sequence star. A popular hypothesis is that mWD inherited their fields from Ap progenitors. Under the assumption of `flux freezing', the magnetic field lines would stay anchored in the star during its evolution through the giant branch, the planetary nebula (PN) phase and the evolution of the remaining core to a WD. This would predict the field strength of an mWD to be $\sim100^2$ higher than the field of the MS stars it started with, which would put it in the observed range. 

The flux inheritance hypothesis is nevertheless somewhat dubious, as it implies that the loss of more than half of the star's mass in the PN phase does not significantly affect its magnetic flux. \blh Figures~\ref{fig:axisym_field} and \ref{fig:x-sec-diag} do suggest that some \bkh fraction of the surface flux of the Ap star \blh may possibly \bkh pass through the core that ends up to form the WD. This flux consists, however, of poloidal field lines. The subsurface torus surrounding this poloidal flux is what maintains the stability of the configuration. It would be lost during the ejection of the envelope in the PN phase. The \blh remaining poloidal configuration \bkh would be highly unstable via the Flowers--Ruderman mechanism (Sect.~\ref{stability}). Numerical results of this process suggest that not much of the poloidal field would survive \blh from such a configuration\bkh. 

Another way of looking at the field strengths is by comparing magnetic energy density ($B^2/8\pi$) with other energy densities in the star. \blh For a magnetic star to be bound, its magnetic energy must be less than the gravitational binding energy.  Per unit mass this energy \bkh is of the order of the central pressure $p_{\rm c}$. The central pressure in stars of mass $M$ and radius $R$ scales as $M^2/R^4$ \citep[e.g.,][]{Kippenhahn_12}.  Comparing the two predicts \blh a scaling of the maximum possible field strength, $B_{\rm max}\sim M/R^2$. \bkh In terms of the magnetic flux: $\Phi_{\rm max}\sim B_{\rm max}R^2\sim M$. As the mass of a WD is only a few times smaller than an Ap star, \blh the similarity of the maximum magnetic flux in the two cases \bkh is effectively the same as predicted by `flux freezing'.  Magnetic field strengths from a flux freezing argument can therefore not be distinguished from arguments relating field strength to energy density (which imply very different physics).

The maximum values of the surface field strength observed are of order $2 \times 10^4$, $5 \times 10^8$ and $10^{15}$~G for Ap stars, mWD and neutron stars, respectively. These numbers are a factor $10^{4}$--$10^{3}$ smaller, respectively than the maximum field strengths allowed by equipartition with the gravitational binding energy of the stars. \blh It is not clear if this factor, or its similarity between the different types of star, has a special meaning. In any case these maxima do not constrain theories for the origin of the field very much.\bkh

\subsection{The nature of magnetic fields in radiative stars}
\label{sec:fossils}

Since the discovery of magnetic fields in Ap stars, there have been two theories to explain their presence: the core-dynamo theory and the fossil-field theory. According to the former, the convective core of the star hosts a dynamo of some kind, which sheds magnetic field into the overlying radiative layer. This magnetic field then rises, under the action of buoyancy, towards the surface. The reason for this buoyant rise can be understood as follows. A magnetic feature (e.g., a flux tube) must be in total pressure equilibrium with its non-magnetised surroundings: the sum of gas, radiation and magnetic pressures inside the feature must be equal to the sum of gas and radiation pressures in the surroundings. To prevent buoyant rise, the density of the feature must be the same as that of the surroundings, which is only possible if the temperature is lower. This causes heat to diffuse into the magnetic feature, causing it gradually to rise.  However, it turns out that the timescale for this buoyancy process is longer than the main-sequence lifetimes of these stars unless very small flux tubes can be generated \citep{Parker79c,Moss89, MacGregorC03}. This does not agree with the observations, which suggest mainly large-scale structure at the surface. In addition, the retreating convective core (in mass coordinates) leaves behind a strong composition gradient in the radiative layers, enormously slowing down the escape process \citep{MacDonaldM04}. Also puzzling in this theory is the enormous range of field strengths between different stars which are predicted to have similar convective cores.

The fossil field theory, on the other hand, appears better able to explain the observations, in particular the large-scale geometry, and large field strengths. The basic idea is that instead of being continually regenerated in some ongoing dynamo process feeding off the star's luminosity, the field is in a stable equilibrium in a static radiative zone. In MHD, the magnetic field $\mathbf{B}$ evolves according to the induction equation
\beq\label{eq:induction}
\frac{\partial\mathbf{B}}{\partial t} = \curl(\mathbf{u} \times \mathbf{B} - \eta\curl\mathbf{B})\,,
\eeq
where ${\bf u}$ is the fluid velocity and $\eta$ is the magnetic diffusivity, the reciprocal of the electrical conductivity. In turn, the velocity field is related to the forces acting on the gas, i.e., the pressure gradient, gravity and Lorentz forces via the momentum equation:
\beq\label{eq:momentum}
\frac{{\rm d}\mathbf{u}}{{\rm d} t} = -\frac{1}{\rho}\del P + \mathbf{g} + \frac{1}{4\pi\rho}(\curl \mathbf{B}) \times \mathbf{B}\,,
\eeq
where $P$, $\rho$ and $\mathbf{g}$ are pressure, density and gravity. In an equilibrium unmagnetised star, the pressure gradient and gravity balance each other. Upon the addition of an arbitrary magnetic field, the Lorentz force will cause the gas to move at approximately the Alfv\'en speed $v_{\rm A}=B/\sqrt{4\pi\rho}$, and the system evolves on an Alfv\'en timescale $\tau_{\rm A}=R/v_{\rm A}$ (where $R$ is the radius of the star) which in a star with a 1~kG field is around ten years. Eventually one might hope to reach an equilibrium -- a so-called fossil field -- where the three forces balance each other and $\mathbf{u}=\mathbf{0}$, so that the first term on the r.h.s.\ of the induction equation disappears and the field evolves only on a diffusive timescale $R^2/\eta$. \citet{Cowling45} first realised that this timescale is of order $10^{10}$ years in the radiative core of the Sun and that a field in equilibrium there could therefore persist for the entire lifetime of the star. The magnetic equilibrium must also be stable, however, since instability time scales are of the order of the Alfv\'en timescale, which as mentioned above can be as short as a few years.

Much effort has been put into finding such stable equilibria, which was historically also motivated by the need for magnetic plasma confinement in nuclear fusion devices. Unfortunately, it turns out to be a difficult problem to solve, and with analytic techniques the existence of these stable equilibria was never convincingly demonstrated. This was historically a major weakness of the fossil field theory but, in light of the weaknesses of the core dynamo theory (including the discovery of magnetic fields with similar properties in white dwarfs, which contain no convective core), there was a widespread feeling that stable equilibria must exist.

Using analytic techniques, attempts to find such equilibria are split into two parts: first finding an equilibrium, and then checking its stability. The first step should not, \emph{prima facie}, represent any major problem. To see this (without an actual proof) consider the following. The magnetic field has two degrees of freedom (reduced from three by the ${\bm\nabla}\cdot\mathbf{B}=0$ constraint) and so the Lorentz force should also have two degrees of freedom. Ignoring thermal diffusion, the thermodynamic state of the gas also has two degrees of freedom. Where the magnetic field is weak -- in the sense that the plasma-$\beta=8\pi P/B^2\gg1$ -- equilibrium can be obtained by suitable adjustments of the thermodynamic variables, say pressure and entropy\epubtkFootnote{This is not possible in a hypothetical convectively neutral star where entropy is constant. In this case the field must be of the much more restricted class for which the Lorentz force has a potential. See also Sect.~\ref{recentanalytic}.}. Where the field strength is not small in this sense, for instance close to or above the surface of a star, an equilibrium must be close to a force-free configuration. We can therefore divide the star conceptually into two domains: the interior of the star where $\beta\gg1$ and  the allowed range of field geometries is not strongly constrained  (as long as equilibrium is the only concern and stability is ignored), and the exterior where $\beta\ll1$ and the field is close to force-free. Near the photosphere of the star the gas pressure scale height is much smaller than the length scales on which the magnetic field changes, and so for any Ap star the $\beta=1$ surface will be fairly close to the photosphere. In fact if the field strength at the photosphere is 300~G, as in the weakest-field Ap stars, then $\beta=1$ coincides with the photosphere; in Ap stars with stronger fields this surface is a little lower down.

\blh In addition to the  MHD processes above, field generation by microscopic processes have been considered: the `Biermann battery', \citet{Biermann}, cf.  \citet{Kulsrud05}, and the thermoelectric effect \citet{Dolg80}, \citet{Urp80}. In the stellar environment, both operate very slowly, as they depend on diffusion. For them to be effective, a very stable environment is required. In and A star, such an environment can be supplied by a stable fossil field, but appealing to this would obviously cause circular reasoning. The situation is better for neutron star crusts, where its solid state can provides a stable environment there. It has been concluded that the process is unlikely to be effective in this case, however, since the field produced in this way early in the life of the neutron star would decay when it cools (Blandford et al.\ \citeyear{Bland83}). \bkh

\subsection{Stability of fossil equilibria}
\label{stability}

To check the \emph{stability} of an equilibrium using analytic techniques is rather trickier. Most studies use the energy method of \citet{Bernstein_58} where the energy change in a configuration is calculated as a displacement perturbation is applied. If the energy change is positive for all possible perturbation fields, the configuration is stable. In the stellar context, this method was more successful in uncovering instabilities than in demonstrating stability. For reasons of tractability, almost all effort has been concentrated on axisymmetric equilibria. Tayler (\citeyear{Tayler73}, see also Spruit \citeyear{Spruit99}) looked at purely toroidal fields, that is, fields that have only an azimuthal component $B_\phi$ in some spherical coordinate frame $(r,\theta,\phi)$ with the origin at the centre of the star. He derived necessary and sufficient stability conditions for adiabatic conditions (no viscosity, thermal diffusion or magnetic diffusion). The main conclusion was that such purely toroidal fields are always unstable to adiabatic perturbations at some place in the star, in particular to perturbations of the $m=1$ form ($m$ being the azimuthal wavenumber). Numerical simulations have also been used to look at the stability of purely toroidal fields \citep{Braithwaite06b}, reproducing many of the analytic results.

\begin{figure*}[htb]
\vspace{2\baselineskip}
\center{\includegraphics[width=0.9\textwidth]{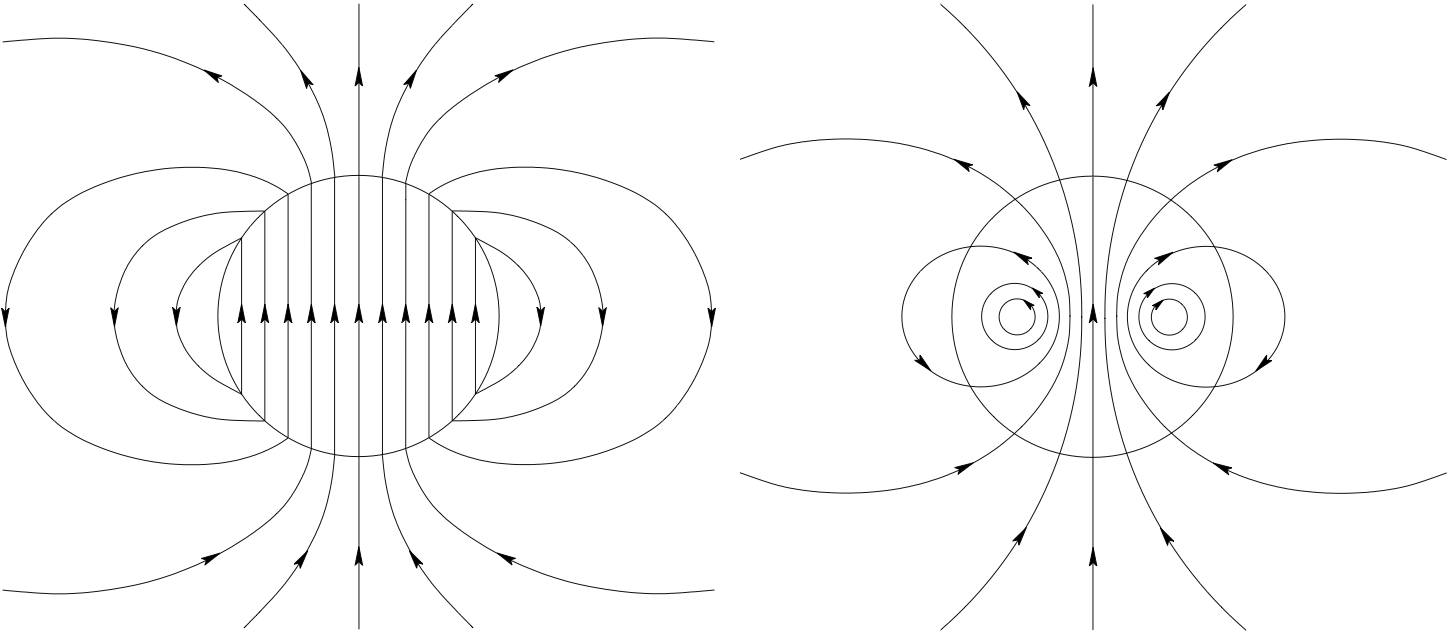}}
\caption{Poloidal field configurations. Left: all field lines close outside the star, this field is unstable by an argument due to \citet{FlowersR77}. For the case where some field lines are closed inside the star, instability was demonstrated by  \citet{MarkeyT73, MarkeyT74} and \citet{Wright73}.}
\label{fig:initfield}
\end{figure*}

The opposite case is a field in which all field lines are in meridional planes ($B_\phi=0$, see Figure~\ref{fig:initfield}). In subsequent papers \citet{MarkeyT73, MarkeyT74} and independently \citet{Wright73} studied the stability of axisymmetric poloidal fields in which (at least some) field lines are closed within the star (right-hand side of Figure~\ref{fig:initfield}). These fields were again found to be unstable.

A case not covered by these analyses was that of a poloidal field in which \emph{none} of the field lines close within the star. An example of such a field is that of a uniform field inside, matched by a dipole field in the vacuum outside the star (left-hand side of Figure~\ref{fig:initfield}). This case has been considered earlier by \citet{FlowersR77} who found it to be unstable, by the following argument. Consider what would happen to such a dipolar field if one were to cut the star in half (along a plane parallel to the magnetic axis), rotate one half by $180^\circ$, and put the two halves back together again. The magnetic energy inside the star would be unchanged, but in the atmosphere, where the field can be approximated by a potential field, i.e., with no current, the magnetic energy will be lower than before. This process can be repeated \emph{ad infinitum} -- the magnetic energy outside the star approaches zero and the sign of the field in the interior changes between thinner and thinner slices. \citet{Marchant_11} present a more rigorous analysis of this instability. The reduction of the external field energy is responsible for driving the instability. Since the initial external field energy is of the same order as the field energy inside the star, the initial growth time of the instability is of the order of the Alfv\'en travel time through the star, as in the cases studied by Markey \& Tayler and Wright. 

\citet{Prendergast56} showed that an equilibrium can be constructed from a linked poloidal-toroidal field, but stopped short of demonstrating that this field could be stable. Since both purely toroidal fields and purely poloidal fields are unstable, an axisymmetric stable field configuration, if one exists, should presumably be such a linked poloidal-toroidal shape. \citet{Wright73} showed that a poloidal field could be stabilised by adding to it a toroidal field of comparable strength. However, the result was again somewhat short of a proof.

These results were all valid only in the absence of dissipative effects and rotation.  The only case in which dissipative effects have been investigated in detail is that of a purely toroidal field  \citet{Acheson78}, (see also Hughes \& Weiss \citeyear{1995JFM...301..383H}, where it is found, for instance, that some combinations of diffusivities can have a destabilising effect on configurations which are stable to the non-diffusive Parker instability). These effects have still to be investigated in a more general geometry such as that of the mixed poloidal-toroidal equilibria. 

The effect of rotation was investigated by \citet{PittsT85} for the adiabatic case (i.e., without the effects of viscosity, magnetic and thermal diffusion). These authors reached the conclusion that although some instabilities could be inhibited by sufficiently rapid rotation, other instabilities were likely to remain, whose growth could only be slowed by rotation -- the growth timescale would still be very short compared to a star's lifetime.  Also, diffusion can reduce the stabilising effect of rotation \citep{Ibanez15}.  Importantly though, rotation does not introduce any new instabilities: a configuration which is stable in a non-rotating star should also be stable in a rotating star.

\subsection{Numerical results}
\label{numerical}

More recently, it has become possible to find stable equilibria using numerical methods. Various equilibria are found. The simplest equiibrium consists of a single flux tube around the magnetic equator of the star, surrounded by a region of poloidal field. More complex equilibria (Section~\ref{nonax}) can have more than one tube in various arrangements. From observations of magnetic A, B and O stars, we see that both the simple and the more complex equilibria do occur in nature. 

\begin{figure}[htb]
\centerline{
\includegraphics[width=0.45\textwidth]{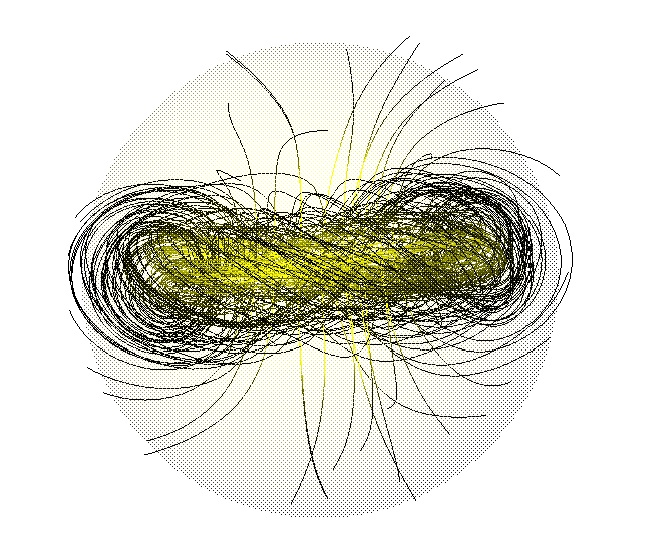}}
\centerline{
\includegraphics[width=0.45\textwidth]{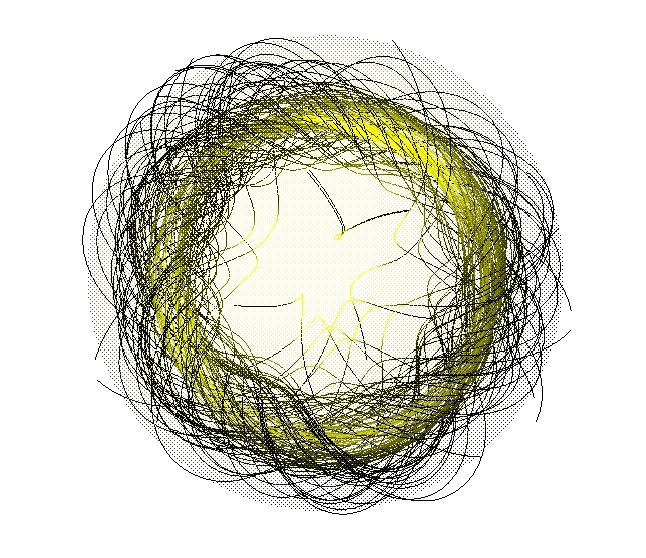}
}
\caption{The shape of the stable twisted-torus field in a star, viewed along and normal to the axis of symmetry. The transparent surface represents the surface of the star; strong magnetic field is shown with yellow field lines, weak with black. Figure from \citet{BraithwaiteN06}.}
\label{fig:axisym_field}
\end{figure}

Essentially, the numerical method consists in evolving the MHD equations in a star containing initially some arbitrary field. \citet{BraithwaiteS04} and \citet{BraithwaiteN06} modelled a simplified radiative star: a self-gravitating ball of gas with an ideal gas equation of state, ratio of specific heats $\gamma=5/3$ and a stratification of pressure and density as in a polytrope of index $n=3$, embedded in an atmosphere with low electrical conductivity. Over few Alfv\'en timescales, the field organises itself into a roughly axisymmetric equilibrium with both toroidal and poloidal components in a twisted-torus configuration, illustrated in Figure~\ref{fig:axisym_field}. This corresponds qualitatively to equilibria suggested by \citet{Prendergast56} and \citet{Wright73}.
  
The stability of these axisymmetric fields, and in particular the range of possible ratios of toroidal to poloidal field strength, was examined further by \citet{Braithwaite09} with a mixture of analytic and numerical methods. It was found that the fraction of energy in the poloidal component must satisfy $a(E/E_{\rm grav}) < E_{\rm p}/E \lesssim 0.8$ where $E$ and $E_{\rm grav}$ are the total magnetic energy and the gravitational energy, $E_{\rm p}$ is the energy of the poloidal field and $a$ is some dimensionless factor of order unity. \citet{Akguen_13} get the same results with more analytic methods. To give some numbers, for an A star the dimensionless factor $a\sim 15$ and the ratio $E/E_{\rm grav}$ is only about $10^{-6}$ even in the most strongly magnetic Ap star ($B\approx$ 30 kG), so that in this star we require for stability  $10^{-5} \lesssim E_{\rm p}/E \lesssim 0.8$. In a neutron star $a\sim 400$ \citep{Akguen_13}, and assuming a magnetar field strength of $10^{15}$~G, the condition is $4 \times 10^{-4} \lesssim E_{\rm p}/E \lesssim 0.8$. In stars with weaker fields of course the lower limit to the ratio $E_{\rm p}/E$ is even lower.
    
Physically, the upper limit comes from the need for a comparable-strength toroidal field to stabilise the instability of a purely poloidal field, and is in rough agreement with the result of \citet{Wright73}. The lower limit is different because the instability of a purely toroidal field, unlike that of a poloidal field, involves radial motion and so the stable entropy stratification has a stabilising effect, preferentially on the longer wavelengths which involve greater radial motion. The poloidal field stabilises preferentially the shorter wavelengths, and at sufficient poloidal field strength the two effects meet in the middle and all wavelengths are stabilised. The stable stratification is more effective if the field is weaker, hence the presence of the total field energy in the threshold. Since $E/E_{\rm grav}$ is always a very small number, only a relatively small poloidal field is required.  See \citet{Braithwaite09} for a more thorough explanation.

   This result is of particular interest in the context of neutron stars, where the deformation of the star from a spherical shape depends crucially on this ratio $E_{\rm p}/E$ since poloidal field makes the star oblate and toroidal field prolate. In the presence of a suitable mechanism to damp torque-free precession, a prolate star should `flip over' until the magnetic and rotation axes are perpendicular, which is the state in which the rotational kinetic energy is at a minimum, given a constant angular momentum. In this state the star emits gravitational waves \citep[see e.g.][]{Stella05}.

\subsection{Non-axisymmetric equilibria}
\label{nonax}

\begin{figure}[htb]
\centerline{
\includegraphics[width=0.23\textwidth]{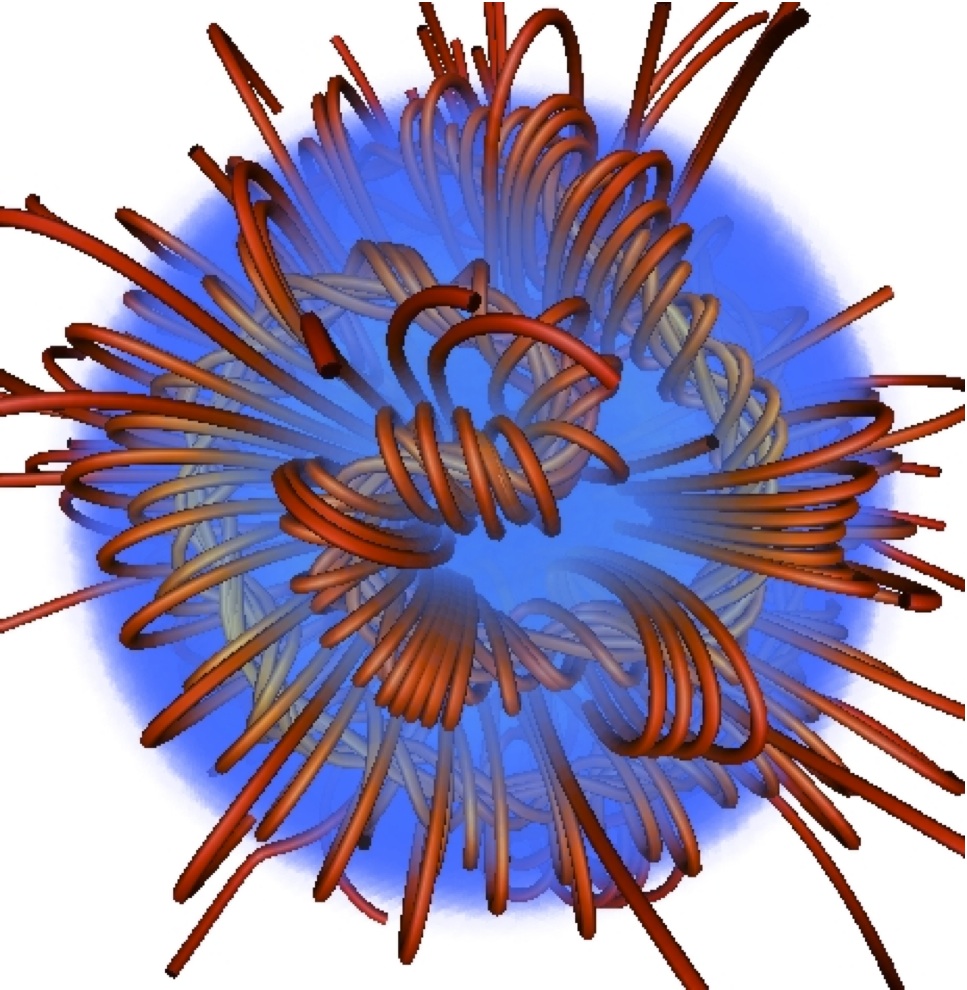}
\includegraphics[width=0.23\textwidth]{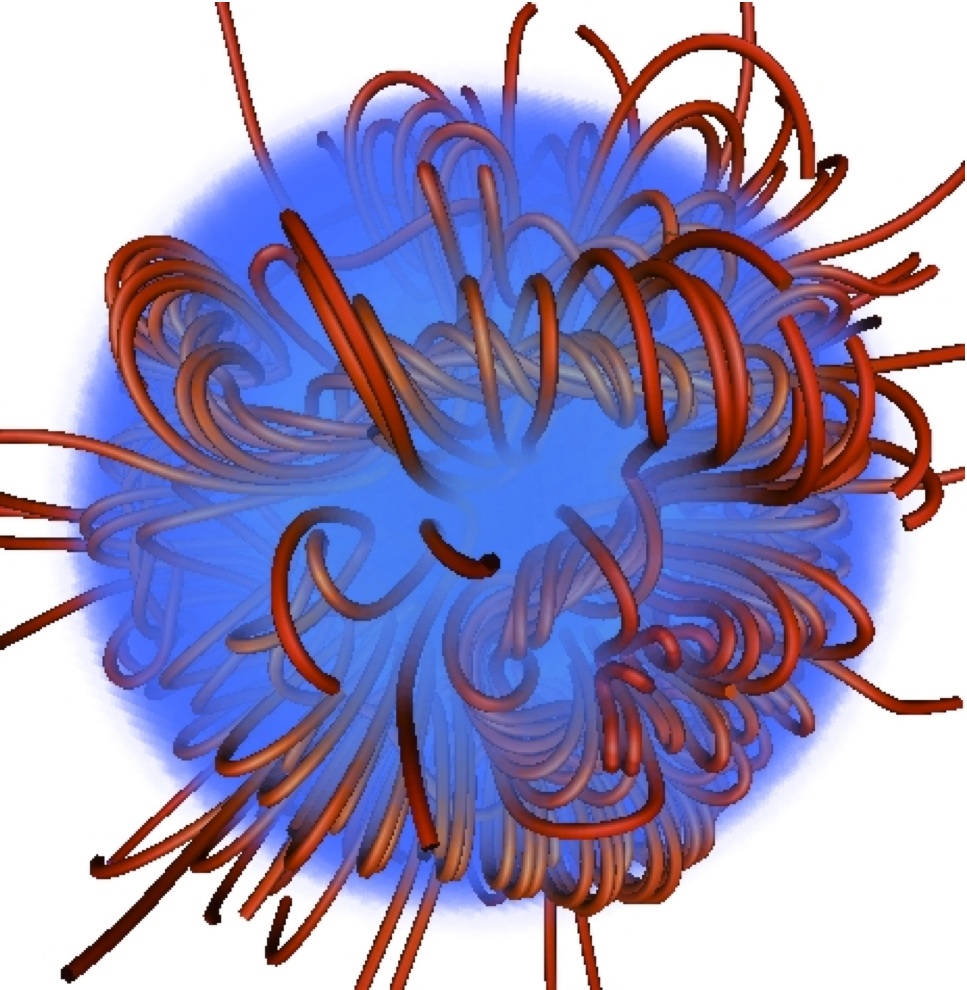}
}
\caption{A typical non-axisymmetric equilibrium as found in simulations, viewed from both sides of the star. This corresponds qualitatively to those observed on stars such as $\tau$ Sco (see Figure~\ref{tauSco}). Figure from \citet{Braithwaite08}.}
\label{fig:nonaxisym_field}
\end{figure}

From further simulations \citep{Braithwaite08} it became clear that depending on the initial conditions, a wide range of equilibria can form, including non-axisymmetric equilibria -- see Figure~\ref{fig:nonaxisym_field}. Crucial is the distribution of magnetic energy and the amount of flux passing through the stellar surface to the low-conductivity medium outside. It is important to note that during relaxation to equilibrium there is essentially no radial transport of gas and magnetic flux, fluid motion being confined to spherical shells, so the total unsigned flux through any spherical shell $\oint |\mathbf{B}\cdot \rd \mathbf{S}|$ can only fall. Therefore an initial field which is buried in the interior of a radiative star or zone evolves into a similarly buried equilibrium. It turns out that in this case, an approximately axisymmetric field forms. At the opposite end of the spectrum, an initial field with a flat radial field-strength profile with a finite amount of magnetic flux at the surface evolves into a non-axisymmetric equilibrium -- see Figure~\ref{fig:equil_seq}. It seems that both axisymmetric and non-axisymmetric equilibria do form in nature, and both types can be found amongst A, B and O main-sequence stars as well as amongst white dwarfs: see Figures~\ref{a2cvn} and \ref{tauSco}. 

\begin{figure}[hb]
\centerline{\includegraphics[width=0.45\textwidth]{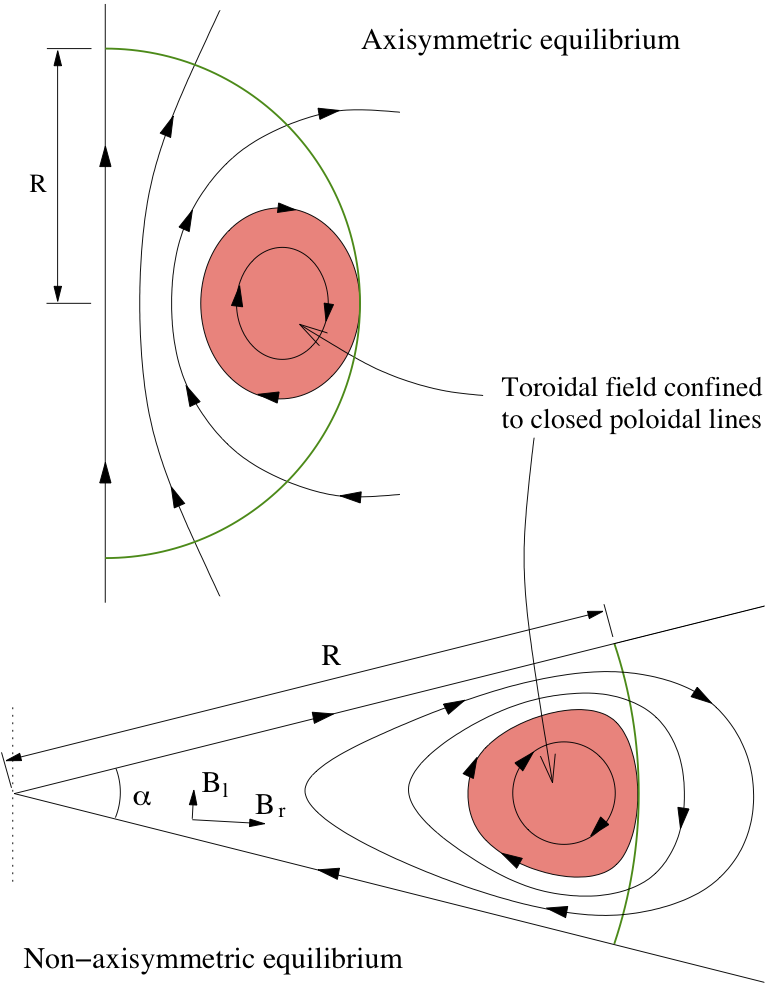}}
\caption{Cross-sections of magnetic equilibria, both of which contain a twisted flux tube surrounded by a volume containing just poloidal field. The stellar surface is shown in green and poloidal field lines (in black) are marked with arrows. The toroidal field (direction into/out of the paper, red shaded area) is confined to the poloidal lines which are closed within the star. (Toroidal field outside this area would unwind rather like a twisted elastic band that is not held at the ends.) Above: the axisymmetric case where the flux tube lies in a circle around the magnetic equator (corresponding to Fig.~\ref{fig:axisym_field}).   Below: a narrower flux tube (corresponding to Fig.~\ref{fig:nonaxisym_field}). In this case, the flux tube must be longer than in the axisymmetric case in order to occupy the whole stellar volume; it meanders around the star in an apparently random fashion, and there may also be two or more such tubes.  Figure from \citet{Braithwaite08}.}
\label{fig:x-sec-diag}
\end{figure}

\begin{figure}[ht]
\centerline{
\includegraphics[width=0.154\textwidth]{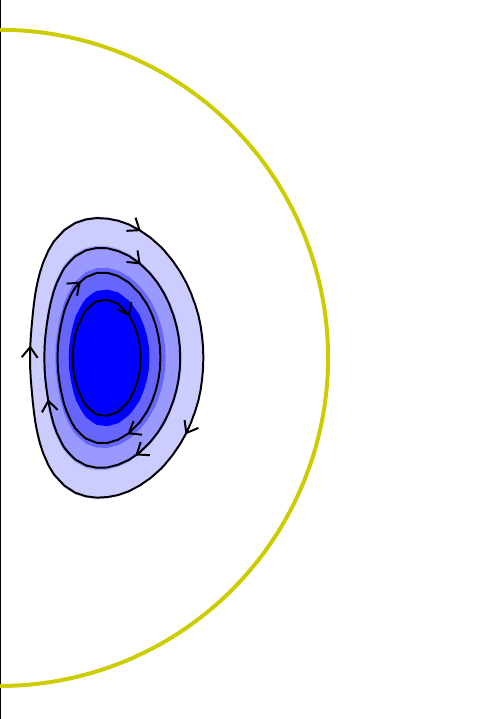}
\includegraphics[width=0.14\textwidth]{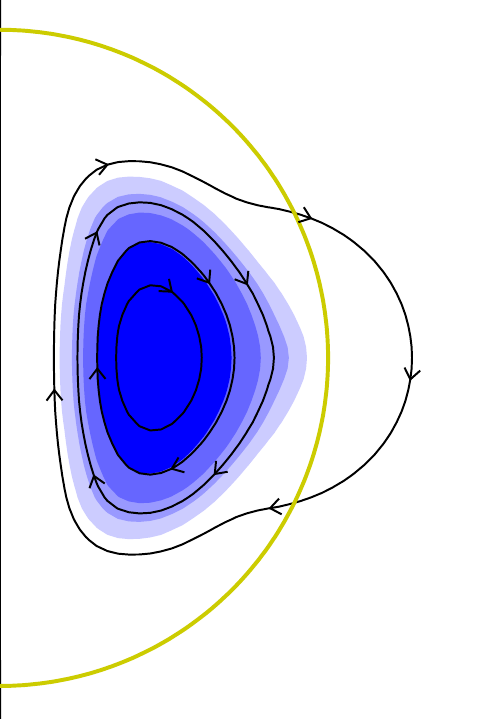}
\includegraphics[width=0.14\textwidth]{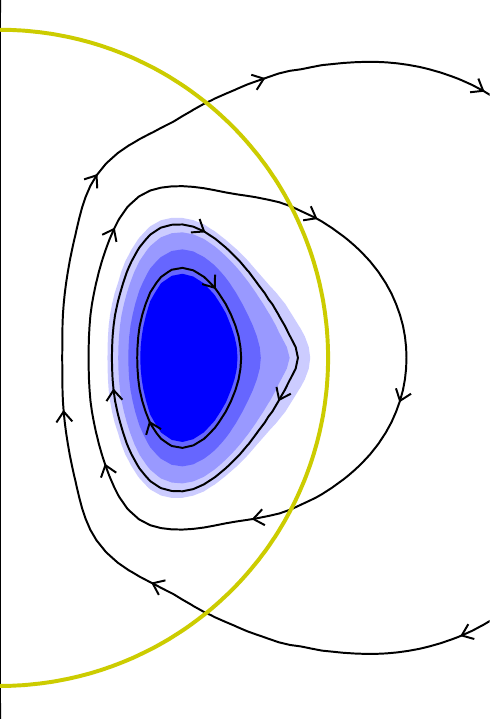}
}
\vskip 1\baselineskip\centerline{
\includegraphics[width=0.23\textwidth]{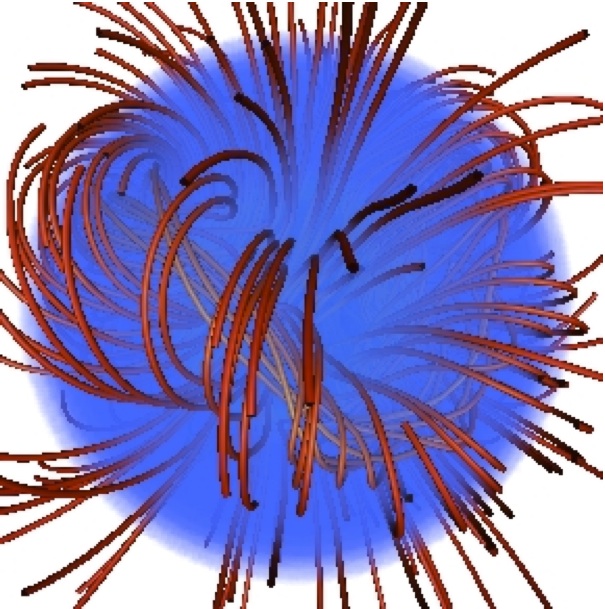}
\includegraphics[width=0.23\textwidth]{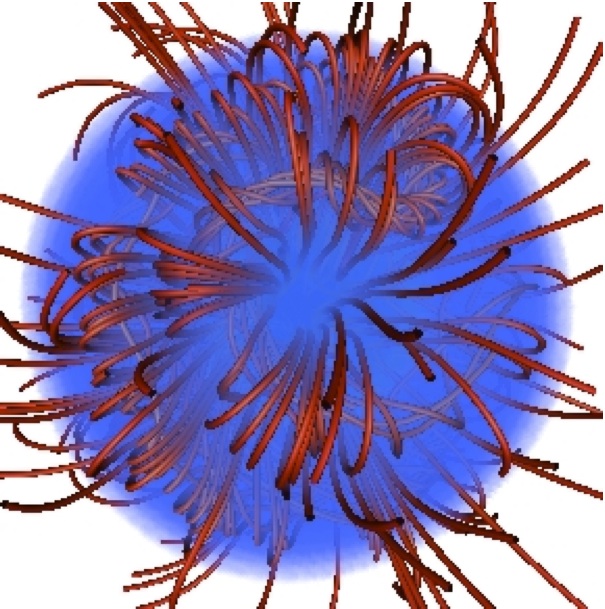}
}
\caption{A sequence of equilibria resulting from initial  conditions  with different degrees of central concentration of the magnetic energy -- above left, centrally concentrated initial conditions; above right, medium concentration (still leading to a roughly axisymmetric equilibrium); and below, flatter distributions; below right the radial energy distribution is completely flat. For the three axisymmetric equilibria, the (relatively small) non-axisymmetric component is ignored: plotted are contours of the flux function of the poloidal field, which are also poloidal field lines, and the shading represents the toroidal field multiplied by the cylindrical radius. For the two non-axisymmetric equilibria shown, field lines are plotted in red and the surface of the star is shaded blue. Note that these two equilibria have flux tube(s) of different widths (corresponding to the angle $\alpha$ in Figure~\ref{fig:x-sec-diag}). Figures from \cite{Braithwaite08, Braithwaite09}.}
\label{fig:equil_seq}
\end{figure}

The geometries of these various equilibria have one feature in particular in common, namely that they can be thought of in terms of twisted flux tubes surrounded by regions of purely poloidal field. The simple axisymmetric equilibrium can be thought of as a single twisted tube wrapped in a circle (a `twisted torus'); passing through this circle are poloidal field lines which pass through the stellar surface. The more complex equilibria contain one or more twisted flux tubes meandering around the star in apparently random patterns; these do not touch each other but are surrounded by regions where the field is perpendicular to the flux-tube axis. In the equilibria found thus far, the meandering is done at roughly constant radius a little below the surface. Equilibria where the flux tubes do not lie at constant radius seem plausible but it also seems plausible that they are difficult to reach from realistic initial conditions, especially in view of the restriction of motion to spherical shells. Figure~\ref{fig:x-sec-diag} shows cross-sections of both the axisymmetric and more complex equilibria. The toroidal field is confined to the region where the poloidal field lines are closed within the star -- this region resembles a twisted flux tube -- toroidal field outside this region would `unwind' through the atmosphere and disappear.

\begin{figure}[htb]
\vskip 0.5\baselineskip
\centerline{\includegraphics[width=0.5\textwidth]{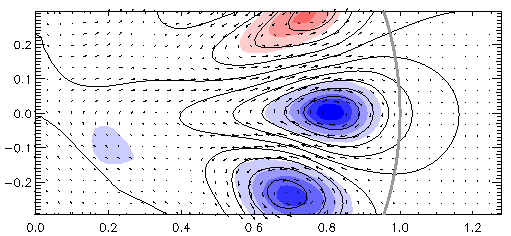}}
\caption{Cross-section of a non-axisymmetric equilibrium. The curved grey line towards the right is the surface of the star and the centre of the star is on the left; the coordinate system used for the plot is cylindrical. The blue/red shading represents the toroidal field component (out of/into the page) multiplied by the cylindrical radius. The poloidal component (in the plane of the page) is represented by the arrows and by contours of its scalar potential, calculated by ignoring spatial derivatives perpendicular to the plane. In fact it can be seen that the arrows are very nearly parallel to the contours of the scalar potential, showing that the length scale of variation in the direction perpendicular to the page is much greater than the length scale in the plane of the page, i.e., that the flux tubes meander around the star over scales much greater than their width. Note also that neighbouring flux tubes can have toroidal field in either the same or opposite directions: since the toroidal field is absent from the space between the tubes, one tube is not `aware' of the direction of toroidal field in its neighbours, so the equilibrium and stability properties are independent of its direction. Figure from \citet{Braithwaite08}.}
\label{fig:x-sec}
\end{figure}

Qualitatively the difference in equilibria resulting from differing radial energy distributions in the initial conditions can be understood in the following way. If the magnetic energy distribution is flatter, the axis of the circular flux tube in an axisymmetric equilibrium will be closer to the surface of the star and the bulk of the poloidal flux goes through the surface, leaving a smaller volume in which the toroidal component can reside. Beyond some threshold, this means that the toroidal field is not able to fulfill its role in stabilising the instability seen in purely poloidal fields, and the field buckles, the flux tube first attaining a shape reminiscent of the seam on a tennis ball and then something more complex. This lengthening of the flux tube (at constant volume) increases the toroidal (axial) field strength and decreases the poloidal field strength, until stability is regained. The same process can also be thought of in terms of the tension in a flux tube, which is equal to $T=(2B_{\rm ax}^2-B_{\rm h}^2)a^2/8$ where $a$ is the radius of the tube, and $B_{\rm ax}$ and $B_{\rm h}$ are the r.m.s.\ axial (toroidal) and hoop (poloidal) components of the field \footnote{In the literature one often finds that the factor of 2 inside the brackets is missing.}. In the simple axisymmetric equilibrium, the tension in the tube must be positive. If the tube is too close to the surface, there is not enough space for the toroidal field and the tension can become negative, causing the tube to buckle into a more complex shape until the lengthening of the tube causes the tension goes to zero. A fuller discussion is given in \citet{Braithwaite08}.
Figure~\ref{fig:x-sec} shows a cross-section of a non-axisymmetric equilibrium found in a simulation. Note how in this figure, as well as in the cross-sections of axisymmetric equilibria in Figure~\ref{fig:equil_seq}, we can see that the poloidal field is parallel to contours of the toroidal field multiplied by cylindrical radius. This condition was derived analytically in the case of axisymmetric equilibria from the need for the toroidal part of the Lorentz force to vanish \citep{Mestel61, Roxburgh66}, and the same condition applies in non-axisymmetric equilibria.

\subsection{Recent analytic work}\label{recentanalytic}

There has recently been renewed interest in finding stable equilibria with analytic and semi-analytic methods. In stably-stratified stars, the range of possible equilibria is large or perhaps even essentially unlimited within the zero-divergence constraint (as described above in Section~\ref{sec:fossils}), although of course only some subset will be stable. To find analytic equilibria, various assumptions and choices are made to constrain the solutions. For instance, all analytic works so far have assumed axisymmetry.

Furthermore, except for a few recent papers (see below), all works have assumed a barotropic equation of state \citep[e.g.,][]{Yoshida_06, Ciolfi_10, Lyutikov10, Fujisawa_12}, which represents a significant restriction on the range of equilibria, as explained in Section~\ref{sec:fossils}.  The barotropic assumption is also quite artificial: regions in stars are either stably stratified or convective. In a convective region in a non-rotating star the stratification is nearly barotropic, but convective flows are incompatible with static field configurations. In a rotating star, convective zones are not even nearly barotropic and thermal winds arise.  The parameter space in between convection and stable stratification is a `set of measure zero',  of rather academic interest. Pursuing the line of thought nevertheless, taking  the momentum equation~(\ref{eq:momentum}), setting the left-hand side to zero (as in equilibrium) and taking the curl, gives
\beqa\label{eq:zerocurl}
\mathbf{0}& = \curl \left(-\frac{1}{\rho}\del P\right) + \curl\mathbf{g} + \\
&\curl\left(\frac{1}{4\pi\rho}(\curl \mathbf{B}) \times \mathbf{B}\right).
\eeqa
Gravity being a conservative force, the curl of $\mathbf{g}$ vanishes. With a barotropic equation of state $\rho=\rho(P)$, the pressure gradient force $-(1/\rho)\del P=-\del h$ where $h=h(P)$ is a new variable; the curl of this force is then obviously zero. We are left with a condition of curl-free Lorentz acceleration:
\beq
\curl[(\curl\mathbf{B}) \times \mathbf{B}/\rho]=\mathbf{0}.
\eeq
In other words, the Lorentz acceleration must have vanishing curl because it is balanced by two other forces with vanishing curl. Thus the barotropic equation of state imposes a restriction on the equilibrium which does not exist with an equation of state $\rho=\rho(P,T)$. 
 Assuming axisymmetry and a barotropic EOS, finding an equilibrium is a matter of solving the Grad-Shafranov equation, derived from the equilibrium condition that comes from setting the LHS of (\ref{eq:momentum}) to zero.
 
\begin{figure*}[htb]
\centerline{\includegraphics[width=0.372\textwidth]{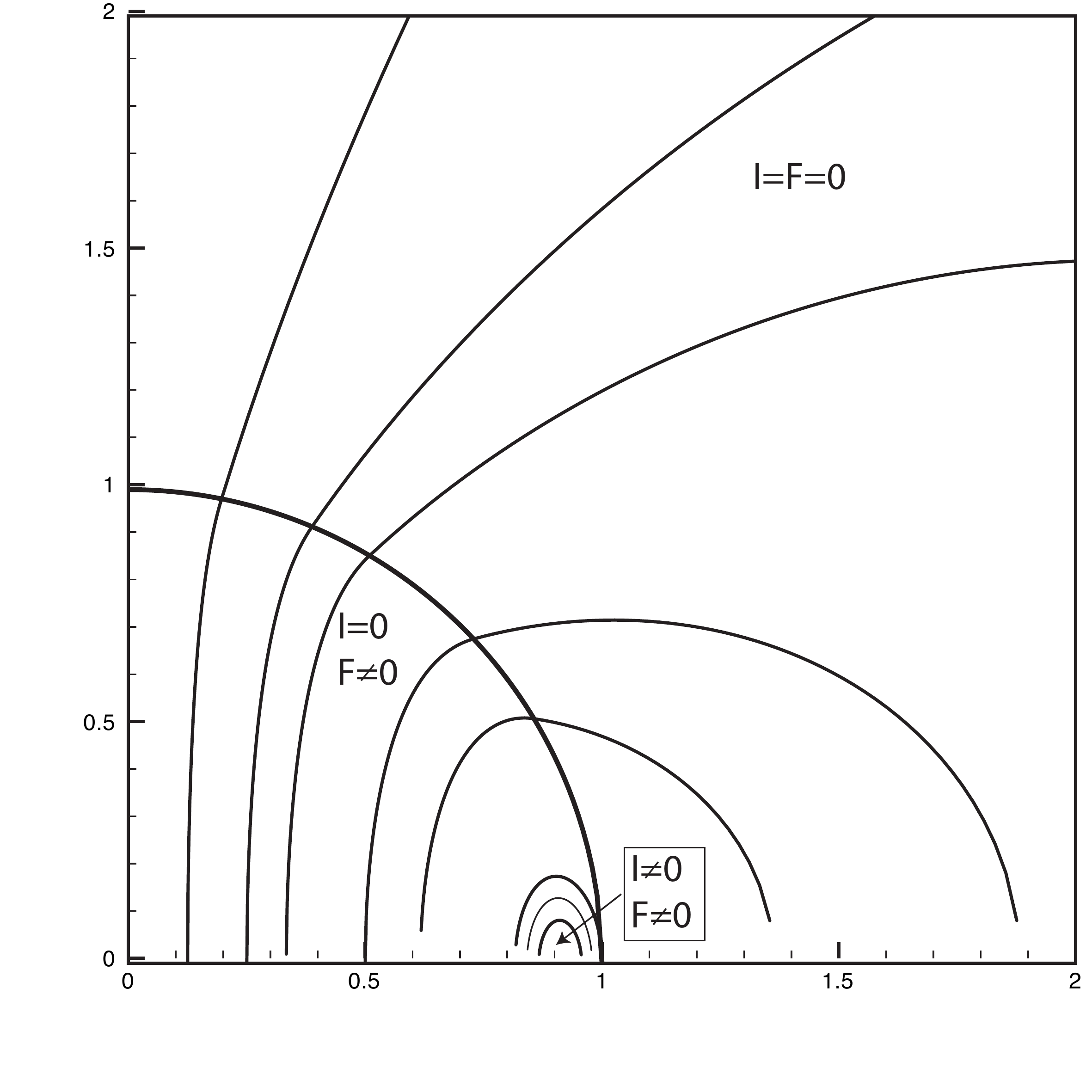}
\includegraphics[width=0.52\textwidth]{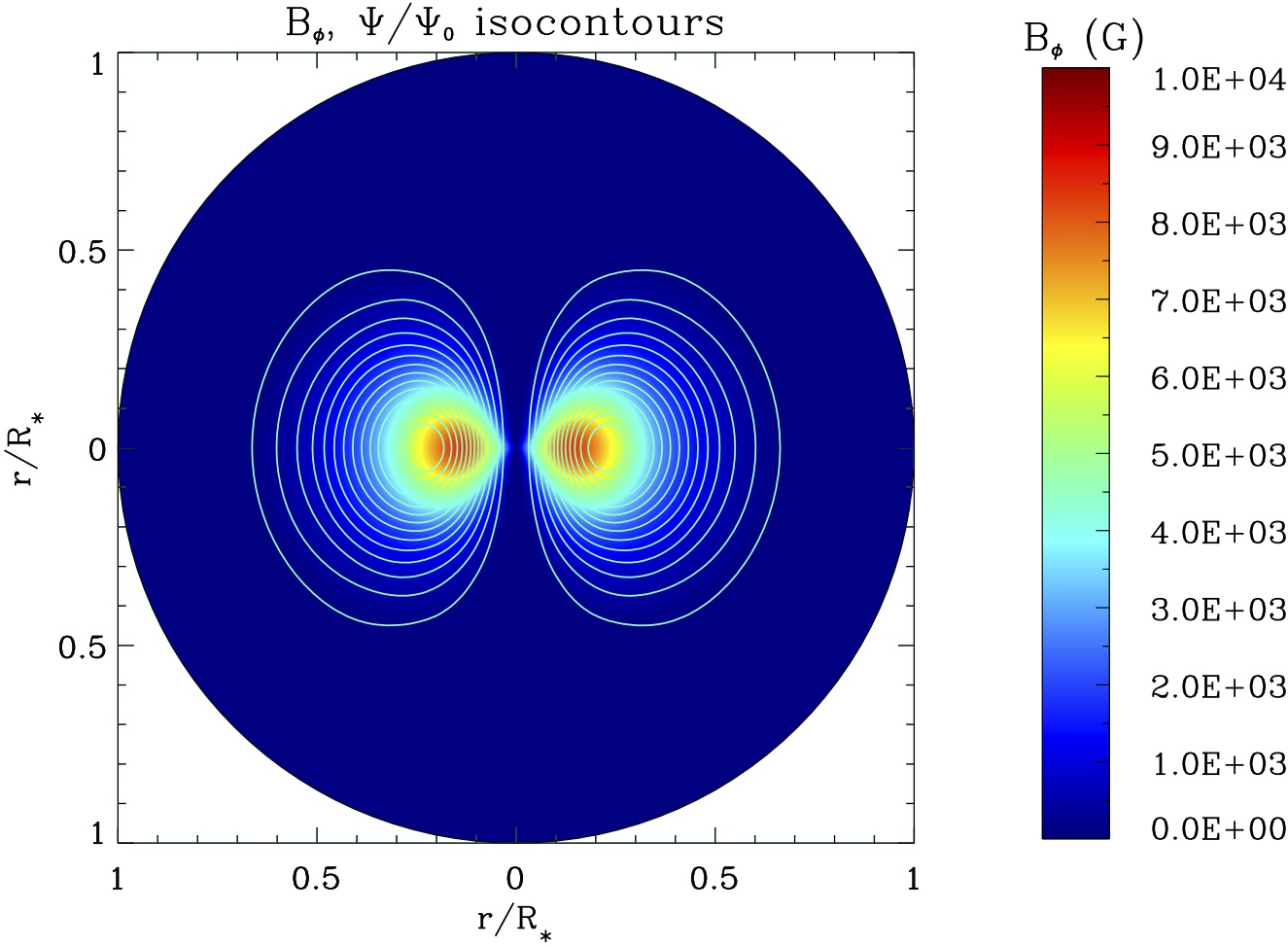}}
\caption{ Magnetic equilibria found analytically. On the left, an equiibrium from \citet{Lyutikov10} where the toroidal field occupies a volume around the neutral line, and on the right, an equiibrium from \citet{Duez_10} where the magnetic field is confined entirely to the star. Both studies use a barotropic equation of state and Newtonian gravity. }
\label{fig:analytic}
\end{figure*}

Most works have constructed simple twisted-torus equilibria of the form in the upper part of Fig.~\ref{fig:x-sec-diag}, with some more complex axisymmetric equilibria with two or more tori. Some examples are shown in Fig.~\ref{fig:analytic}.

A common feature of the simple equilibria is that the volume containing closed poloidal lines and toroidal field is rather small, the neutral line (where the poloidal field vanishes) being close to the stellar surface \citep[e.g.][see fig.~\ref{fig:analytic}]{Lyutikov10}. This is possibly something to do with the requirement that the equilibrium can be expressed mathematically as the sum of low-order spherical harmonics -- indeed the interior field is often matched to a pure dipole field outside the star. Physically, an equilibrium can form out of an initial field which has most of its flux buried away from the stellar surface and so more deeply buried equilibria must be possible, and such buried equilibria are found in simulations (see Fig.~\ref{fig:equil_seq}). Alternatively this may have to do with the use of a barotropic EOS. 

Other authors \citep[e.g.,][]{Haskell_08, Duez_10} have found equilibria where the poloidal field does not penetrate through the stellar surface at all, so that toroidal field occupies the entire volume of the star.  Strictly speaking,  these are of course of academic interest if the application is to objects with an observed field at the surface,  but equilibria in nature may have only a modest fraction of their flux passing through the surface, if the magnetic energy is relatively concentrated in the middle of the star. 

 Several models \citep[inter alia]{BroderickN08, 2004ApJ...600..296I, 2008MNRAS.385.2080C} include a current sheet at the stellar surface. This means that the Lorentz force is infinite, which is impossible in nature, especially at a location where the fluid density goes to zero. In for instance \citet{BroderickN08} the current sheet is an unavoidable consequence of the assumption made that the field in the interior is force-free (i.e., $\mathbf{j \times  B}=\mathbf{0}$).  To see this, recall a classical result:  the `vanishing force free field theorem'. It says that \emph{a magnetic field which is force free everywhere in space vanishes identically}. Force-free fields can exist only by virtue of a surface where the Maxwell stress in the field is taken up.  (For the 3-line proof  see \citealp{Roberts67}, also reproduced in \citealp{Spruit13}). The Lorentz force density (the divergence of this stress) may vanish; the stress itself however vanishes only when the field itself vanishes.

Whilst most studies use the approximations that (a) the field is too weak to have a significant effect on the shape of the star, (b) the star is not rotationally flattened and (c) general relativistic effects can be ignored, some authors drop these assumptions. For instance \citet{Fujisawa_12} have a rotating model with a strong magnetic field and \citet{Ciolfi_09, Ciolfi_10} include general relativity (relevant in the context of neutron stars). Perhaps reassuringly, including these effects does not seem to result in any qualitative difference to the geometry of the equilibria found. Also interesting in the context of neutron-star magnetic equilibria is the Hall effect, which is essentially an extra term in the induction equation~(\ref{eq:induction}) to account for the velocity difference between the electron fluid, to which the magnetic field is ÔfrozenÕ, and the bulk flow: see, e.g., \citet{Gourgouliatos_13}.

\subsubsection{Stability}\label{stability2}

Having found an equilibrium, one needs to check its stability. Some authors have tried to ensure the stability of their equilibria by finding an energy minimum with respect to some invariants. A popular invariant to use is magnetic helicity $H=\int \mathbf{A}\cdot\mathbf{B} \,\rd V$ where $\mathbf{A}$ is the vector potential defined by $\curl\mathbf{A}=\mathbf{B}$, which is approximately conserved in a highly-conducting fluid. For instance, \citet{Ciolfi_09} use a maximum helicity argument to find the ratio between poloidal and toroidal field strengths. This approach was also used by, for instance, \citet{BroderickN08} and \citet{DuezM09}, who also introduce further higher-order invariants. The efficacy of these higher-order invariants is not completely established; indeed it is not even certain to what degree helicity should be conserved when a significant fraction of the flux passes through the stellar surface, above which helicity conservation breaks down.

As mentioned above, almost all of the analytic work has assumed a barotropic equation of state, such as would apply to a convective star. This is done mainly for mathematical convenience; however it may be of physical relevance in neutron stars, where beta processes eliminate the stable stratification over some timescale 
 \citep[see][]{Reisenegger09}. In a barotropic star, one can imagine that an equilibrium magnetic field might be able to `hold itself down' against magnetic buoyancy somehow by means of magnetic tension. This would presumably only work if the magnetic field has organised itself globally, with buoyancy acting in opposite directions on opposite sides of the star. Physically relevant of course is not just the question of whether equilibria are possible in principle, but whether they can actually form from realistic initial conditions. It was suggested by \citet{Braithwaite12} that buoyancy, acting on a disorganised initial magnetic field, pushes the magnetic field to the surface faster than it is able to organise itself into an equilibrium. Using star-in-box numerical methods 
 \citet{Mitchell14} have also investigated this issue. They use an ideal gas EOS with a heating/cooling term which maintains a uniform entropy in the star's interior. This removes the stabilising effect of the stratification and so is equivalent to a barotropic equation of state, but is numerically easier to implement. It is found that the small-scale random magnetic field which does evolve into a stable equilibrium in a stably stratified star does \emph{not} reach an equilibrium in an isentropic star.

It seems possible in principle however that an equilibrium could form while the neutron star is non-barotropic, and could somehow adjust to the changing structure of the star, remaining in a quasi-statically evolving stable equilibrium as the star becomes barotropic. 
\citet{2015MNRAS.447.1213M} construct various mixed toroidal-poloidal axisymmetric torus fields in a barotropic star, and use numerical methods to test their stability in the linear regime. All of the equilibria constructed prove to be unstable, and the authors tentatively suggest that stable equilibria might not exist in barotropic stars. 
The decay involves global-scale modes and happens on an Alfv\'en timescale. The condition for instability is simply that the buoyancy frequency is less than the inverse Alfv\'en timescale. In light of these results, it is probably safe to assume that barotropic stars cannot host MHD equilibria. Since neutron stars do host magnetic fields, it seems they cannot be perfectly barotropic, or that the crust plays an important role.

\bigskip
\subsection{The `failed fossil' hypothesis}
\label{failed}

Above it was described how an arbitrary magnetic field evolves towards an equilibrium on an Alfv\'en timescale $\tau_{\rm A}$, but the discussion ignored any (solid-body) rotation of the star, which adds a Coriolis acceleration $-2\,\boldsymbol{\Omega} \times \mathbf{u}$ to the momentum equation~(\ref{eq:momentum}). As a general principle in MHD, it can be shown \citep{FriemanR60} that if the rotation is slow, such that $\Omega\tau_{\rm A}\ll 1$, the rotation has little effect on the evolution and stability of magnetic fields; this can also be seen from a comparison of the relative sizes of terms in the momentum equation. If however $\Omega\tau_{\rm A}\gg 1$ then one expects the Lorentz force to be balanced not by inertia but by the Coriolis force, and a comparison of the sizes of these two terms shows that the evolution timescale is no longer equal to the Alfv\'en timescale $\tau_{\rm evol}\sim\tau_{\rm A}$ but instead $\tau_{\rm evol}\sim\tau_{\rm A}^2\Omega$. This general principle is seen in various contexts in MHD; an early reference is \citet[Sections~84 and after, and Figure~101]{Chandrasekhar61} and it is seen for instance in the growth rates of various instabilities such as the Tayler instability \citep{PittsT85,Ibanez15}. The idea of the Coriolis force, rather than inertia, balancing whatever is driving fluid motion, is of course also well known from atmospheric physics (quasi-geostrophic balance, see, e.g., \citealp{Pedlosky82}).

As described by \citet{BraithwaiteC13}, this principle should also apply to the relaxation of an arbitrary initial magnetic field in a star towards an equilibrium. In a non-rotating star the field evolves on a timescale $\tau_{\rm A}$, its energy falling as it does so. If its energy falls by a large factor as it does so (as seems likely; see discussion in Section~\ref{sec:variations}) then $\tau_{\rm A}$ will increase by a large factor as the relaxation progresses, and the time taken to reach equilibrium can be approximated simply to the Alfv\'en timescale at equilibrium. In a rotating star the evolution timescale becomes $\tau_{\rm A}^2\Omega$, and increases in the same way as relaxation progresses. Putting in some numbers, it takes one year to form an equilibrium of strength 10 kG in a non-rotating star, but $5 \times 10^{11}$ yr to form an equilibrium of 1 G in a star rotating with a period of 12 hours. It may be that in stars lacking strong fields, the magnetic fields are still evolving `dynamically' towards a weak equilibrium. During this time, the evolution timescale $\tau_{\rm A}^2\Omega$ will be roughly equal to the age of the star, and so given the age and rotation period of a star, it should be possible to calculate the field strength. With Vega, assuming a rotation period of 12~hours and an age of $4 \times 10^8$ years, this argument predicts a field strength of 20~gauss; for Sirius it predicts about 7~gauss. It may well be then that Vega and Sirius (see Section~\ref{failed}) contain a non-equilibrium fields undergoing dynamic evolution. It is easy to reconcile the predictions with observed field strengths of 0.6 and 0.2~G: the observations will underestimate the strength of a smaller-scale field, and one would naturally expect the field strength on the surface to be lower than the volume-average predicted by the theory.
 
Finally, if it can be assumed in all stars which hosted a pre-main-sequence convective dynamo, that the dynamo leaves behind a magnetic field, then a magnetic field of this order of magnitude should be visible at the surface of all such stars during the main sequence. However this assumption is not certain -- the slow retreat of the pre- main-sequence convective envelope containing a time-dependent dynamo will leave a field of small radial length scale, causing the field to decay more quickly via magnetic diffusion.  \blh Its survival will also be affected by processes like meridional circulation and differential rotation.  \bkh see Section \ref{decay} for a more detailed discussion of this point.

\bigskip

 \subsection{The evolution of fossil fields}\label{decay}

As mentioned in Sections \ref{herbig} and \ref{OBstars}, the strength of the fields observed in magnetic early-type stars falls during the main-sequence, and it falls faster than expected from simple flux conservation while the star's radius increases. Various explanations spring to mind, the most obvious of which is Ohmic diffusion. The global timescale for Ohmic diffusion is of order $10^{10}$ years, somewhat longer than the main-sequence lifetime of the least massive stars in question. It may however be possible to massage this timescale downwards, perhaps by making use of the lower conductivity near the surface of the star -- conductivity goes as $T^{3/2}$. In the absence of other effects, one would expect the electric current associated with the magnetic field to die away reasonably quickly in the outer part of the star, so that after some time the field at the surface is simply a potential-field extrapolation of the field further inside. Depending on the initial geometry and radial distribution of the magnetic field, this could cause the surface field either to rise or fall during the main sequence. This will depend on the origin of the magnetic field. The main weakness of finite conductivity in explaining this decay though is that it is also observed in massive stars with much shorter main-sequence lifetimes and somewhat longer Ohmic timescales than intermediate-mass stars.

Another possibility is the combination of buoyancy and thermal diffusion. In short, a magnetic feature is in pressure balance with its less-strongly magnetised surroundings, and so its gas pressure must be lower; to avoid moving on a dynamic timescale its temperature must therefore be lower than its surroundings. Heat consequently diffuses into it, resulting in a buoyant rise to the surface 
 (\citealt{Parker79c,1979SoPh...62...23A, 1995JFM...301..383H, MacGregorC03, Braithwaite08}). 
 Note that this mechanism is distinct from the so-called buoyancy instability (or Parker instability) where diffusion is not required.  In the low-density environment near the surface of a star where heat diffusion is very efficient, the rise is limited by aerodynamic drag and takes place at the Alfv\'en speed (see Section \ref{subsurf}). Deeper down, the process is limited by heat diffusion, and for a global magnetic field structure its timescale can be expressed in terms of the Kelvin-Helmholtz timescale and the plasma-$\beta$ as $\tau_{\rm decay}\sim\beta\,\tau_{\rm KH}$ \citep{Braithwaite08}. On the one hand, this would immediately explain why a similar flux decay is seen in A, B and O stars, since the thermal timescale is roughly the same fraction of the main-sequence lifetime and $\beta$ falls in roughly the same range at all spectral types. On the other hand, it might be tricky to get this process to work fast enough: even assuming that the interior field is ten times stronger than the surface field, the timescale for the most strongly magnetised stars (e.g. an Ap star with a $30$ kG field) would be comparable to the main-sequence lifetime. Since the timescale goes as $B^{-2}$, the effect in stars with much weaker fields would be negligible. However, as with Ohmic diffusion, it may be possible to massage the numbers in light of the fact that the thermal diffusion timescale is much shorter near the surface of the star.

Another possibility is meridional circulation. 
 Its characteristic time scale is the Eddington-Sweet time, $\tau_{\rm decay}\sim \tau_{\rm KH}E_{\rm grav}/E_{\rm rot}$, making it occupy roughly the same range, as a fraction of main-sequence lifetime, at all masses. The geometry of this flow has been clarified in the seminal work of \citet{Zahn92}. 
 Note the similarity with the timescale of diffusive buoyancy in the previous paragraph -- the only difference is that rotational energy has replaced magnetic energy. Since the ratio of energies in this case can be much closer to unity than in the previous case, the timescale can also be much smaller, indeed the fast-rotating magnetic stars should experience relatively fast decay, unless the magnetic field finds some way to coexist alongside the meridional flow. It may be though that this flow is simply inhibited by the \blh magnetic stress in the \bkh fossil field.
 Essentially nothing is present in the literature regarding how this inhibition might work. A first guess might be that the magnetic energy would have to be greater than the circulation kinetic energy, leading to the stability condition $E_{\rm mag} > M R^2 \tau_{\rm KH}^{-2} E_{\rm grav}^{-2} E_{\rm rot}^2$, or
\begin{equation}
\beta < \frac{\tau_{\rm KH}^2}{\tau_{\rm rot}^2}\frac{E_{\rm grav}^3}{E_{\rm rot}^3}.
\end{equation}


Alternatively an interaction with the convective core could be the crucial process -- the convective motion and waves that it sends into the radiative envelope could somehow result in an enhanced diffusion. In this case, one would certainly expect a correlation with mass, since the core is very small in late A stars but reaches to around a third of the stellar radius in O stars. This topic is explored in Section~\ref{convcore}. In any case, important clues will come if/when correlations are observed between flux decay and mass, rotation and field strength.

\bigskip
\section{\large The Origin of Fossil Fields}
\label{sec:variations}

The discussion in the previous section begs a question: where does the variation in magnetic fields in otherwise similar stars come from?
An important clue must be the observed extreme range of field strengths in A stars, with roughly `equal numbers per decade' between 200 and $2 \times 10^4$~G, as well as the bimodality, with no stars having fields between a few gauss and 200~gauss \citep{Auriere_07}. The existence of very weak fields such as in Vega and Sirius (Section~\ref{vega}) highlights the suspicion that the field in the protostellar cloud from which a star forms is not particularly relevant in determining the fields observed in stars (contrary to the classical view of the origin of fossil fields). 
 There are various ways in which one might explain the observed range in field strength; the relevant processes are discussed below approximately in order of decreasing scale and/or increasing time.
 
\subsection{Variations within the ISM}\label{ism}

According to the traditional model, variations in magnetic field strength in the interstellar medium (ISM) are simply carried forwards into the star. In light of recent results though, including the very weak fields in Vega and Sirius, this scenario, at least in its simplest form, now looks very unlikely -- the range in field strengths in stars is far greater than that in the ISM. In addition, this model requires an additional ingredient to produce the observed bimodality between Ap and other A stars. It is perhaps a clue that the lower threshold of 200~G in Ap stars is the same as the equipartition field strength at the photosphere -- in the merger hypothesis (Section~\ref{mergers}) this might have to be a coincidence. Also, this model is compatible with the lack of magnetic stars observed in binaries, since collapsing cloud cores with a strong magnetic field will spin down efficiently, whereas cores lacking a strong field will retain too much angular momentum to form a single star and so form a binary. This effect has been seen in simulations \citep{Machida_08}. To summarise, the simple ISM-variation model ignores much of the star formation process and so alone, it will not explain what we see in stars. It may play some role however; in any case it is worthwhile to take a more detailed look at star formation from the perspective of magnetic field evolution.

\subsection{Core collapse}

In star formation, the relative strength of the gravitational and magnetic fields is often expressed as a dimensionless mass-to-flux ratio, defined as $\lambda\equiv 2\pi G^{1/2} M/\Phi$, where $M$ and $\Phi$ are the mass and magnetic flux, or locally in a disc context as $2\pi G^{1/2}\Sigma/B_z$ where $B_z$ and $\Sigma$ are the field normal to the disc, and surface density; this ratio is conserved if flux freezing is valid. It is related to the gravitational and magnetic energies by (ignoring factors of order unity) $\lambda^2 = |E_{\rm grav}|/E_{\rm mag}$. A cloud with $\lambda\gtrsim1$ is said to be `magnetically supercritical' and will collapse, in the absence of significant thermal or rotational energy. Conversely a cloud with $\lambda\lesssim1$ is `magnetically subcritical' and the magnetic field supports the cloud against gravity. Once a cloud has become supercritical, it can collapse dynamically. Magnetic braking becomes ineffective once the collapse is super-Alfv\'enic, so the rotational energy becomes larger in relation to the other energies. Normally this leads to formation of a disc of radius 100\,--\,1000 astronomical units.

\subsection{The role of the accretion disk}\label{accretiondisc}

There is strong evidence that discs contain strong, ordered, net poloidal flux -- there are direct measurements of the magnetic field in protostellar discs at various radii from 1000~AU right down to 0.05~AU \citep{Vlemmings_10, LevyS78, Donati_05} and the mass-to-flux ratio is always found to be of order unity.\epubtkFootnote{Note that $\beta \sim \lambda^2 (h/r) M_\ast/M_{\rm disc}$ where $\beta=8\pi P/\bar{B^2}$; in a disc therefore it is possible to have both $\beta>1$ and $\lambda<1$ at the same time, in constrast to stars where $\beta\sim\lambda^2$.}

There is additional evidence for the presence of ordered magnetic fields in discs. Systems ranging from protostars to active galactic nuclei usually (not always,  and not all of them) show evidence of a fast outflow in the form of a collimated jet. The default model for its origin is the rotation of an ordered magnetic field in the inner regions of the disk. That is, a field crossing the disk with a uniform polarity over a significant region around the central object. Models assuming the existence of such an ordered field (as opposed to the small scale field of mixed polarities generated in MRI turbulence) have been particularly successful in producing fast magnetically driven outflows. Accepting this as evidence of the existence of such ordered fields, they might also be the fields that are accreted to form magnetic A, B and O stars. 

The origin of the ordered field in disks is less certain. An important constraint on the possibilities is the fact that the net magnetic flux crossing an accretion disk is a conserved quantity (a direct consequence of div $\mathbf{B}=0$). It can change only by field lines entering or leaving the disk through its outer boundary, i.e., any net flux in a disc must be accreted from the ISM.

Given the strong ordered fields in discs, it is something of a puzzle  that in stars we observe mass-to-flux ratios $\lambda\sim10^3$ in the most strongly magnetised Ap stars and ratios up to at least $10^8$ in other stars. In other words, we have an extra phenomenon to explain: why even the most strongly magnetised stars have such weak fields. If in the steady state the star is accreting mass and flux in the ratio $\lambda_\ast$, then mass and flux must be passing through each surface of constant radius in the disc in the same ratio (ignoring outflow), even though the local ratio will in general be much lower $\lambda(r)\ll\lambda_\ast$, requiring almost perfect slippage at all radii. Either there is a fundamental problem accreting flux through a disc, or there is a bottleneck further in through which magnetic flux cannot be accreted, located either in the star's magnetosphere or in the star itself. 

This may be related somehow to the fact that accretion disks are turbulent. Simple estimates show that accretion of an external field is very inefficient if the disk has a magnetic diffusivity similar to the turbulent viscosity that enables the accretion \citep{vanBallegooijen89}. Numerical simulations \citep{FromangS09} show that this is in fact a good approximation for magnetorotational turbulence. Though intuitively appealing, accretion of the field of a protostellar cloud as the source of Ap star fields is therefore not an obvious possibility. 

The flux bundles that drive jets from the inner regions of the disk, inferred indirectly from observations,  must somehow be due to a more subtle process. Numerical simulations for the case of accretion onto black holes \citep[cf.][and references therein]{Tchekhovskoy_11} have shown that `a flux bundle' of uniform polarity in the inner disk can persist against outward diffusion in a turbulent disk. These results are still somewhat artificial since the flux of the bundle in such simulations depends on the magnetic field assumed in the initial conditions. The speculation is \citep[cf.][]{Igumenshchev_03} that this flux itself starts as a random magnetic fluctuation further out in the disk, which somehow is advected inward with the accretion flow \citep[e.g.,][]{SpruitU05}. 

It may be then that magnetic stars have accreted random magnetic features from the accretion disc. Whether such a scenario is realistic is an open question \citep[for recent suggestions in this direction see][]{Sorathia_12}. If something like this happens, it might explain the apparently unsystematic presence of jets from accretion disks. In the context of A stars, the assumed randomness of the sources in the disk on a range of length and time scales, might perhaps explain the range in field strength of magnetic Ap stars. 

\subsection{Destroying flux in or near the star}\label{destroyflux}

Given the evidence for strong ordered fields in the inner part of the accretion disc, it might be necessary to prevent the bulk of it from entering the star. Star-disc interaction seems to be intrinsically complex and is poorly understood; and it may be difficult to accrete flux from the inner edge of the disc. What flux does reach the star, however, is certainly not obliged to remain there in its entirety -- excess flux can easily be destroyed after it reaches the star \citep{Braithwaite12}. If the protostar is convective, magnetic field is prevented by its own buoyancy from penetrating into the star, and if the star is already radiative then the quantity of magnetic helicity is crucial.

Magnetic helicity is a global scalar quantity defined as $H=\int \mathbf{A}\cdot\mathbf{B} \,\rd V$ where $\mathbf{A}$ is the vector potential defined by $\curl\mathbf{A}=\mathbf{B}$. It can be shown \citep{Woltjer58} that this quantity is conserved in the case of infinite conductivity. In plasmas of finite but high conductivity, it has been demonstrated that it is \emph{approximately} conserved, for instance in the laboratory \citep{ChuiM95, HsuB02} and the solar corona \citep{ZhangL03}.

If the magnetic field in a radiative star is allowed to relax, it will evolve into an equilibrium (Section~\ref{sec:theory}). 
 Once an equilibrium has been reached, energy and helicity are related by $H=EL$ where $L$ is some length scale which is comparable to the size of the system or, in this context, to the size of the star. Therefore, one can predict the magnetic energy of the equilibrium from the helicity present initially -- helicity is a more relevant quantity than initial magnetic energy. The observations then imply that there is an enormous range in the magnetic helicity which stars contain at birth, as well  as  a possible bimodality.

In the symmetrical `hourglass model' of star formation, the helicity is zero. Any field accumulated from such an hourglass should therefore decay to zero energy (cf. Flowers \& Rudermann instability, Section~\ref{stability} and Figure~\ref{fig:initfield}). In reality of course, one expects some asymmetry. However, an unrealistically high degree of symmetry would be required to  be left with  a field of only $1$~G from an initial mass-to-flux ratio of order unity.

\subsection{Pre-main-sequence convection}\label{preMSdynamo}
As they decend from the birth line down the Hayashi track, intermediate-mass stars are fully convective, before turning onto the Henyey track (leftwards on the HR diagram), during which time the convective zone retreats outwards and disappears (see e.g. \citealt{2005fost.book.....S}). The situation with stars above very approximately $6$ $M_\odot$ is less well understood; they may have a radiative core throughout the pre-main-sequence. In any case, as suggested by \citet{Braithwaite12} and shown more thoroughly by \citet{2015MNRAS.447.1213M} a star with a constant entropy -- as in a convective star -- cannot hold onto any previous magnetic field: it rises buoyantly towards and through the surface on a Alfv\'en timescale. This process does not actually require the convective motion itself, just the flat entropy profile it creates. Instead of a pre-existing field, the star would have a convective dynamo field.

As the convective zone retreats towards the surface, it should leave something behind in the radiative core. Assuming that the dynamo fluctuates, as in the case of the Earth and the Sun, layers of alternating polarity are deposited into the growing radiative core, rather like we see in the rocks of the mid-Atlantic ridge, which keep a memory of the polarity of the Earth's magnetic field at the time when it solidified from magma. However, given that the star is fluid, and evolves on a thermal timescale which is very much greater than the dynamo fluctuation timescale, the layers will be very thin and should annihilate each other by finite resistivity almost immediately. It is therefore not clear whether anything could be left behind at all; the best one might hope for is some net north-south asymmetry from $\sqrt N$ statistics. One way out of this may be if the convective region retreats more rapidly. This could happen in the inner part of the star: the Schwarzschild criterion is breached first at some intermediate radius and a thin radiative layer forms, cutting off the supply to the interior of deuterium from the ongoing accretion, and resulting in a relatively fast transition to a radiative core \citep{1993ApJ...418..414P}. This would not work though in the more massive stars, in which deuterium burning is less important to the stellar structure.

The most obvious serious weakness of pre-main-sequence convection as the origin of Ap-star magnetic fields is that one would expect it to produce a range of magnetic fields strengths according to the rotation rate of the star; it is difficult to produce any bimodality. Bimodality could though perhaps be produced afterwards, if some mechanism existed which destroys magnetic fields below some threshold. \blh This highlights the problem that it is not enough to have a theory, whatever it is, that explains a certain typical field strength. The enormous scatter in the observed field values is the actual challenge to theory here. \bkh

\subsection{Mergers}\label{mergers}
Mergers are a strong candidate to create strong fossil fields in a subset of stars \citep[e.g.,][]{Bidelman02, ZinneckerY07, Maitzen_08}. \citet{BogomazovT09} suggest that Ap stars may be the result of mergers between binary stars with convective envelopes whose orbits shrink as a result of magnetic braking. This is seems more likely on the pre-main-sequence. \citet{Ferrario_09} point out that the observed correlation between mass and magnetic fraction \citep{Power_07} 
 could be explained by the need for the merger product to be radiative, i.e., on the Henyey track. Exactly how a merger should produce a fossil field is not understood, but we can at least expect plenty of free energy in the form of differential rotation. This model would also explain the lack of close binaries containing an Ap star, although there are one or two peculiar counter examples to this observational result, for instance the binary HD~200405 with a period of 1.6~days. Even periods of 3~days would be tricky to explain as the result of a merger in a triple system; it might be necessary to reduce the orbital period of the resulting binary after the merger. This might be related to a similar issue for the merger hypothesis, namely that a merger product will initially be rotating close to break-up but that magnetic stars are observed to rotate slowly. In both cases, angular momentum must be extracted. One could imagine perhaps that the material ejected in a merger, which is estimated to be around 10\% of the total mass of the two merging stars, could absorb angular momentum as it flows outwards. Some form of magnetic coupling between the stars and the circumstellar material might be important, just as is thought to work to slow down the rotation of single magnetic stars, as discussed in the next section.

\subsection{The rotation periods}

Apart from the large range in field strengths in radiative stars, we also want to explain the range in rotation periods. On the one hand we need an explanation of how sufficient angular momentum can be removed to form a star at all, but on the other hand we need to form some stars rotating at close to break-up and others with periods of several decades or more. The correlation between the presence of a significant magnetic field and rotation period is strong, but there is still a large range in rotation period amongst stars with similar field strengths. How is it possible at all to spin a star down to a rotation period of 50~years, or in other words up to a Keplerian co-rotation radius of $\approx$~17~AU (assuming $M=2\,M_{\odot}$)? Assuming a stellar radius $R=2\,R_\odot$, a surface dipole of 3~kG and the standard $r^{-3}$ radial dependence, we arrive at just $0.5$ $\mu$G at the co-rotation radius -- it is obviously a major challenge for any kind of disc-locking model, where the co-rotation and magnetospheric (Alfv\'en) radii are comparable, for a magnetic field weaker than the galactic average to be in equipartition with gas of much greater density than the galactic average.

\bigskip
\section{\large Magnetic Fields and Differential Rotation}
\label{diffrot}

Since strong stable fields are already found among pre-MS (Herbig Ae-Be-) stars, their origin must lie in earlier phases of star formation, when the protostar was in a state of rapid, possibly differential, rotation. The sequence of events that led to the formation of a stable magnetic field is not known, but may have involved processes of interaction between magnetic field and differential rotation. This interaction takes different paths depending  on the relative strength of initial field and initial rotation. Since we do not know the star formation process well enough, we have to consider diffferent possibilities.   

Take as a measure of the strength of the magnetic field the Alfv\'en frequency $\omega_{\rm A}=\bar v_{\rm A}/R$, with $ \bar v_{\rm A}=B/(4\pi\bar\rho)^{1/2}$, where $B$ is a representative field strength, $\bar\rho$ the star's average density and $R$ its radius. If $\omega_{\rm A}$ is large compared with (a representative value of) the differential rotation rate $\Delta\Omega$, the evolution of the field configuration under the Lorentz forces is fast and the field will relax to a stable equilibrium, if it exists. On top of this, there will be oscillations with frequencies of order $\omega_{\rm A}$, reflecting the aftermath of the relaxation process and the differential rotation that was present initially. 

\subsection{Phase mixing}
These oscillations are then damped by a process of \emph{phase mixing}. As a model for this damping consider an idealised case, where the field is in a stable equilibrium to which an azimuthal flow in the form of differential rotation has been added as an initial condition. The deformation of the field lines in this flow reacts back on the flow; the result is an Alfv\'enic oscillation. The oscillation period is given by an Alfv\'en travel time. Since the energy of Alfv\'en modes travels along field lines, neighbouring magnetic surfaces oscillate independently of each other.  Their frequencies are in general different, with the result that neighboring surfaces get out of phase. The  length scale in the flow in the direction perpendicular to the surfaces decreases linearly with time. The result is damping of the oscillation on a short time scale. For more detailed discussions of this process, see e.g.\ \citet{HeyvaertsP83} and the references in \citet{Spruit99}. 
 
\subsection{Rotational expulsion}
In the opposite case $\omega_{\rm A}\ll \Delta\Omega$, the differential rotation flow is initially unaffected by the field. The non-axisymmetric component of the field (with respect to the axis of rotation) gets `wrapped up', such that lines of opposite direction get increasingly close together, increasing the rate of magnetic diffusion. As a result magnetic diffusion cancels opposite directions in a finite time. The nonaxisymmetric component decays, it is effectively expelled from the region of differential rotation. The process (`rotational expulsion', \citealp{Raedler80}) is similar to the evolution of a weak field in a steady convective cell   (`convective expulsion', see Section~\ref{expuls} below).

This process, if allowed to proceed to completion, will therefore `axisymmetrise' the initial field configuration. In this idealised form, it is probably somewhat academic, however, since the wrapping process increases the field strength linearly with time (the non-axisymmetric as well as the axisymmetric component). It may well happen that magnetic forces become important before magnetic diffusion has become effective  \citep[for discussion see][]{Spruit99}.  Magnetic instabilities, Tayler instability being the first to set in, then take over and determine the further evolution (Section~\ref{varint}).  As discussed below, this can  have important consequences for the rotational properties of the star.

\subsection{Angular momentum transport in radiative zones}

In A stars, the detection limit for large-scale magnetic fields is of order few gauss, and somewhat lower in very bright stars like Vega and Sirius where subgauss fields have been found (Section~\ref{vega} above). It is reasonable to assume that the internal field strengths of these stars is rather higher than the  measured value at the surface, since stronger smaller-scale fields at the surface would escape detection and since it would not be surprising if the magnetic field were weaker at the surface than deeper inside the star. The internal field strength even in these `non-magnetic' stars is likely to be of order 10 G or perhaps more (see also Section~\ref{failed}).

Even fields below current detection limits for these stars can have dramatic effects though on the internal rotation of stars. If $r$ is the distance from the center of the star, the torque exerted by Maxwell stresses in a field with toroidal (azimuthal) component $B_\phi$ and radial component $B_r$ is of the order $r^3 B_\phi B_r$. The torque in a `geometric mean' field $\bar B=\langle B_\phi B_r \rangle^{1/2}$ of the order 1 gauss is sufficient to redistribute angular momentum on a time scale of the star's main sequence life time, and to keep the core corotating with its envelope as the star spins down by a stellar wind torque (a classical argument dating from the 1950s).

This idea may however be a little over-simplistic.  As mentioned above, the MHD instabilities are expected to set in. This could result in a magnetic dynamo.

\subsection{Dynamos in radiative zones}
\label{varint}

Fields generated by some form of dynamo activity have traditionally been associated with convective envelopes, to the extent that dynamo action in stars was considered equivalent with a process of interaction between magnetic fields, convective flows, and differential rotation. This is only one of the possibilities, however. In fact, differential rotation alone is sufficient to produce magnetic fields.  The most well known example is that of magnetorotational fields generated in accretion disks \citep{BalbusH91}. The idealization of an accretion disk in this case is a laminar shear in a rotating flow, with a rotation rate  $\Omega$ decreasing with distance $r$.  Dynamo action triggered by magnetorotational instability (MRI) quickly (10\,--\,20 rotation periods). It generates a fluctuating field with a small scale radial length scale ($l\sim c_{\rm s}/\Omega$, where $c_{\rm s}$ is the sound speed, comparable to the thickness of the accretion disk).

Given a sufficiently high magnetic Reynolds number, the energy source of differential rotation is sufficient for field generation, in the presence of a dynamical instability of the magnetic field itself. In disks the magnetic instabilities involved in `closing the dynamo cycle' are  magnetorotational  instability \citep{BalbusH91} and magnetic buoyancy instability \citep{Newcomb61, Parker66}. In the case of the solar cycle, the phenomenology strongly indicates magnetic buoyancy as the main ingredient in closing the dynamo cycle  (see review by \citealt{2009LRSP....6....4F}). This contrasts with conventional turbulent mean-field views of the solar cycle (e.g.\ \citealt{2010LRSP....7....3C}); for a critical discussion see \citealt{Spruit11, Spruit12}. 

In the radiative interior of a star, the high stability of the stratification allows buoyant instability only at very high field strengths \citep[cf.\ the review in][]{Spruit99}. Instead, in such a stable stratification a pinch-type instability is likely to be the first to set in. A dynamo cycle operating on differential rotation combined with this `Tayler instability' has been described in \citet{Spruit02}. Its application to the solar interior predicts field generation at a level just enough to exert the torques needed to keep the core in nearly uniformly rotation and to transport the angular momentum extracted by the solar wind  \citep{Spruit02, 2005A&A...440L...9E} 
 The magnetic field generated in this model is extremely anisotropic: in the radial ($r$) direction, the length scale for changes of sign of field line direction is very small. It should thus be regarded as a `small scale dynamo' instead of a global one\footnote{The two uses of the term `dynamo' in the literature may lead to confusion here. The process is not a global dynamo as envisaged in `mean field' models.}. This reflects the dominant role of the stable stratification, but also the nature of Tayler instability, which is a local one in the $r$ and $\theta$ directions. In the azimuthal direction, however, its length scale is large, dominated by the fastest growing nonaxisymmetric Tayler mode, $m=1$. 

Tests of this dynamo cycle through some form of numerical simulation would be desirable. In a proof-of-principle simulation by \citet{Braithwaite06a}, a dynamo cycle in a differentially rotating stable stratification including thermal diffusion was observed with properties as predicted. Simulations for realistic stellar conditions present serious obstacles, however. The predicted fields are very weak compared with the stability of the stratification, while the cycle time is very large and its length scale small compared with the other physical scales of the problem. A widely cited simulation by \citet{2007A&A...474..145Z}, claimed to be valid for the physical conditions in the Sun, did not yield dynamo action. Inspection of the parameter values actually used in this simulation shows that the negative result is caused by damping of magnetic perturbations by the high magnetic diffusivity assumed (orders of magnitude off). With the thermal and magnetic diffusivities used, the differential rotation in this simulation is actually three orders of magnitude below the threshhold for dynamo action (eq.\ 27 in \citealt{Spruit02}). The case studied in \citet{2007A&A...474..145Z} is therefore not relevant, neither for questions of existence or otherwise of differential rotation driven dynamos in stably stratified zones of stars, nor as a test of a given dynamo model.

The large range in length and time scales involved makes simulations for realistic conditions in stellar interiors impossible to achieve at present. The dynamo mechanism should be directly testable, however, by appropriate simulations which lie in a parameter space that satisfies the minimum criteria (such as those that were reported already in \citealt{Braithwaite06a}).

%
%

 Recently \citet{2015A&A...575A.106J} performed simulations of a differentially-rotating star with an incompressible constant-density equation of state, finding that the MRI is the dominant dynamo process. In a more realistic stably-stratified star, it is not immediately obvious whether the MRI or the Tayler instability should dominate. Whilst under adiabatic conditions the Tayler instability is the first to set in, the situation may be different once thermal diffusion is included. 

\subsection{Observational clues}
\label{clurot}

Only very indirect observational evidence is available for or against the existence of magnetic fields in radiative interiors. The nearly uniform rotation of the solar interior, as well as its corotation with the convective envelope, have long posed the strongest   constraints on possible angular momentum transport mechanisms\footnote{The standard recipes used in stellar evolution calculations ``with rotation" in fact fail this constraint rather spectacularly when applied to the Sun.}.  In addition, the rotation rates of the end products of stellar evolution, the white dwarfs and neutron stars, may provide clues.  To the extent that  the rotation of these stars  descends directly from their progenitors (AGB stars and pre-supernovae), they also contain information regarding the degree to which the cores of the progenitors were coupled to their envelopes.  The very high effectiveness of Maxwell stress at transporting momentum makes magnetic fields the natural candidate.  Transport of angular momentum by internal gravity waves   (e.g.\ \citealt{2005Sci...309.2189C}) may also be an important mechanism, however (for recent theoretical developments see \citealt{2013sf2a.conf...77A}).

\begin{figure}[htb]
\vspace{2\baselineskip}
\centerline{\includegraphics[width=0.5\textwidth]{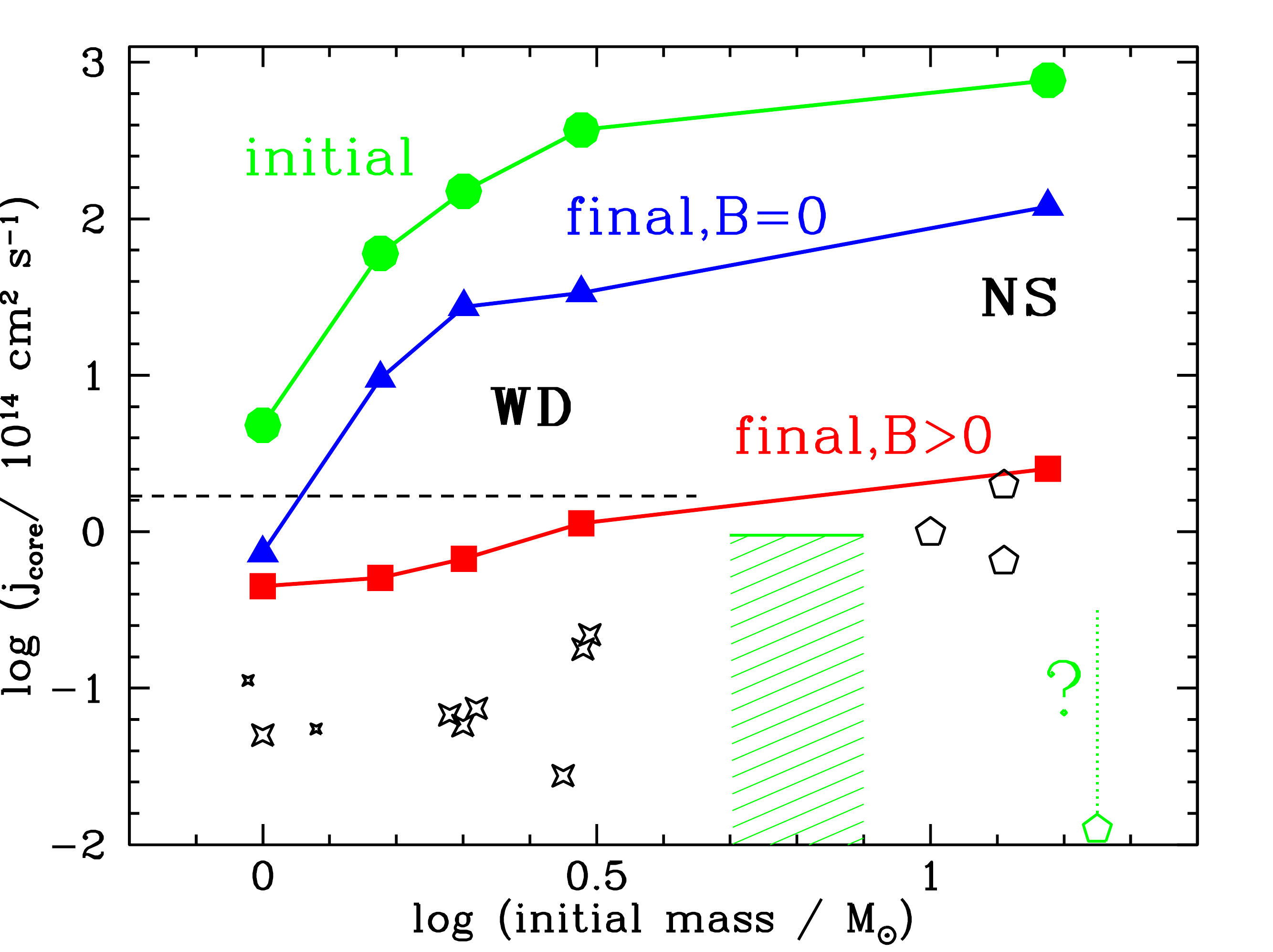}}
\caption{Specific angular momentum of white dwarf and neutron stars as a function of initial mass of the progenitor star, as computed by \citet{Suijs_08}. Green  circles: initial core angular momentum, blue  triangles: including known hydrodynamic angular momentum transport mechanisms, red  squares: including the Tayler-Spruit magnetic torque prescription.  The dashed horizontal line indicates the spectroscopic upper limit on the white dwarf spins obtained by \citet{Berger05}. Star symbols represent astroseismic measurements from ZZ Ceti stars and the green hatched area is populated by magnetic white dwarfs. The three black open pentagons correspond to the youngest Galactic neutron stars, while the green pentagon is thought to roughly correspond to magnetars, where the vertical-dotted green line indicates the possibility that magnetars are born rotating faster. See \citet{Suijs_08} for more details.}\label{suijs}
\end{figure}

It is somewhat uncertain, however, whether there is a direct connection between the end products and the internal rotation of their progenitors. The observed asymmetries in planetary nebulae, for example, indicate highly asymmetric mass loss in the final evolution stages of AGB stars. Such asymmetric `kicks' may have reset the angular momentum of the cores, such that the rotation rates of WD actually reflect the `kicks' imparted by the last few mass loss events rather than the initial core rotation \citep{Spruit98}. Supernova kicks \citep{Wongwathanarat_13} may also be the main process determining of the rotation of neutron stars \citep{SpruitP98}.

Leaving these complications aside, stellar evolution calculations  are used to  make predictions of the rotation rates of end products by including (parametrisations of) known mechanisms of angular momentum transport such as meridional circulation \citep[e.g.,][]{Zahn92} and hydrodynamic instabilities. Figure~\ref{suijs} shows results from evolution calculations of massive and intermediate-mass stars. Including the hydrodynamic processes predicts rotation rates that are too high by $\sim2 $ orders of magnitude, clearly indicating that much more effective processes of angular momentum transport must be present in stars. When magnetic torques according to \citet[see also Section~\ref{varint}]{Spruit02},  are included, agreement with observations is better, but discrepancy  of 1 order  of magnitude nevertheless is still present. 

\subsubsection{Asteroseismic results}\label{astero}

\begin{figure}
\centerline{\includegraphics[width=0.5\textwidth]{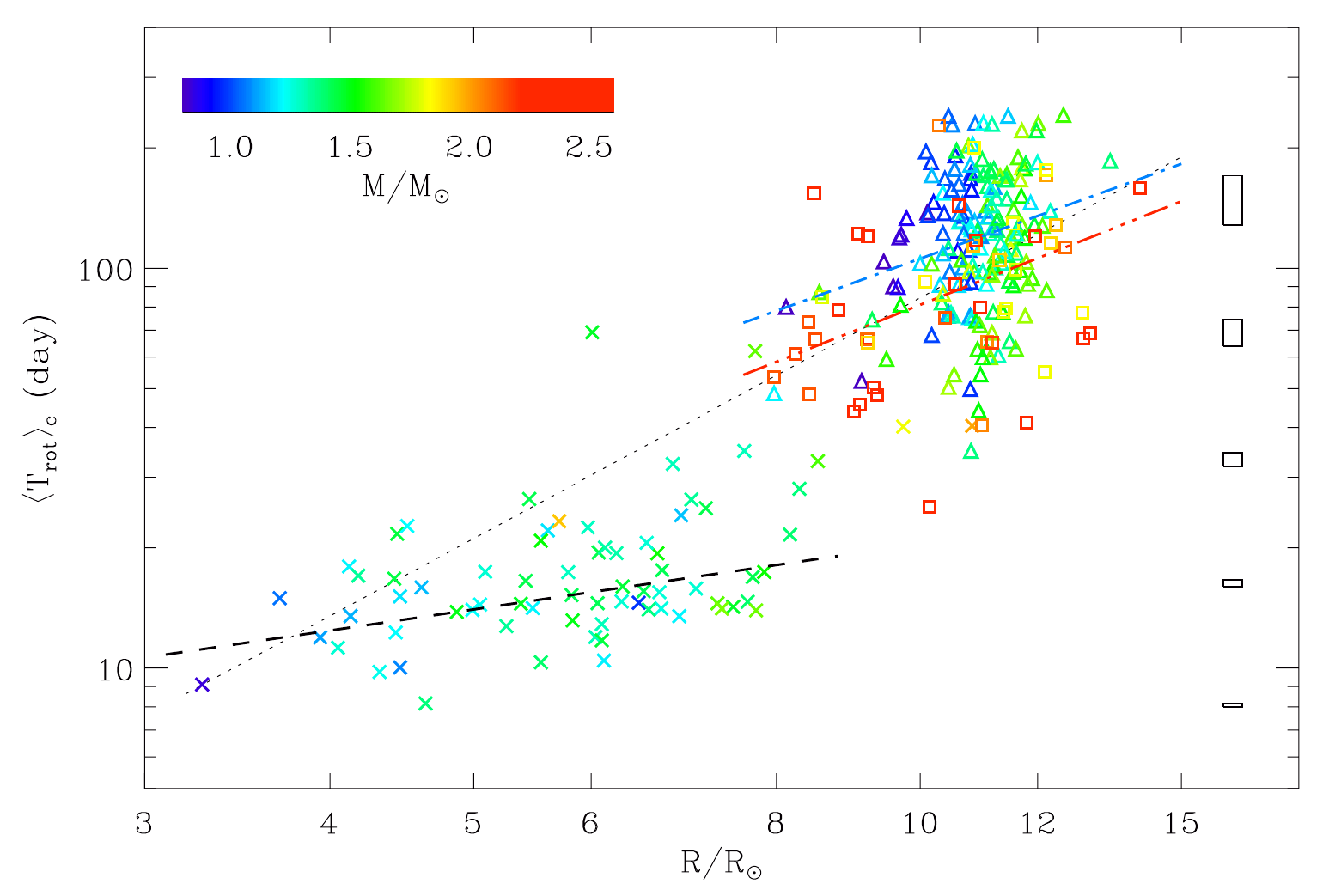}}
\caption{Mean period of core rotation as a function of the asteroseismic stellar radius, in log-log scale. Crosses correspond to RGB stars, triangles to clump stars, and squares to secondary clump stars. The color code gives the mass estimated from the asteroseismic global parameters. The dotted line indicates a rotation period varying as $R^2$. The dashed (dot-dashed,
triple-dot-dashed) line indicates the fit of RGB (clump, secondary
clump) core rotation period. The rectangles in the right side
indicate the typical error boxes, as a function of the rotation
period. From \citet{Mosser_12}.}
\label{Kepler}
\end{figure}

The new asteroseismic results of giants and subgiants from the Kepler mission have greatly expanded the evidence on  the  internal rotation of stars other than the Sun \citep{Mosser_12}. The cores of these stars rotate faster than their envelopes, with typical periods of 10  to 200 days (see Figure~\ref{Kepler}). This shows a degree of decoupling between envelope and interior. The torques required to explain these rotation rates are still much stronger than can be explained by the known non-magnetic processes, however (with the possible exception of angular momentum transport by internal gravity waves, cf.\ \citealp{OgilvieL07, Mathis_08, BarkerO11, Rogers_13}).
 Results from stellar evolution calculations using the estimate in \citet{Spruit02} can be compared with the Kepler rate of Figure~\ref {Kepler}. As with the rotation rates of the end products, the predicted rotation rates are up to a factor of 10 too high \citep{Cantiello14}).  (The fact that the disagreement is by a similar factor in both cases may be a coincidence.)

\section{\large Interaction between Convective and Radiative Zones }
\label{conv}

In this section we explore the interaction of steady magnetic fields with convection. In particular, we look at the possibility of steady fields in the radiative interiors of solar-type stars, \blh shielded from becoming observable at the surface \bkh (sect. \ref{shield}) and how these might interact with the convective envelope above (sect. \ref{convenv}).  In \ref{convcore} we look at at how fossil fields in the radiative envelopes of early-type stars might interact with their  convective core; and finally we take a brief look at fully convective stars and the possibility of producing magnetic fields in subsurface convective layers in early-type stars.

In contrast with A, B and O stars, among which a significant minority has a detectable field and the rest of the population has at most very weak fields, \blh  the cooler stars (those with convective envelopes) \bkh seem \emph{all} to display magnetic fields, but never of the stable kind seen in A, B and O stars. \blh In addition,  there is apparently some difference among the cooler stars, between \bkh magnetic fields in stars such as the Sun \blh which have \bkh a radiative core and a convective envelope, and in fully convective stars (this is looked at below in Section~\ref{fullyconv}).

 For observation results, the reader is referred to recent spectropolarimetric surveys \citep{2014MNRAS.444.3517M, 2014MNRAS.441.2361V} which have provided excellent results on magnetic properties of late-type stars. Also \citet{2003A&A...397..147P,2011ApJ...743...48W} obtained interesting results which link dynamo activity to rotation.  In solar-type stars the magnetic fields vary much like the Sun's field, without a consistent steady component; the small-scale field is much stronger than the large-scale field. Upper limits on a steady dipole component on the Sun (which would be observable as a North-South asymmetry during the solar cycle), for example, are probably of the order of only a few gauss. 

 In these stars a field anchored in the radiative core could in principle be present, but if it is, it seems not to manifest itself at the surface. Two kinds of explanation suggest themselves. A convective envelope might somehow be incompatible with the presence of a fossil field, either by preventing it from forming during star formation, or by destroying it soon after (Section~\ref{convenv} below). Alternatively, it might be that a fossil field is actually still present in (some) solar type stars, but somehow confined or `shielded' by the convection zone (Section~\ref{shield}).

\subsection{Confinement of steady fields in the interior of sunlike stars}
\label{shield}

To the extent that a steady internal field connects to the convective envelope it might become detectable at the surface as a time independent component, superposed on the cyclic dynamo-generated field characteristic of stars with convective envelopes. The Sun provides some limits on this possibility. Its cyclic dipole component has an amplitude of about 20 G at the magnetic poles \citep{Petrie12}. In most cycles some asymmetries are seen between the north and south hemispheres, but no signal of a long-term average polarity has been reported. The implied detection limit is probably on the order of a few gauss. This indicates either that a steady field in the solar interior has a dipolar component below a few gauss, or that a (possibly stronger) field is somehow actively shielded by the convection zone.

A steady, shielded field (called `inevitable' by its authors) was discussed by \citet{GoughM98}. Their model assumes a meridional circulation near the base of the convection zone, downward at the poles and the equator and upward at mid-latitudes. A magnetic field in the radiative zone, such that the field lines are parallel to its interface with the convective zone, is in contact with this circulation. The authors find a steady solution to a set of reduced equations, with the property that the interior field does not spread into the convection zone.

This is a somewhat surprising result. At its upper boundary, field lines from the interior spread upward into the convection zone by magnetic diffusion. At the poles and equator of the model, the downward advection of field lines by the circulation opposes this spreading. At its mid-latitude, however, advection in the model is upward, into the convection zone, carrying the field lines in the {\em same} direction as the spreading by diffusion. This does not depend on the particular configuration of the circulation: there is always at least one region where diffusion and advection both act to spread the field into the flow. A circulation therefore cannot prevent a field in the radiative interior from diffusing into the convective envelope. The conflicting result by Gough and McIntyre is  a  consequence of the reductions made to the induction equation\epubtkFootnote{The model does not include the radial advection and diffusion of the poloidal component of the magnetic field ($B_{\rm p}$). It is instead replaced by an assumed fixed value $B_0$. The model includes an equation for diffusion of the azimuthal field component ($B_\phi$),  but leaves  out its radial advection by the circulation.}.

The conclusion from the above is that a field in the radiative interior is inevitably connected to some degree with the convection zone by field lines crossing the interface, even if it does not extend far enough into the convective envelope to become visible at the surface.
An internal field can be {\em shielded} from the visible surface  in this way, while at the same time  remaining {\em  mechanically  coupled}  to flows in the envelope, the differential rotation for example. The construction in Gough \& McIntyre makes the assumption that the interior field is both shielded and decoupled. 

 Observations of the interior rotation of the Sun through helioseismology show that the radiative interior has approximately the same specific angular momentum as the convective envelope.  Since the Sun spins down through angular momentum loss in the solar wind, this observation indicates the existence of an efficient coupling between interior and envelope, on an evolutionary time scale. The coupling resulting from magnetic  interaction between interior and envelope naturally fits this observation. 

  Decoupling is a far stronger assumption than shielding. The contrasting conclusions reached in numerical simulations \citep{Strugarek11, Acevedo-Arreguin13} can be traced to confusion of these two  concepts.   

\subsubsection{ Shielding:   convective expulsion}
\label{expuls}

Shielding of a sufficiently weak internal field could be achieved by a process of \emph{convective expulsion}  \citep{Zeldovich56, Parker63}. The first numerical study  of expulsion, idealised as a steady circulating flow interacting with a magnetic field was by \citet{Weiss66}.  It shows how a convective cell can create a field-free region by pushing the field lines passing through it to the sides of the cell. This happens in two stages. First the field lines in the cell are `wrapped around' by the flow, changing the field into a complex configuration with changes of direction on small scales. This happens on a short time scale, a few turnover times of the cell. In the second stage magnetic diffusion reconnects field lines of different direction until the wrapped field has disappeared from the cell.  In the field which is now concentrated at the boundaries of the cell the flow is suppressed. The process is effective up to a certain maximum field strength roughly given by equipartition between the magnetic and convective energy densities. In more realistic, time-dependent convective flows, the separation between flows and magnetic fields is observed to be stable once established. An example is the highly fragmented magnetic field seen at the solar surface  (\citealt{2004ApJ...610L.137C, 2014ApJ...789..132R} and references therein).

In the case of shielding of a field in the radiative interior of stars with convective envelopes, both convective flows and meridional circulation would contribute to shielding. But as argued in the above and in \citet{Strugarek11} both would also lead to mechanical coupling of the envelope to the interior by the poloidal field component. 

\subsubsection{ Coupling}\label{coupling}

 The above (Section~\ref{expuls}) shows that by  a convective expulsion process the envelope may be able to effectively shield a field in the radiative interior from manifesting itself at the surface. This does not imply, however, that the  envelope and interior are also mechanically decoupled.  There is always some connection between the two through field lines diffusing from the interior into the envelope (see \ref{shield}). 

The  simulations by \citet{Acevedo-Arreguin13} focus on the shielding aspect. As in \citet{Strugarek11}, however, their simulations also show a poloidal field connecting the interior to the envelope. 
The consequence  of this connection  is that flows in the envelope, whether in the form of convection or differential rotation, keep the field in the interior in a time-dependent state. It excludes the steady internal field  assumed by Gough \& McIntyre. 

The time dependences that can be covered in numerical simulations are on the order of tens of rotation periods of the star. This is to be compared with the  age of the fossil field (nine orders of magnitude longer).
A poloidal field component that is inconsequential  on numerically accessible time scales can  wreak havoc on the internal field already on time scales that are very short compared with the age of the star. 

The differential rotation of the envelope, for example, stretches the connecting poloidal field lines into an azimuthal component. Mestel's  well known estimate \citep[][remark on p735]{Mestel53} shows that  even a field as weak as a few microgauss would,  on an evolution time scale, be amplified to a strength sufficient to affect the rotation of the interior. Long before this happens, however, the wound-up azimuthal field will develop magnetically driven instabilities.  As argued in \citet{Spruit02}, this sequence of events is likely to lead to a self-sustained, time dependent magnetic field independent of the initial field configuration.

\subsection{Strong magnetic fields below  convective envelopes}
\label{convenv}
 Discussions  of magnetic field evolution are easiest when a `kinematic' view is  (implicitly) assumed: when the fields considered are sufficiently weak that their Lorentz forces can be ignored to first order. One may wonder what stars with convective envelopes would look like if they contained magnetic fields as strong as those of Ap stars in their radiative interiors \blh and Lorentz forces cannot be ignored. \bkh

The type of behavior of a magnetic field in a flow (convective or otherwise) depends on its strength relative to the kinetic energy density in the flow. Field lines connecting the interior with a convective envelope are subject to advection (i.e., being moved around) by the differential rotation $\Delta\Omega$ between pole and equator that takes place in the envelope.  Take  the Sun as a representative example, where the differential rotation rate is $\Delta\Omega$ is $\sim 10^{-6}$ s$^{-1}$, and the density $\rho$ at the base of the convection zone is $0.2$ g/cm$^3$. Equipartition of magnetic energy density $B^2/8\pi$ with the energy density $\frac{1}{2}\rho(r\Delta\Omega)^2$ in this differential rotation corresponds to a field strength of about $10^5$~G. Above this strength the field would be able to affect differential rotation in the convection zone. Equipartition with the typical energy density in convective flows at the base of the solar convection zone on the other hand would be somewhat less, a few kG.

The typical Ap star field (several kG) would therefore significantly interfere with convection throughout a convective envelope. Even in such a strong field, however, some form of reduced energy transport is likely to exist, since the convective flow and the magnetic field can disengage from each other by the  convective  expulsion process discussed above. The field gets concentrated in narrow bundles in which flows are suppressed. In the gaps between the bundles the fluid is in a nearly field-free convective state\footnote{This separation is reminiscent of the formation of flux bundles in a superconductor in an imposed magnetic field, cf.\ \url{http://hyperphysics.phy-astr.gsu.edu/hbase/solids/scbc.html}}. 

An example of how this might work is seen in sunspots, which have surface field strengths well above equipartition with the surrounding convective flows. Below the visible surface of a sunspot a `splitting' process is present in the field configuration, extending to just the visible surface. It produces gaps through which convective flows can transport heat. This explains both the inhomogeneities observed in sunspots, and the relatively large heat flux in sunspots \citep{Parker79a, SpruitS06}. It has convincingly been seen in operation in realistic radiative MHD simulations of sunspots \citep{SchuesslerV06, Heinemann_07, Rempel11}.

Stars with strong, stable fields and with such sunspot-like phenomenology are not known. The process of accumulation of magnetic flux from a protostellar disk might be different in stars with a final mass like the Sun compared with more massive stars, since the star's magnetic field could affect the accretion process. However, it seems more likely that survival of a fossil field somehow is not compatible with a convective envelope. If this turns out to be the case, it also raises the possibility that the convective core of an Ap star may affect the evolution of its fossil field.  This is discussed further in Section~\ref{convcore}. 

Convective motions in the envelope would impose random displacements of the field lines extending through the interior. Since convective flows have length scales smaller than the stellar radius, the displacements of field lines by these motions are incoherent between their points of entry and exit from the interior. This `tangles' the field lines in the interior: neighboring field lines get wrapped around each other.  This raises the issue of {\em reconnection} : if in the course of tangling neighboring field lines can exchange paths by reconnection, the cumulative effect of many such events on small scales would act like an effective diffusion process, allowing field lines to drift at a rate much higher than resulting from microscopic resistivity.

\subsubsection{Reconnection}

The consequences of this wrapping process have been studied extensively in the context of the solar corona by \citet[and refs. therein]{Parker72, Parker79a, Parker12}, who finds that it leads to rapid formation of current sheets (on the length and time scale of the displacements), through which reconnection takes place. Numerical simulations of this process have been done for the case of coronal heating of the Sun driven by convective displacements of the footpoints of the coronal field in the photosphere \citep{GalsgaardN96}. Using simple MHD simulations, \citet{Braithwaite15} finds this phenomenon not only in a low-$\beta$ plasma (as in the corona) but also in order-unity (e.g., the ISM) and high-$\beta$ (e.g., stellar interiors) plasmas.   Particularly relevant for the present case of reconnection by small scale flows in a strong background field is the extensive study  by \citet{Zhdankin13}. 

The tangling process transports a certain amount of energy from the convection zone into the interior, but more importantly the continued reconnection of field lines in the interior driven by flows in the envelope effectively also acts as an enhanced diffusion of those field lines that are affected by the tangling process. 

If the effect is large enough, the field could already disappear from the star when accretion of magnetic flux ends, toward the end of the star's formation. There is some observational evidence relevant for this, since large scale fields have been observed also in Herbig Ae-Be stars (cf.\  Section~\ref{herbig} above). Though only a few have been found so far, they occupy the same range of surface temperatures as the main sequence magnetic Ap stars. The onset of efficient convection around F0 coincides with the disappearance of the Ap phenomenon \citep[e.g.,][]{Landstreet91}. In the above interpretation it would also mark the onset of enhanced magnetic diffusion. The observations would then imply that the decay of the Ap-type field  from a star with a convective envelope  is in fact effective on a time scale noticeably less than the pre-main sequence life of an F0 star, i.e., less than about $10^7$ yr, or about a factor 100 shorter than the decay time expected from purely Ohmic diffusion.

\subsection{Convective cores}
\label{convcore}

The above line of thought about enhanced diffusion  by `tangling'  is also relevant to convection in the cores of the magnetic early-type stars, especially the more massive ones. Going from essentially fully radiative at around $1.5\,M_{\odot}$, the convective core extends to around $r/R\approx 0.16$ in late B stars, around $5\,M_{\odot}$, and up to around $r/R\approx 0.3$ at even higher masses. 
 In this case some of the field lines emerging at the surface could pass through the core. In the simple dipolar configuration of Figures~\ref{fig:axisym_field}, \ref{fig:x-sec-diag} and \ref{fig:equil_seq} these field lines would populate the magnetic poles. The azimuthal field torus that stabilises the configuration as a whole (Section~\ref{stability}) needs to be located in the stably stratified radiative zone outside the core. Shuffling of field lines by convection could keep the polar field region in a somewhat time-dependent state.  This tangling by convective motion may have a similar effect to the tangling of field lines in the solar corona,  where it keeps the field above  the convective zone close to a potential field.   Unlike the case of a convective envelope, however, the stabilizing part of the field is not connected to the convective region; it is unaffected by reconnection processes taking place on the polar field lines. We hypothesise that this explains how Ap star magnetic fields can coexist with a convective core.

Convection in the core would have the secondary effect of exciting some level of internal gravity waves in the surrounding radiative envelope  (see e.g. \citealt{Rogers_13}). Waves can also increase the rate of magnetic diffusion, but, being periodic, their effect (at the same velocity amplitude) is not comparable with the reconnection processes resulting from the wholesale tangling of field lines discussed above. The magnetic diffusion time scale by Ohmic diffusion alone is of the order $10^{10}$ yr in an Ap star, their main sequence life time of the order $10^8$ yr. A possible wave-induced increase of the diffusion rate by a factor 10--100 would still be compatible with the fields seen in Ap stars.

Another possibly important difference between convective core and convective envelope is the direction of gravity at the boundary between radiative and convective zones. In solar-type stars, cold, fast downflows should penetrate some distance into the radiative zone, but in contrast rising bubbles in convective cores are not expected to overshoot significantly. 
 In addition to this, magnetic fields have an inherent buoyancy since they provide pressure without contributing to density, and have a tendency to rise, either on some thermal timescale (which in most contexts is very long) or on a dynamical timescale if conditions for buoyancy instability \citep{Newcomb61, Parker66} are met. Both of these effects would tend to make it easier to expel a large-scale field from a convective core, and keep it from re-entering, than to prevent the field in a radiative core from interacting with a convective envelope.  It may be that fossil fields do not enter the convective core. The structure and stability of the non-axisymmetric equilibria in particular (Figure~\ref{fig:x-sec-diag}, lower panel) would be little affected by expulsion from the core.

 Convective cores are expected to host a dynamo; this has been seen in the simulations of \citet{Browning_04} and \citet{Brun05}. Interestingly, a fossil field in the radiative zone might have an important effect on the nature of the core dynamo. \citet{Featherstone09} performed simulations of a core dynamo in the case where the surrounding radiative zone contains a fossil field. Without a fossil field, an equipartition field is generated in the core; the addition of a fossil field (significantly weaker than this equipartition strength) switches the dynamo to a different regime in which the field generated is much stronger, which not surprisingly  changes the properties of the convection. This would be analogous to the situation in accretion discs, where the presence or absence of even a weak net flux through the disc appears to have a fundamental effect on the nature of the dynamo. This result could be relevant for any compact stellar remnant born out of the core, as it would enable neutron stars and perhaps also white dwarfs to inherit in some way the magnetic properties of their progenitors. 


\subsection{Fully convective stars}\label{fullyconv}

\begin{figure}[t]
\centerline{\includegraphics[width=0.55\textwidth]{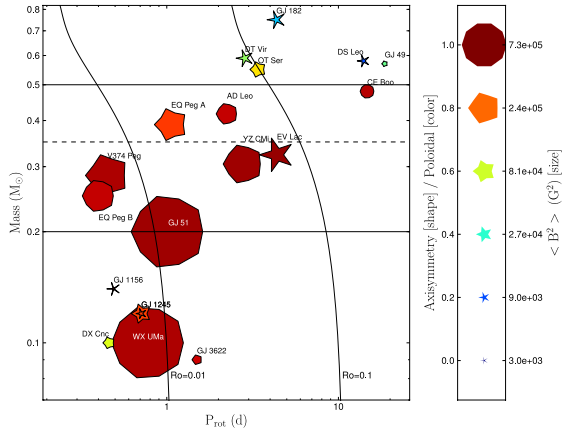}}
\caption{Properties of the surface distribution
 of the magnetic field (derived from Zeeman-Doppler Imaging)
of the M dwarfs observed with the spectropolarimeters ESPaDOnS (Canada-France-Hawaii Telescope) and NARVAL (T\'elescope Bernard Lyot) as a function of rotation
period and mass. Larger symbols indicate stronger fields, symbol shapes depict
the degree of axisymmetry of the reconstructed magnetic field (from decagons
for purely axisymmetric to sharp stars for purely non axisymmetric), and
colours the field configuration (from blue for purely toroidal to red for purely
poloidal). 
 Solid  curves  represent contours of constant Rossby number Ro =
0.1 (saturation threshold) and 0.01. The theoretical full-convection limit ($0.35\,M_{\odot}$) is plotted as a horizontal dashed line, and the approximate
limits of the three stellar groups discussed in the text are represented as
horizontal  solid  lines. Compiled from the studies by \citet{Morin_08a, Morin_08b, Morin_10, Donati_08} and \citet{Phan-Bao_09}.}
\label{Morin}
\end{figure}

 In the light of interesting recent observations of pre-main-sequence stars and low-mass stars, in this section we deviate from the main focus of this article -- non-convective zones -- to discuss briefly magnetic fields in fully convective stars.  In constrast to solar-type stars, both main-sequence stars below about $0.4\,M_{\odot}$ and T Tauri stars often display dipole fields of order 1~kG \citep[see, e.g.,][and references therein]{Morin_10, YangJ-K11, Hussain12}. Some recent results are summarized in Figure~\ref{Morin}. \citet{Gregory_13} find that  fully convective pre-main-sequence stars tend to have strong dipolar fields roughly aligned with the rotation axis. Those which have a small radiative core tend to have both strong dipolar and octupolar components, and those with larger radiative cores have more complex fields and only a weak dipole component. Intuitively, this is perhaps not surprising since it is difficult in a thin convective envelope to get different parts of the envelope to `communicate' with each other and form a global magnetic field. The length scale of sunspot systems is comparable to the depth of the convective layer; extrapolation to a fully convective star might explain the large scale of their field.

\citet{Browning08} performed simulations of a fully convective star, finding that a dipole-like field can indeed be generated. It is thought that the rotation is a key ingredient in producing a coherent large-scale field from smaller-scale motion, as in standard mean field dynamo models.

One can make an analogy here with planetary magnetic fields. The Earth, Jupiter and Saturn all contain convective, conducting fluids: the Earth (the outer core) between about 0.19 and $0.44\,R_{\rm E}$, Jupiter from $0$ or $0.1$ out to $0.78\,R_{\rm J}$, and Saturn between about $0.15$ and $0.5\,R_{\rm S}$. Important for the nature of the dynamo is the ratio between these inner and outer radii, and in that sense these planets are analogous to stars with small radiative cores. Looking at the results of \citet{Gregory_13} we should expect Jupiter to have a predominantly dipolar field aligned with the rotation axis, which it does. We expect Earth and Saturn to have probably also a significant octupole component, but since this drops off faster with radius in the overlying non-conducting fluid than the dipole component, we would still expect the field to be predominantly dipolar at the surface and approximately aligned with the rotation axis, which is also consistent with observations.

By analogy though with the Earth's dynamo, where the polarity of the field changes, we may speculate that the same happens in low-mass stars, in which case even the strong dipole fields observed would not be steady like those in radiative stars.

\subsection{Subsurface convection in early type stars}
\label{subsurf}

\begin{figure}[htb]
\centerline{\includegraphics[width=0.47\textwidth]{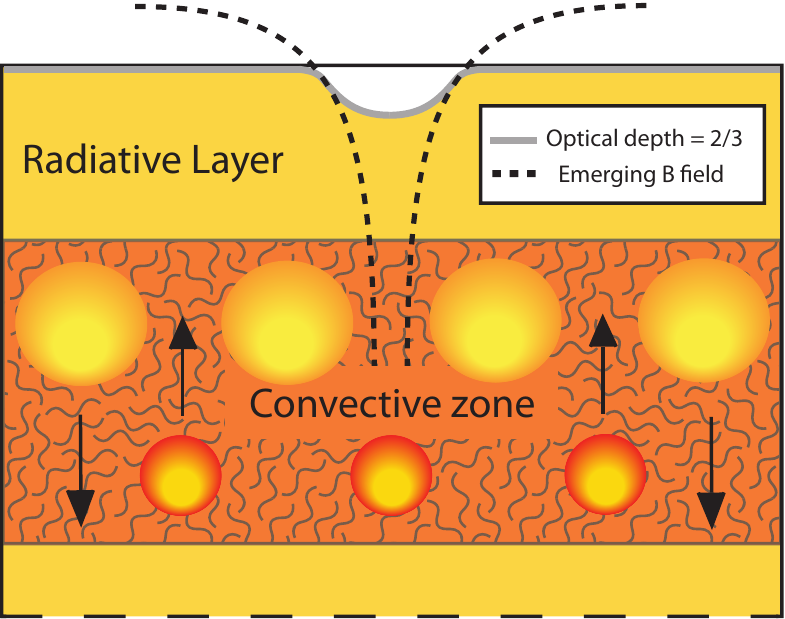}}
\caption{Schematic of the magnetic field generated by dynamo action in a subsurface convection zone. Note that magnetic features should appear as bright spots on the surface, rather than as dark spots as in stars with convective envelopes. From \citet{CantielloB11}.}
\label{fig:subsurf}
\end{figure}

In early-type stars, whilst it seems very unlikely that a magnetic field generated by a dynamo in the convective core could rise all the way to the surface on a sensible timescale, these stars also have convective layers close to the surface which could in principle generate fields which then rise to the surface. The convection is driven by bumps in the opacity at certain temperatures, caused by the ionisation of iron, helium and hydrogen. Massive stars (above about $8\,M_\odot$) have two or three such layers, the deepest and energetically most interesting of which is driven by ionization of iron \citep[see][]{Cantiello_09}. If this layer hosts a dynamo, there is no difficulty for the resulting magnetic field to reach the surface very quickly via buoyancy, since the thermal timescale is so short at the very low density in the overlying radiative layer \citep{CantielloB11}; see Figure~\ref{fig:subsurf}. 

 At the surface of the star, magnetic pressure supports magnetic features against the surrounding gas pressure, meaning that at a given height, the gas pressure inside the magnetic features is lower than in the surroundings. Since the photosphere is approximately located where the column density of gas above it has a certain value, the photosphere in magnetic features is lower than in the surroundings. In a radiative star, this means that magnetic spots appear bright. This contrasts to convective stars where the magnetic field has the additional effect of suppressing convection and therefore heat transfer, and magnetic spots are dark.

\begin{figure}[htb]
\centerline{\includegraphics[width=0.47\textwidth]{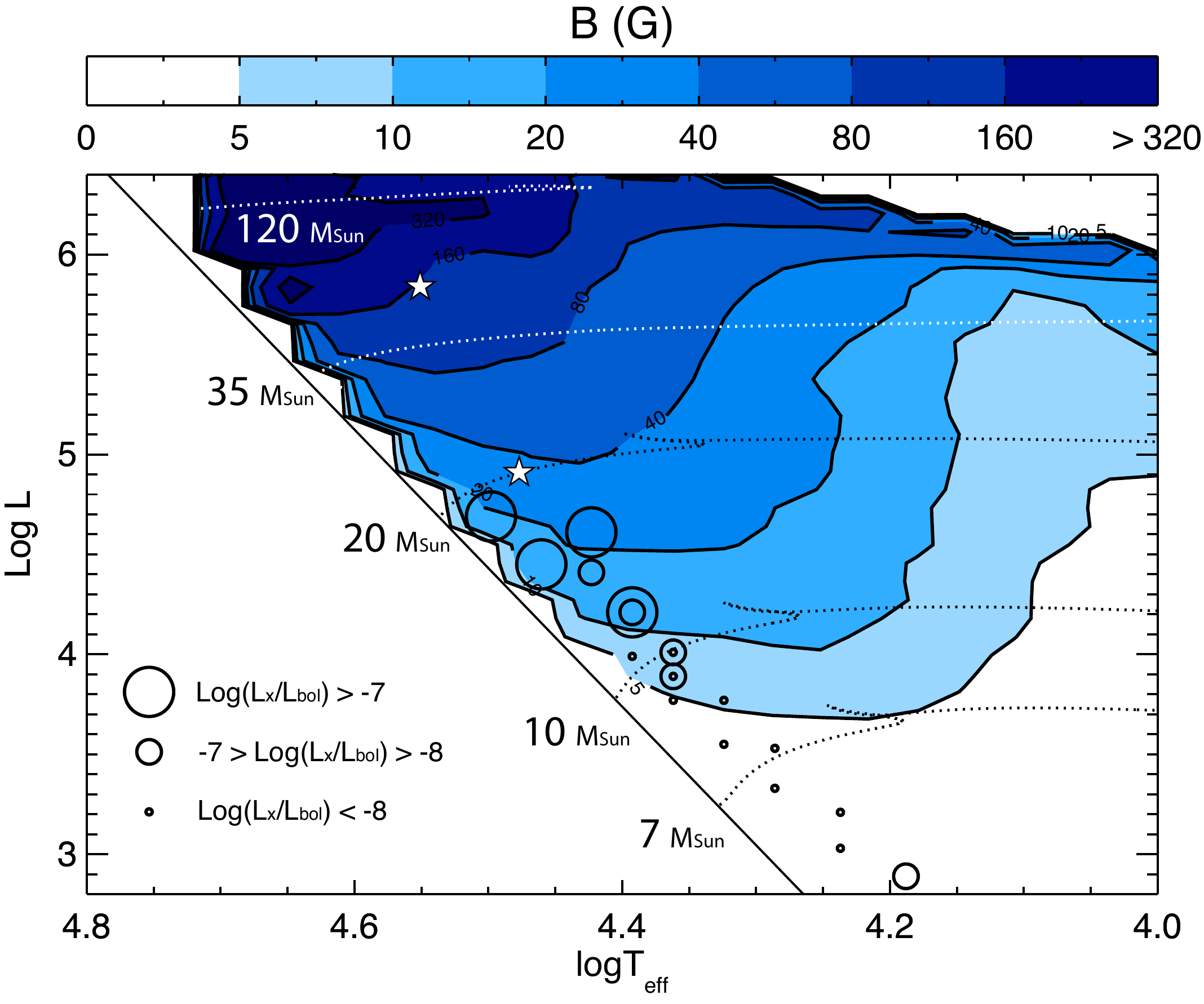}}
\caption{The field strength predicted at the surface of massive stars. The prediction assumes an equipartition dynamo in the convective layer and a $B\propto\rho^{2/3}$ dependence in the overlying radiative layer. Also shown are the X-ray luminosities measured in a number of stars: there does seem to be some connection between subsurface convection and X-ray emission. The dotted lines are evolutionary tracks of stars of various initial masses. From \citet{CantielloB11}.}
\label{fig:bsur}
\end{figure}

For solar metallicity field strengths of approximately 5 to 300~G are predicted -- as depicted in Figure~\ref{fig:bsur}. The field strength depends on the mass and age of the star: higher fields in more massive stars and towards the end of the main sequence. These fields are expected to dissipate energy above the stellar surface and could give rise to, or at least play some role in, various observational effects such as line profile variability, discrete absorption components, wind clumping, solar-like oscillations, red noise, photometric variability and X-ray emission \citep[see, e.g.,][for a review of these phenomena]{Oskinova_12}. Indeed, if the X-ray luminosities of various main-sequence stars are plotted on the HR diagram (Figure~\ref{fig:bsur}) a connection with subsurface convection does seem apparent. Of course, we cannot be certain that it is a magnetic field which is mediating transfer of energy and variability from the convective layer to the surface; one could also imagine that internal gravity waves are involved. Simulations of the generation and propagation of such gravity waves were presented by \citet{Cantiello_11}.

\subsubsection{Intermediate-mass stars}

Stars below about $8\,M_{\odot}$ have no such iron-ionisation-driven convective layer, but do have similar layers caused by helium and hydrogen ionisation. In intermediate-mass stars such as Vega and Sirius (spectral types A0 and A1 respectively) a dynamo-generated field could float to the surface from a helium-ionization-driven convective layer beneath the surface \citet{CantielloB11}. Field strengths of a few gauss are predicted. With the observations done so far though, it might be difficult to distinguish between this and the failed fossil hypothesis \citep{BraithwaiteC13} described above in Section~\ref{failed}. The main difference would be the length scales: the convective layer is very thin and it would be difficult to generate magnetic features large enough to be detectable as a disc-averaged line-of-sight component \citep[cf.][]{KochukhovS13}. In comparison to more massive stars, late-B and early-A stars are very quiet in X rays, and so far only upper limits on X-ray emission have been established. For instance, \citet{Pease_06} obtained an upper limit of $3 \times 10^{25}$ erg s$^{-1}$ in X rays from Vega. Pre-main-sequence stars in this mass range are known to emit X rays, but \citet{Drake14} obtain a limit of $1.3 \times 10^{27}$ erg s$^{-1}$ from the 8~Myr old A0 star HR~4796A, demonstrating that this X-ray activity shuts down roughly when the star reaches the ZAMS. This is consistent with weak dynamo activity; \citet{Drake14} predict an X-ray luminosity of very approximately $10^{25}$ erg s$^{-1}$ from magnetic activity in Vega, and lower luminosites in slower rotators. Cooler than around type A5 ($T_{\rm eff}\sim 10\,000\mathrm{\ K}$), X rays become detectable again, presumably owing to convection at the surface; this convection is also detected more directly as microturbulence \citep{Landstreet_09b}.

\subsubsection{Interaction with fossil fields}\label{subsurffossil}

The various observational phenomena in massive stars, listed above, seem to be ubiquitous. Notably, they are present also in stars in which strong large-scale fields have been detected. This means that if these phenomena are caused by subsurface convection, that this convection is not disturbed significantly by the fossil field. However, a fossil field of order 1~kG is in equipartition with the predicted convective kinetic energy and should at least have some effect on it \citep[see, e.g.,][]{CantielloB11}. The magnetic field may simply force the entropy gradient to become steeper until convection resumes. 
 This is often assumed in parametrisations of the effect of magnetic fields in stellar evolution calculations (e.g.\ \citealt{Feiden13}). As discussed above (\ref{convenv}), however, interaction of convection with fields at strengths of order equipartition is very inhomogeneous, and even small gaps between strands of strong field can allow a nearly unimpeded convective heat flux. The effects of such inhomogeneous fields on stellar structure are much smaller than in conventional parametrisations based on average field strengths  \citep{SpruitW86, Spruit91}. 

 In the case of  the early type stars (\ref{subsurf}),  subsurface convection  only transports a modest fraction of the total energy flux and one could imagine it is easier to suppress. There is recent observational evidence that a very strong magnetic field can indeed suppress the subsurface convection: \citet{Sundqvist_13} measure macroturbulent velocities in a sample of magnetic OB stars, finding that one star in the sample (NGC~1624-2), which has a field of around 20~kG, lacks significant macroturbulence, whilst the rest, which have fields up to around 3~kG, have vigorous macroturbulence of over 20 km s$^{-1}$. The thermal energy density in the convective layer corresponds to an equipartition field of around 15~kG, so this results appears to confirm that convection can be suppressed only by a field comparable in energy density to the thermal -- rather than convective kinetic -- energy density, even when convection is weak. Obviously it would be useful to improve the observational statistics, and to look at stars with fields between 3 and 20~kG. Finally, whilst it seems likely that macroturbulence (the part thereof which is not the result of stellar rotation) is essentially gravity waves produced by subsurface convection, the origin of the various other observational phenomena in massive stars is less certain; it would therefore be a very useful to determine whether strongly magnetic stars lacking macroturbulence display these other phenomena. 

In intermediate-mass stars, there is also some evidence that fossil fields suppress convection. Although late-A stars normally display microturbulence, it seems to be lacking in the magnetic subset of the population (Shukyak, priv. comm.) However, this is perhaps not directly comparable to massive stars because the surface convection is very weak, and all fossil fields are at least an order of magnitude stronger than equipartition with the convective kinetic energy.

\bigskip
\section{\large Neutron Stars}
\label{neutron}

\begin{figure*}[tbp]
\centerline{\includegraphics[width=0.65\textwidth]{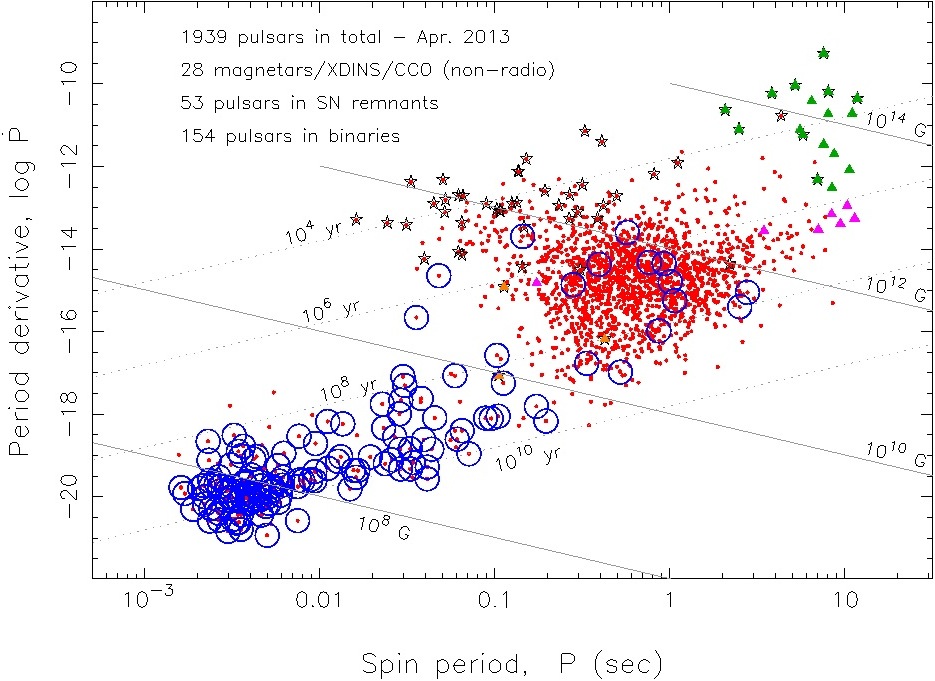}}
\caption{The well-known $P-\dot{P}$ diagram on which two readily measurable quantities are plotted, period derivative against period, together with inferred magnetic dipole field strengths and spindown ages (diagonal lines). The main clump is the radio pulsars (red dots), of which around 2000 are known. At the top right are the magnetars (green circles), and on the lower left we have old neutron stars with very weak dipole fields which have been spun up by accretion. Blue circles represent binary systems and stars supernova remnants. The so-called magnificent seven are the pink triangles on the right.  Figure provided by T.\ M.\ Tauris.}
\label{PPdot}
\end{figure*}

Field strengths of neutron stars observed as pulsars and magnetars span the range of  $10^{10}$  -- $10^{15}$~G (see Fig.\ \ref{PPdot}). 
 (In addition there are the `recycled millisecond pulsars', with field strengths around $10^8$~G. Their fields are believed not to be representative of their formation, instead representing a process related to the recycling. See \citealp{Harding13} for a review of different classes of neutron star.)  The width of this range: several decades,  is similar to that  seen in mWD and Ap stars. The presence of a solid crust makes a difference compared with the other classes, however, since it can anchor fields that otherwise would be unstable. Before this anchoring takes place, the magnetic field of the proto-neutron star is subject to the same instabilities and decay processes as in Ap stars. This is discussed below in Section~\ref{fluid}.

\begin{figure*}[t]
\centerline{\includegraphics[width=0.8\textwidth]{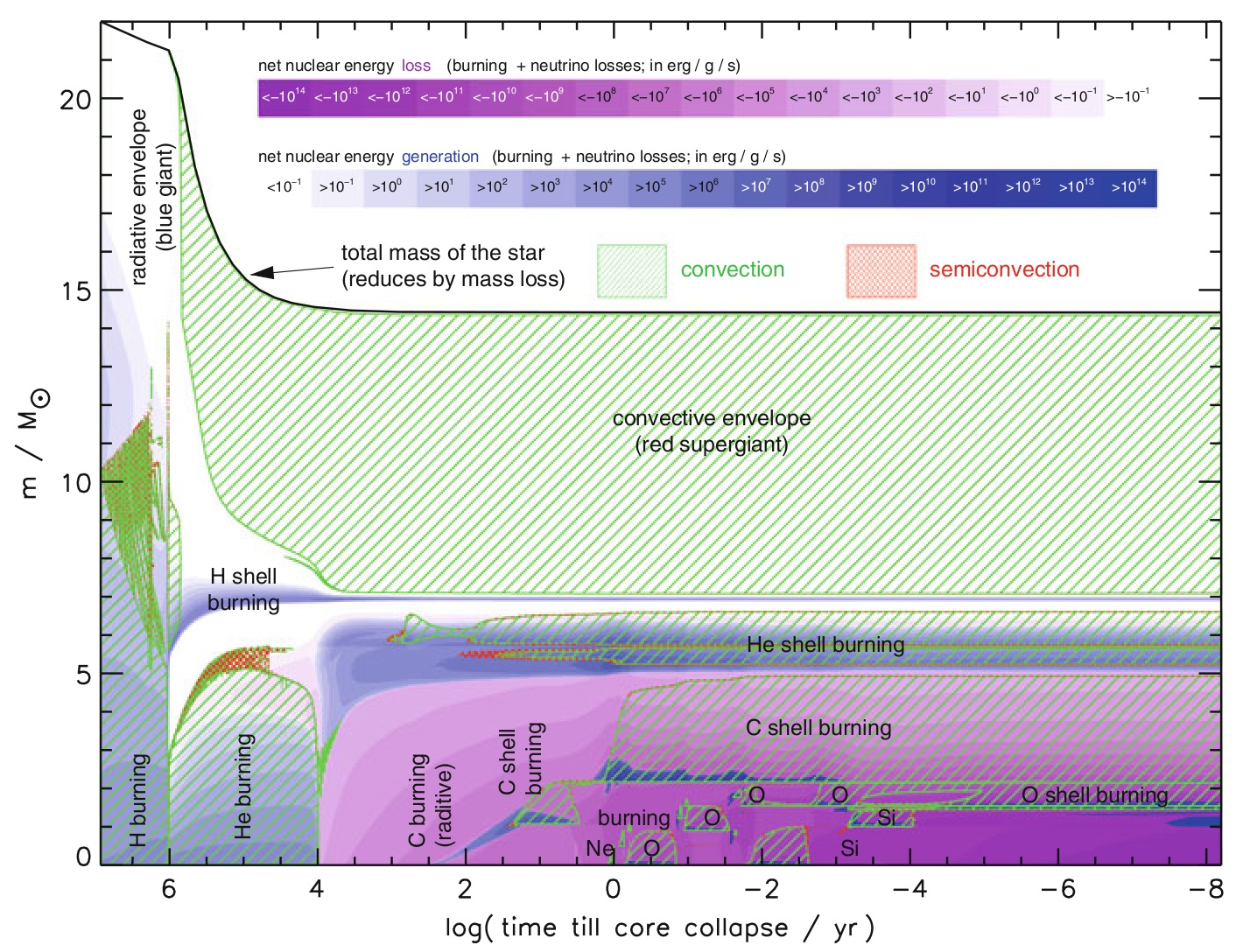}}
\caption{{ A Kippenhahn diagram of a star with an initial mass $22\,M_{\odot}$. Along the horizontal axis is log time before the supernova, so the main sequence takes up only a short space on the left. Note the appearance and disappearance of convective layers. Figure provided by A. Heger.}}
\label{Kipp_Heger}
\end{figure*}

\subsection{Mechanisms of field generation}\label{NSgen}

Three mechanisms have been proposed. The field could be inherited from the pre-collapse core of the progenitor star (the `fossil' theory\epubtkFootnote{There is some inconsistency in the literature concerning the meaning of `fossil field'. In the Ap-star context it is normally taken to mean a field left over from an earlier epoch, for instance the pre-main-sequence or the parent ISM cloud, rather than being the result of some contemporary dynamo process. In the neutron star context it means not just this but also that the earlier epoch is not the proto-neutron star phase but the pre-collapse progenitor.}), it could be generated by a convective dynamo  during (or shortly after) core collapse, or it could be generated by differential rotation alone, via a magnetorotational mechanism like that operating in accretion disks.  This mechanism looks especially promising for field generation during or shortly after collapse of a rotating core (cf.\ \citealt{Spruit08}). 

\subsubsection{Inheritance}
\label{inheritance}

In this scenario the field is simply a compression of what was present in the progenitor, which may be either an evolved high-mass star or an accreting white dwarf. We discuss first the pre-collapse evolution of massive stars.

During the main sequence the core of the star is convective, and afterwards various convective zones appear and disappear at different locations as burning moves in steps to heavier elements, as is illustrated in Figure~\ref{Kipp_Heger} (Heger 2013, personal communication). These convective zones may become relevant to the magnetic field of the star, especially in the core, out of which the neutron star will eventually form. During the main sequence, we expect an active dynamo in the core, perhaps similar to those in low-mass stars. For instance, \citet{Browning_04, Brun05, Featherstone09} performed simulations of an A-star convective core, finding a sustained dynamo. After the main sequence the star moves onto the red-giant branch. Magnetic fields have been observed in several red giants, such as EK Boo, an M5 giant, where a field of $B \sim 1-8$~G was detected \citep{Konstantinova-Antova_10}. Another example is Arcturus, a K1.5 III giant, where a subGauss field has been found \citep{SennhauserB12}. The nature of these fields is poorly constrained at present but they are presumably generated by a dynamo in the convective envelope, and may or may not be directly relevant for the neutron star's magnetic field.  

Under flux freezing (see also Section~\ref{freeze}), the field in a  pre-collapse core of radius $R_0\sim 3 \times 10^8$ cm collapsing to a neutron star of radius $R=10^6$, will be amplified by  factor $10^5$. To explain magnetar strength fields of $10^{15}$~G (more realistically, internal fields of $10^{16}$, see Section~\ref{internal}) requires an initial field $B_0$ of $10^{10}$~G. This is more than a factor 10 larger than the largest field strengths seen magnetic white dwarfs (which have about the same size and mass as such cores). The question what could lead to such a strong field in the progenitor core is still open.

 In the accretion-induced collapse (AIC) scenario, a white dwarf passes over the Chandrasekhar mass limit and collapses into a neutron star, the composition of the star being such that there is insufficient nuclear energy to produce a supernova. Accreting white dwarfs have a variety of magnetic properties, with large-scale fields observed with strengths up to at least $10^8$~G. This does not seem to be quite sufficient to explain fields in magnetars.

\subsubsection{Convection}

\citet{ThompsonD93} proposed that magnetic fields in neutron stars are generated in the young neutron star (NS) by a dynamo deriving its energy  from the convection with the help of  the differential rotation, estimating an upper limit to the field strength of $3 \times 10^{15}$~G, a little above the highest dipole component measured in soft gamma repeaters (SGRs) and anomalous X-ray pulsars (AXPs). A challenge for the theory is how to transfer this energy from the length scale of convection (around 1~km) to the dipole component on ten times that scale. If magnetic helicity is conserved (see Section~\ref{sec:variations}) then most of the energy should be lost once convection dies away; in fact even if the magnetic field in every convective element is twisted in the same direction in some sense (corresponding to maximum helicity) then one still expects to lose 90\% of the energy (a reduction in field strength of a factor of 3) if the dominant length scale is to rise by a factor of ten. If symmetry breaking doesn't work and convective elements are twisted in random directions then one loses a further factor of $\sqrt{N}$ of the energy. It is a general challenge for dynamo theory to produce large-scale structures,  but the problem does not seem insurmountable: rotating fully-convective stars are observed to have dipolar fields \citep{Morin_10} and the same has been reproduced in simulations (e.g. \citealt{Browning08}); see also Sect.~\ref{fullyconv}.  It may be that magnetars were born very quickly rotating; energetically one is in a much more favourable position if the star is rotating with a period of 3~ms or 1~ms at birth. However, one expects such a fast-rotating, highly-magnetised NS to spin down within a day or so, or even faster if it has a strong wind, injecting $2 \times 10^{51}$ to $2 \times 10^{52}$ erg into the supernova and creating a `hypernova'. There is evidence from supernovae remnants that this is not the case: \citet{VinkK06} compared energies of SN remnants around magnetars and other neutron stars and found no significant difference, concluding that the spin period of magnetars at birth must be at least 5~ms.

\subsubsection{Differential rotation}

 Energy in the differential rotation could also be tapped by the magneto-rotational instability \citep[MRI, ][]{Chandrasekhar60, Acheson78, BalbusH91}. Its importance in the context of neutron star magnetic fields comes from  the exponential growth of MRI on a differential-rotation timescale. It may be most relevant in the proto-neutron star,  after the convection has finished but when the star is still differentially rotating.
 
 In main-sequence stars,    the dynamo mechanism proposed by Spruit (\citeyear{Spruit02}; see also \citealp{Braithwaite06a})  should dominate over the MRI since it is inhibited less by the strong entropy stratification and works better than the MRI when differential rotation is weak. In proto neutron stars, which probably have quite strong differential rotation, it  could also convert some of the energy in the differential rotation into magnetic form, but it works more slowly because it involves an initial amplification stage where the field strength increases only linearly with time. The time available in the  collapse and immediate post-collapse phase is likely to be insufficient for this process. In proto neutron stars, field generation from the magnetorotational instability is a more likely process.

 In any case, the fact that older neutron stars are known to have magnetic fields, despite having used up their natal energy reservoir of differential rotation, brings us to the next section. 
 
\subsubsection{Magnetic helicity}\label{NShelicity}

Important for generation of a magnetar field will be the magnetic helicity present, as when the driving processes switch off, the star will try to relax into a minimum energy state for its level of helicity. The process that determines the helicity produced by the combination of a field amplification process, as well as the buoyant instabilities that bring the field to the surface of the star, are unclear. This is just as in the case of the Ap and mWD stars.

Helicity need not be conserved if the field is brought to the surface since we only expect conservation in a highly-conducting medium, so it may still be possible to build or destroy helicity  during the star's early evolution.  To build helicity, the symmetry (between positive and negative helicity, or in other words between right- and left-handed twist), has to be broken.

In terms of explaining the apparent diversity in neutron star properties, beyond the two degrees of freedom which place the star on the $P-\dot{P}$ diagram, it helps to think in terms of the range in available magnetic field configurations (Section~\ref{internal}).

\subsection{Field evolution before crust formation}
\label{fluid}

When neutron stars form they are differentially rotating and convective in most of the volume. As the star cools by neutrino emission the stratification becomes stable.  Some time later -- estimates vary from 30 seconds to 1 day after formation -- a solid crust forms \citep[see, e.g.,][and references therein]{Suwa13}. At some stage (either before or after crust formation) any field-generating dynamo present will die away, after which time the field will relax towards an MHD equilibrium; this happens on a dynamical timescale, the Alfv\'en timescale $\tau_{\rm A}$. This time scale is short, ranging from 3 hours down to 0.1~s for field strengths ranging from $10^{10}$ to $10^{15}$~G respectively, so in a non-rotating star one would expect that in most cases an MHD equilibrium  must be established  before the crust forms. Rotation however should slow down the formation of equilibrium to a timescale given by $\tau_{\rm evol}\sim \tau^2_{\rm A}\Omega$ where $\Omega$ is the angular rotation velocity of the star \citep[see][and Section~\ref{failed} for a discussion of this effect in A stars]{BraithwaiteC13}. Assuming an initial rotation period of 30~ms, an equilibrium would then take 600~years to form for a field of $10^{10}$~G or 2~seconds for a field of $10^{15}$~G. So it may be that only the magnetars really find an MHD equilibrium before the crust forms. Making an estimate of the strength of a crust (shear modulus and breaking strain,  see e.g. \citealt{Horowitz09,Hoffman12}),   it seems that a non-equilibrium magnetar-strength field would not be stopped from evolving anyway, in contrast to radio-pulsar-strength fields which could be held in position `against their will', so to speak.  Of course it might not be a coincidence that most neutron stars have fields at about the crust-breaking threshold; the magnetic field decays until it reaches this threshold and is prevented from decaying further. In this picture the magnetars would, for some reason, be born with a greater magnetic helicity than the other neutron stars; however the so-called central compact objects (CCOs) with fields of $10^{10}$~G would indicate intrinsically less efficient field generation. In any case, the location of the crust-breaking threshold is subject to large uncertainty, because of our poor understanding of the properties of the crust and perhaps more importantly of the geometry of the magnetic field and crustal fracturing. 

\subsection{The energy budget of magnetars}
\label{internal}

There is a consensus that magnetars are powered by the decay of their magnetic field, whereby it is necessary for energy to be dissipated in or above the crust rather than the interior, to avoid losing the energy to neutrinos \citep[e.g.,][]{Kaminker_06}. A magnetar with an r.m.s.\ strength $10^{15}$ gauss in its interior contains around $3 \times 10^{47}$ erg in magnetic energy, enough to maintain a mean luminosity of $3 \times 10^{35}$ erg s$^{-1}$ for a lifetime of $3 \times 10^4$ yr \citep[see, e.g.,][for a review of the observations]{Mereghetti08}. However, we have now seen giant flares in three objects, the most energetic of which is thought to have released (very approximately) $3 \times 10^{46}$ erg in just a few seconds; if it is not to be a coincidence to have observed this many giant flares then the energy source needs to be larger. In addition, most SGRs and AXPs have somewhat weaker measured dipole fields than $10^{15}$ gauss, some are below $10^{14}$ gauss. It looks as if the field strength in the interior of the star needs to be greater than we infer from the spindown rate, which gives us just the dipole component at the surface.

Fortunately this seems very possible. It may be that the magnetars have a strong toroidal field in relation to the poloidal component which emerges through the surface; \citet{Braithwaite09} and \citet{Akguen_13} found that for a given poloidal field strength a much stronger toroidal field is permitted, since the stratification hinders radial motion and therefore also the instability of a toroidal field. Also the magnetar fields could be more complex than a simple dipole; \citet{Braithwaite08} found a range of non-axisymmetric equilibrium where a measurement of the dipole component gives an underestimate of the actual field strength. Alternatively the field could be largely buried in the stellar interior with only a small fraction of the total flux actually emerging through the surface; whether this is possible depends both on where the field is originally generated and on diffusive processes which bring the field towards the surface over long timescales \citep[see][]{Braithwaite08, Reisenegger09}. In any case, there are various degrees of freedom available; see Section~\ref{sec:theory}.

\bigskip
\section{\large Summary and open questions}\label{summary}

The subject of magnetic fields in the interior of stars inevitably relies to a large extent on  theoretical developments. The increasing quality and quantity of observational constraints, however, is providing more clues and constraints on theory than ever before. In parallel, the increasing  realism of numerical MHD simulations makes them an effective  and indispensable means of testing theoretical speculation. An example is the subject of `fossil fields'  that have become the preferred interpretation of the steady magnetic fields seen on Ap, Bp and O stars. Here numerical MHD has not only convincingly reproduced the range of observed surface distributions of these stars, but also provided physical understanding of their stability and internal structure.  Axisymmetric purely toroidal or purely poloidal fields are unstable, and so they must exist together, in a certain range of strength ratios; the two components can either be comparable to each other, or the toroidal field can be stronger. The average interior field of a neutron star could be much higher than the surface dipole component inferred from spindown, which would explain how magnetars appear to have a more generous energy budget than one would estimate from their dipole components alone.

 A key to understanding the nature of such stable equilibria is magnetic helicity (Sections \ref{stability2}, \ref{destroyflux} and \ref{NShelicity}). To the extent that it is conserved during relaxation of a field configuration it guarantees the existence of stable equilibria of finite strength: a vanishing field has vanishing helicity.  An open  issue, however, concerns {\em helicity generation}: the question of which mechanism(s) has/have given the field the magnetic helicity that is essential for its long-term survival.  Perhaps stochastical coincidences during the dynamical phases of star formation and evolution (Section~\ref{accretiondisc}) may be all that is needed.

The recent finding that the fields in Ap stars appear to have a minimum of about 200 G (Section~\ref{apstars}) may be a  clue, still to be deciphered, for the formation mechanism of the fields.  The much lower fields of the order of a gauss discovered in the two brightest A stars, on the other hand, may just be the result of an initially stronger field that is presently still in the process of decaying (Section~\ref{vega}). 

Stars in the A-B-O range have convective cores, which are likely to interact to some extent with the stable fossil field in (the remainder of) the star. This is still a very open question. Some inconclusive speculations on the physics that may be involved in such interaction were given in Section~\ref{convcore}. The question may have an observational connection, however. The distribution of field strengths of ABO stars across the HRD has recently been shown to indicate decay on a somewhat shorter time scale than can be attributed to finite (`Ohmic') resistivity, perhaps an indication of enhanced diffusion related to interaction with the convective core, or with a convective envelope developing as the star evolves off the main sequence (Section~\ref{decay}, \ref{convcore}).   In stars going through a fully convective phase, any fossil field (such as that inherited from a molecular cloud) should be lost by magnetic buoyancy.
Common to ABO stars, white dwarfs, and neutron stars is the large range in fields strengths in the population, and the ratio of magnetic to gravitational energy, which ranges from about $10^{-16}$ to $10^{-6}$ in all three classes of star. The size of this range is a puzzle for any theory of the origin of these magnetic fields (Sections \ref{apstars}, \ref{ism}). 
 
Solar type stars do not show magnetic fields remotely resembling the stable Ap configurations and strengths. This has led to the speculation that the radiative interiors of these stars might still have a fossil field, but shielded from the surface by the convective envelope. Such a field can then be invoked to explain the near-uniform rotation of the Sun's radiative interior. 
A clear distinction has to be made here between {\em shielding} by the convective envelope and {\em decoupling} from it (Section~\ref{shield}). Convective processes are known that could shield an internal field from becoming observable at the surface, but in the presence of magnetic diffusion it is impossible to avoid mechanical coupling across the boundary between interior and envelope. It causes the internal field to evolve on time scales governed by interaction with the differential rotation of the envelope; the result will probably look more like the differential rotation-driven dynamo process discussed in Section~\ref{varint}. 

Further clues on magnetic fields in the interiors of stars come from asteroseismic results on the internal rotation of giants and subgiants (Section~\ref{astero}). The coupling between core and envelope deduced from these results  is far stronger than can be explained with existing hydrodynamic coupling processes, such as shear instabilities. In stably stratified zones of stars, time dependent self-sustained magnetic fields powered only by differential rotation and magnetic instabilities (i.e. ones governed by the Maxwell stress) are likely to operate, except in very slowly rotating stars. The favored scenario for such a dynamo process (Section~\ref{varint}) fares much better in matching the asteroseismic observations, but still misses the target by a significant factor.  Magnetic fields, while probably involved in stably stratified zones of stars, may    have modes of behavior not yet recognised. 

Future progress on the questions raised by the observations is likely to benefit increasingly from numerical simulations. The main obstacle is the fact that in almost all cases simulations for the actual physical conditions in stellar interiors will remain out of reach in the foreseeable future, irrespective of expected increases in computing power.  Experience shows that extrapolation over the missing orders of magnitude in physical parameter space cannot be done simply from the simulations themselves. Extrapolation needs physical understanding formalised in models that cover the asymptotic conditions encountered in stars. The validity of such models can often be tested well with targeted simulations of reduced scope (i.e. not with `3D stars'). The formulation of the models themselves requires more classical style theoretical effort, however (compare the discussion on p.\ 11 in \citet{schwarzschild70}.


\medskip\medskip
\bibliography{refs}

\begin{thebibliography}{99}
 \parskip=-0.3\baselineskip
\bibitem[Abt \& Morrell(1995)]{AbtM95}  \href{http://adsabs.harvard.edu/abs/1995ApJS...99..135A}{Abt, H.~A., \& Morrell, N.~I., 1995}, \apjs, 99, 135

\bibitem[Abt \& Snowden(1973)]{AbtS73} \href{http://adsabs.harvard.edu/abs/1973ApJS...25..137A}{Abt, H.~A., \& Snowden, M.~S., 1973}, \apjs, 25, 137
\bibitem[Acevedo-Arreguin et al.(2013)]{Acevedo-Arreguin13} \href{http://adsabs.harvard.edu/abs/2013MNRAS.434..720A}{Acevedo-Arreguin, L.~A., Garaud, P., \& Wood, T.~S.\ 2013}, \mnras, 434, 720 
\bibitem[Acheson(1978)]{Acheson78}\href{http://adsabs.harvard.edu/abs/1978RSPTA.289..459A}{Acheson, D.~J., 1978}, Royal Society of London Philosophical Transactions Series A, 289, 459
\bibitem[Acheson(1979)]{1979SoPh...62...23A}\href{http://adsabs.harvard.edu/abs/1979SoPh...62...23A}{Acheson, D.~J.\ 1979}, \solphys, 62, 23 
\bibitem[Akg{\"u}n et al.(2013)]{Akguen_13}\href{http://adsabs.harvard.edu/abs/2013MNRAS.433.2445A}{Akg{\"u}n, T., Reisenegger, A., Mastrano, A., \& Marchant, P., 2013}, arXiv:1302.0273 
\bibitem[Alecian et al.(2013a)]{Alecian_13a}\href{http://adsabs.harvard.edu/abs/2013A\%26A...549L...8A}{Alecian, E., Neiner, C., Mathis, S., et al., 2013a}, \aap, 549, L8 
\bibitem[Alecian et al.(2013b)]{Alecian_13b}\href{http://adsabs.harvard.edu/abs/2013MNRAS.429.1027A}{Alecian, E., Wade, G.~A., Catala, C., et al., 2013b}, \mnras, 429, 1001
\bibitem[Alecian et al.(2013c)]{Alecian_13c} \href{http://adsabs.harvard.edu/abs/2013MNRAS.429.1001A}{Alecian, E., Wade, G.~A., Catala, C., et al., 2013c}, \mnras, 429, 1027
\bibitem[Alvan et al.(2013)]{2013sf2a.conf...77A}\href{http://adsabs.harvard.edu/abs/2013sf2a.conf...77A}{Alvan, L., Brun, A.-S., \& Mathis, S.\ 2013}, SF2A-2013: Proceedings of the Annual meeting of the French Society of Astronomy and Astrophysics, 77 
\bibitem[Auri{\`e}re et al.(2007)]{Auriere_07}\href{http://adsabs.harvard.edu/abs/2007A\%26A...475.1053A}{Auri{\`e}re, M., Wade, G.~A., Silvester, J., et al., 2007}, \aap, 475, 1053 
\bibitem[Balbus \& Hawley(1991)]{BalbusH91}\href{http://adsabs.harvard.edu/abs/1991ApJ...376..214B}{Balbus, S.~A., \& Hawley, J.~F., 1991}, \apj, 376, 214
\bibitem[Barker \& Ogilvie(2011)]{BarkerO11}\href{http://adsabs.harvard.edu/abs/2011MNRAS.417..745B}{Barker, A.~J., \& Ogilvie, G.~I., 2011}, \mnras, 417, 745 
\bibitem[Berger et al.(2005)]{Berger05}\href{http://adsabs.harvard.edu/abs/2005A\%26A...444..565B} {Berger, L., Koester, D., Napiwotzki, R., Reid, I.~N., \& Zuckerman, B.\ 2005}, \aap, 444, 565 
\bibitem[Bernstein et al.(1958)]{Bernstein_58}\href{http://adsabs.harvard.edu/abs/1958RSPSA.244...17B}{Bernstein, I.~B., Frieman, E.~A., Kruskal, M.~D., \& Kulsrud, R.~M., 1958}, Royal Society of London Proceedings Series A, 244, 17 
\bibitem[Bidelman(2002)]{Bidelman02}\href{http://adsabs.harvard.edu/abs/2002Obs...122..343B}{Bidelman, W.~P., 2002}, The Observatory, 122, 343 
\bibitem[Biermann(1950)]{Biermann}\href{http://adsabs.harvard.edu/abs/1950ZNatA...5...65B} {Biermann, L.\ 1950}, Zeitschrift Naturforschung Teil A, 5, 65
\bibitem[Blandford et al.(1983)]{Bland83}\href{http://adsabs.harvard.edu/abs/1983MNRAS.204.1025B} {Blandford, R.~D., Applegate, J.~H., \& Hernquist, L.\ 1983}, \mnras, 204, 1025
\bibitem[Bogomazov \& Tutukov(2009)]{BogomazovT09} \href{http://adsabs.harvard.edu/abs/2009ARep...53..214B}{Bogomazov, A.~I., \& Tutukov, A.~V., 2009}, Astronomy Reports, 53, 214 
\bibitem[Braithwaite(2006a)]{Braithwaite06a}\href{http://adsabs.harvard.edu/abs/2006A\%26A...449..451B}{Braithwaite, J., 2006a}, \aap, 449, 451
\bibitem[Braithwaite(2006b)]{Braithwaite06b}\href{http://adsabs.harvard.edu/abs/2006A\%26A...453..687B}{Braithwaite, J., 2006b}, \aap, 453, 687
\bibitem[Braithwaite(2008)]{Braithwaite08}\href{http://adsabs.harvard.edu/abs/2008MNRAS.386.1947B}{Braithwaite, J., 2008}, \mnras, 386, 1947
\bibitem[Braithwaite(2009)]{Braithwaite09}\href{http://adsabs.harvard.edu/abs/2009MNRAS.397..763B}{Braithwaite, J., 2009}, \mnras, 397, 763
\bibitem[Braithwaite(2012)]{Braithwaite12}\href{http://adsabs.harvard.edu/abs/2012MNRAS.422..619B}{Braithwaite, J., 2012}, \mnras, 422, 619
\bibitem[Braithwaite(2015)]{Braithwaite15}\href{http://adsabs.harvard.edu/abs/2015MNRAS.450.3201B}{Braithwaite, J.\ 2015}, \mnras, 450, 3201 
\bibitem[Braithwaite \& Spruit(2004)]{BraithwaiteS04}\href{http://adsabs.harvard.edu/abs/2004Natur.431..819B}{Braithwaite, J., \& Spruit, H.~C., 2004}, \nat, 431, 819 
\bibitem[Braithwaite \& Nordlund(2006)]{BraithwaiteN06}\href{http://adsabs.harvard.edu/abs/2006A\%26A...450.1077B}{Braithwaite, J., \& Nordlund, {\AA}., 2006}, \aap, 450, 1077 
\bibitem[Braithwaite \& Cantiello(2013)]{BraithwaiteC13}\href{http://adsabs.harvard.edu/abs/2013MNRAS.428.2789B}{Braithwaite, J., \& Cantiello, M., 2013}, \mnras, 428, 2789 
\bibitem[Broderick \& Narayan(2008)]{BroderickN08}\href{http://adsabs.harvard.edu/abs/2008MNRAS.383..943B}{Broderick, A.~E., \& Narayan, R., 2008}, \mnras, 383, 943
\bibitem[Browning(2008)]{Browning08}\href{http://adsabs.harvard.edu/abs/2008ApJ...676.1262B}{Browning, M.~K.\ 2008}, \apj, 676, 1262 
\bibitem[Browning et al.(2004)]{Browning_04}\href{http://adsabs.harvard.edu/abs/2004ApJ...601..512B}{Browning, M.~K., Brun, A.~S., \& Toomre, J., 2004}, \apj, 601, 512
\bibitem[Brun et al.(2005)]{Brun05}\href{http://adsabs.harvard.edu/abs/2005ApJ...629..461B}{Brun, A.~S., Browning, M.~K., \& Toomre, J.\ 2005}, \apj, 629, 461 
\bibitem[Cantiello et al.(2009)]{Cantiello_09}\href{http://adsabs.harvard.edu/abs/2009A\%26A...499..279C}{Cantiello, M., Langer, N., Brott, I., et al., 2009}, \aap, 499, 279 
\bibitem[Cantiello \& Braithwaite(2011)]{CantielloB11}\href{http://adsabs.harvard.edu/abs/2011A\%26A...534A.140C}{Cantiello, M., \& Braithwaite, J., 2011}, \aap, 534, A140
\bibitem[Cantiello et al.(2011)]{Cantiello_11}\href{http://adsabs.harvard.edu/abs/2011IAUS..272...32C}{Cantiello, M., Braithwaite, J., Brandenburg, A., et al., 2011}, IAU Symposium 272, 32 
\bibitem[Cantiello et al.(2014)]{Cantiello14}\href{http://adsabs.harvard.edu/abs/2014ApJ...788...93C}{Cantiello, M., Mankovich, C., Bildsten, L., Christensen-Dalsgaard, J., \& Paxton, B.\ 2014}, \apj, 788, 93 
\bibitem[Carlsson et al.(2004)]{2004ApJ...610L.137C}\href{http://adsabs.harvard.edu/abs/2004ApJ...610L.137C}{Carlsson, M., Stein, R.~F., Nordlund, {\AA}., \& Scharmer, G.~B.\ 2004}, \apjl, 610, L137 
\bibitem[Carrier et al.(2002)]{Carrier_02}\href{http://adsabs.harvard.edu/abs/2002A\%26A...394..151C}{Carrier, F., North, P., Udry, S., \& Babel, J., 2002}, \aap, 394, 151
\bibitem[Chandrasekhar(1960)]{Chandrasekhar60}\href{http://adsabs.harvard.edu/abs/1960PNAS...46..253C}{Chandrasekhar, S., 1960}, Proceedings of the National Academy of Science, 46, 253
\bibitem[Chandrasekhar(1961)]{Chandrasekhar61}\href{http://adsabs.harvard.edu/abs/1961hhs..book.....C}{Chandrasekhar, S., 1961}, International Series of Monographs on Physics, Oxford: Clarendon, 1961 
\bibitem[Charbonneau(2010)]{2010LRSP....7....3C}\href{http://adsabs.harvard.edu/abs/2010LRSP....7....3C}{Charbonneau, P.\ 2010}, Living Reviews in Solar Physics, 7, 3 
\bibitem[Charbonnel \& Talon(2005)]{2005Sci...309.2189C}\href{http://adsabs.harvard.edu/abs/2005Sci...309.2189C}{Charbonnel, C., \& Talon, S.\ 2005}, Science, 309, 2189 
\bibitem[Choudhuri(1998)]{1998pfp..book.....C}\href{http://adsabs.harvard.edu/abs/1998pfp..book.....C}{Choudhuri, A.~R.\ 1998}, The physics of fluids and plasmas : an introduction for astrophysicists ,  Cambridge University Press
\bibitem[Chui \& Moffatt(1995)]{ChuiM95}\href{http://adsabs.harvard.edu/abs/1995RSPSA.451..609C}{Chui, A.~Y.~K., \& Moffatt, H.~K., 1995}, Proc., Roy., Soc., Lond., A - Mat. 451 (1943), 609
\bibitem[Ciolfi et al.(2009)]{Ciolfi_09}\href{http://adsabs.harvard.edu/abs/2009MNRAS.397..913C}{Ciolfi, R., Ferrari, V., Gualtieri, L., \& Pons, J.~A., 2009}, \mnras, 397, 913 
\bibitem[Ciolfi et al.(2010)]{Ciolfi_10}\href{http://adsabs.harvard.edu/abs/2010MNRAS.406.2540C}{Ciolfi, R., Ferrari, V., \& Gualtieri, L., 2010}, \mnras, 406, 2540
\bibitem[Cowling(1945)]{Cowling45}\href{http://adsabs.harvard.edu/abs/1945MNRAS.105..166C}{Cowling, T.~G., 1945}, \mnras, 105, 166 
\bibitem[Colaiuda et al.(2008)]{2008MNRAS.385.2080C}\href{http://adsabs.harvard.edu/abs/2008MNRAS.385.2080C}{Colaiuda, A., Ferrari, V., Gualtieri, L., \& Pons, J.~A.\ 2008}, \mnras, 385, 2080 
\bibitem[Deutsch(1952)]{Deutsch52}\href{http://adsabs.harvard.edu/abs/1952ApJ...116..536D}{Deutsch, A.~J., 1952}, \apj, 116, 536 
\bibitem[de Val-Borro et al.(2009)]{deValBorro09}\href{http://adsabs.harvard.edu/abs/2009ApJ...700.1148D}{de Val-Borro, M., Karovska, M., \& Sasselov, D.\ 2009}, \apj, 700, 1148 
\bibitem[Donati(2001)]{Donati01}\href{http://adsabs.harvard.edu/abs/2001LNP...573.....B}{Donati, J.-F., 2001}, Astrotomography, Indirect Imaging Methods in Observational Astronomy, Boffin, H.M.J., Steeghs, D., Cuypers, J. (Eds.),  Springer LNP  573, 207
\bibitem[Dolginov \& Urpin(1980)]{Dolg80} Dolginov, A.~Z., \& Urpin, V.~A.\ 1980, \apss, 69, 259
\bibitem[Donati et al.(2009)]{Donati_06}\href{http://adsabs.harvard.edu/abs/2006MNRAS.370..629D}{Donati, J.-F., Howarth, I.~D., Jardine, M.~M., et al., 2006}, \mnras, 370, 629 
\bibitem[Donati et al.(2008)]{Donati_08}\href{http://adsabs.harvard.edu/abs/2008MNRAS.390..545D}{Donati, J.-F., Morin, J., Petit, P., et al., 2008}, \mnras, 390, 545 
\bibitem[Donati \& Landstreet(2009)]{DonatiL09}\href{http://adsabs.harvard.edu/abs/2009ARA\%26A..47..333D}{Donati, J.-F., \& Landstreet, J.~D., 2009}, \araa, 47, 333
\bibitem[Donati et al.(2005)]{Donati_05}\href{http://adsabs.harvard.edu/abs/2005Natur.438..466D}{Donati, J.-F., Paletou, F., Bouvier, J., \& Ferreira, J., 2005}, \nat, 438, 466 
\bibitem[Donati et al.(1997)]{Donati_97}\href{http://adsabs.harvard.edu/abs/1997MNRAS.291..658D}{Donati, J.-F., Semel, M., Carter, B.~D., Rees, D.~E., \& Collier Cameron, A., 1997}, \mnras, 291, 658
\bibitem[Drake et al.(2014)]{Drake14}\href{http://adsabs.harvard.edu/abs/2014ApJ...786..136D} {Drake, J.~J., Braithwaite, J., Kashyap, V., G{\"u}nther, H.~M., \& Wright, N.~J.\ 2014}, \apj, 786, 136 
\bibitem[Duez \& Mathis(2009)]{DuezM09}\href{http://arxiv.org/abs/0904.1568}{Duez, V., \& Mathis, S., 2009}, arXiv:0904.1568 
\bibitem[Duez et al.(2010)]{Duez_10}\href{http://adsabs.harvard.edu/abs/2010ApJ...724L..34D}{Duez, V., Braithwaite, J., \& Mathis, S., 2010}, \apjl, 724, L34
\bibitem[Eggenberger et al.(2005)]{2005A&A...440L...9E}\href{http://adsabs.harvard.edu/abs/2005A\%26A...440L...9E}{Eggenberger, P., Maeder, A., \& Meynet, G.\ 2005}, \aap, 440, L9 
\bibitem[Fan(2009)]{2009LRSP....6....4F}\href{http://adsabs.harvard.edu/abs/2009LRSP....6....4F}{Fan, Y.\ 2009}, Living Reviews in Solar Physics, 6, 4 
\bibitem[Featherstone et al.(2009)]{Featherstone09}\href{http://adsabs.harvard.edu/abs/2009ApJ...705.1000F}{Featherstone, N.~A., Browning, M.~K., Brun, A.~S., \& Toomre, J., 2009}, \apj, 705, 1000
\bibitem[Feiden \& Chaboyer(2013)]{Feiden13}\href{http://adsabs.harvard.edu/abs/2013ApJ...779..183F} {Feiden, G.~A., \& Chaboyer, B.\ 2013}, \apj, 779, 183 
\bibitem[Ferrario et al.(2009)]{Ferrario_09}\href{http://adsabs.harvard.edu/abs/2009MNRAS.400L..71F}{Ferrario, L., Pringle, J.~E., Tout, C.~A., \& Wickramasinghe, D.~T., 2009}, \mnras, 400, L71
\bibitem[Flowers \& Ruderman(1977)]{FlowersR77}\href{http://adsabs.harvard.edu/abs/1977ApJ...215..302F}{Flowers, E., \& Ruderman, M.A., 1977}, Astrophys., J.,  215, 302
\bibitem[Folsom et al. (2013a)]{Folsom_13a}\href{http://adsabs.harvard.edu/abs/2014IAUS..302...87F}{Folsom, C.~P., Bagnulo, S., Wade, G.~A., Landstreet, J.~D., \& Alecian, E., 2013a}, arXiv:1311.1552 
\bibitem[Fossati et al. (2016)]{Fossati16} \href{http://adsabs.harvard.edu/abs/2016A\%26A...592A..84F}{Fossati, L., Schneider, F.~R.~N., Castro, N., et al. 2016},\aap 592, A84
\bibitem[Folsom et al.(2013b)]{Folsom_13b}\href{http://adsabs.harvard.edu/abs/2014IAUS..302..313F}{Folsom, C.~P., Wade, G.~A., Likuski, K., et al., 2013b}, arXiv:1311.1554 
\bibitem[Frieman \& Rotenberg(1960)]{FriemanR60}\href{http://adsabs.harvard.edu/abs/1960RvMP...32..898F}{Frieman, E., \& Rotenberg, M., 1960}, Reviews of Modern Physics, 32, 898
\bibitem[Fromang \& Stone(2009)]{FromangS09}\href{http://adsabs.harvard.edu/abs/2009A\%26A...507...19F}{Fromang, S., \& Stone, J.~M., 2009}, \aap, 507, 19 
\bibitem[Fujisawa et al.(2012)]{Fujisawa_12}\href{http://adsabs.harvard.edu/abs/2012MNRAS.422..434F}{Fujisawa, K., Yoshida, S., \& Eriguchi, Y., 2012}, \mnras, 422, 434
\bibitem[Galsgaard \& Nordlund(1996)]{GalsgaardN96}\href{http://adsabs.harvard.edu/abs/1996JGR...10113445G}{Galsgaard, K., \& Nordlund, \AA. 1996}, J., Geophys., Res., 101, 13445
\bibitem[Gerbaldi et al.(1985)]{Gerbaldi_85}\href{http://adsabs.harvard.edu/abs/1985A\%26A...146..341G}{Gerbaldi, M., Floquet, M., \& Hauck, B., 1985}, \aap, 146, 341 
\bibitem[Goedbloed \& Poedts(2004)]{2004prma.book.....G}\href{http://adsabs.harvard.edu/abs/2004prma.book.....G}{Goedbloed, J.~P.~H., \& Poedts, S.\ 2004}, {\it Principles of Magnetohydrodynamics}.~ISBN 0521626072. Cambridge University Press
\bibitem[Goedbloed et al.(2010)]{2010adma.book.....G}\href{http://adsabs.harvard.edu/abs/2010adma.book.....G}{Goedbloed, J.~P., Keppens, R., 
\& Poedts, S.\ 2010}, {\it Advanced Magnetohydrodynamics}, Cambridge University Press, 2010 
\bibitem[Gough \& McIntyre(1998)]{GoughM98}\href{http://adsabs.harvard.edu/abs/1998Natur.394..755G}{Gough, D.~O., \& McIntyre, M.~E., 1998}, \nat, 394, 755
\bibitem[Gough \& Tayler(1966)]{GoughT66}\href{http://adsabs.harvard.edu/abs/1966MNRAS.133...85G}{Gough, D.~O., \& Tayler, R.~J., 1966}, \mnras, 133, 85
\bibitem[Gourgouliatos et al.(2013)]{Gourgouliatos_13}\href{http://adsabs.harvard.edu/abs/2013MNRAS.434.2480G}{Gourgouliatos, K.~N., Cumming, A., Reisenegger, A., et al., 2013}, \mnras, 434, 2480
\bibitem[Gregory et al.(2013)]{Gregory_13}\href{http://adsabs.harvard.edu/abs/2014IAUS..302...40G}{Gregory, S.~G., Donati,  J.-F., Morin, J., et al., 2013}, arXiv:1309.7556 
\bibitem[Harding(2013)]{Harding13}\href{http://adsabs.harvard.edu/abs/2013FrPhy...8..679H}{Harding, A.~K., 2013}, Frontiers of Physics, 8,679 
\bibitem[Haskell et al.(2008)]{Haskell_08}\href{http://adsabs.harvard.edu/abs/2008MNRAS.385..531H}{Haskell, B., Samuelsson, L., Glampedakis, K., \& Andersson, N., 2008}, \mnras, 385, 531 
\bibitem[Heger(2012)]{Heger12}\href{http://adsabs.harvard.edu/abs/2012ASSL..384..299H}{Heger, A., 2012}, Astrophysics and Space Science Library, 384, 299 
\bibitem[Heger(2013)]{Heger13}{Heger, A., 2013}, private communication 

\bibitem[Heinemann et al.(2007)]{Heinemann_07}\href{http://adsabs.harvard.edu/abs/2007ApJ...669.1390H}{Heinemann, T., Nordlund, \AA., Scharmer, G. B., \& Spruit, H. C. 2007}, ApJ, 669, 1390
\bibitem[Henrichs et al.(2000)]{Henrichs_00}\href{http://adsabs.harvard.edu/abs/2000ASPC..214..324H}{Henrichs, H.~F., de Jong, J.~A., Donati, J.-F., et al., 2000}, IAU Colloq.~175: The Be Phenomenon in Early-Type Stars, 214, 324 
\bibitem[Henrichs(2012)]{Henrichs12}\href{http://adsabs.harvard.edu/abs/2012POBeo..91...13H}{Henrichs, H.~F., 2012}, Publications de l'Observatoire Astronomique de Beograd, 91, 13 
\bibitem[Herbig \& Bell(1988)]{HerbigB88}\href{http://adsabs.harvard.edu/abs/1988cels.book.....H}{Herbig, G.~H., \& Bell, K.~R., 1988}, Third catalog of emission-line stars of the Orion population,  Lick Observatory Bulletin \#1111, Santa Cruz: Lick Observatory, p90
\bibitem[Heyvaerts \& Priest(1983)]{HeyvaertsP83}\href{http://adsabs.harvard.edu/abs/1983A\%26A...117..220H}{Heyvaerts, J., \& Priest, E.~R., 1983}, \aap, 117, 220 
\bibitem[Hoffman \& Heyl(2012)]{Hoffman12}\href{http://adsabs.harvard.edu/abs/2012MNRAS.426.2404H}{Hoffman, K., \& Heyl, J.\ 2012}, \mnras, 426, 2404 
\bibitem[Horowitz \& Kadau(2009)]{Horowitz09}\href{http://adsabs.harvard.edu/abs/2009PhRvL.102s1102H}{Horowitz, C.~J., \& Kadau, K.\ 2009}, Physical Review Letters, 102, 191102 
\bibitem[Hsu \& Bellan(2002)]{HsuB02}\href{http://adsabs.harvard.edu/abs/2002MNRAS.334..257H}{Hsu, S.~C., \& Bellan, P.~M., 2002}, \mnras, 334, 257 
\bibitem[Hubrig et al.(2000)]{Hubrig_00}\href{http://adsabs.harvard.edu/abs/2000ApJ...539..352H}{Hubrig, S., North, P., \& Mathys, G., 2000}, \apj, 539, 352
\bibitem[Hubrig et al.(2013)]{Hubrig_13}\href{http://adsabs.harvard.edu/abs/2014EPJWC..6408006H}{Hubrig, S., Ilyin, I., Schoeller, M., et al., 2013}, arXiv:1308.6777 
\bibitem[Hughes \& Weiss(1995)]{1995JFM...301..383H}\href{http://adsabs.harvard.edu/abs/1995JFM...301..383H}{Hughes, D.~W., \& Weiss, N.~O.\ 1995}, Journal of Fluid Mechanics, 301, 383 
\bibitem[Hussain(2012)]{Hussain12}\href{http://adsabs.harvard.edu/abs/2012AN....333....4H}{Hussain, G.~A.~J., 2012}, Astronomische Nachrichten, 333, 4 
\bibitem[Ib{\'a}{\~n}ez-Mej{\'{\i}}a \& Braithwaite(2015)]{Ibanez15}\href{http://adsabs.harvard.edu/abs/2015A\%26A...578A...5I}{Ib{\'a}{\~n}ez-Mej{\'{\i}}a, J.~C., \& Braithwaite, J.\ 2015}, \aap, 578, A5 
\bibitem[Igumenshchev et al.(2003)]{Igumenshchev_03}\href{http://adsabs.harvard.edu/abs/2003ApJ...592.1042I}{Igumenshchev, I.~V., Narayan, R., \& Abramowicz, M.~A., 2003}, \apj, 592, 1042
\bibitem[Ioka \& Sasaki(2004)]{2004ApJ...600..296I}\href{http://adsabs.harvard.edu/abs/2004ApJ...600..296I}{Ioka, K., \& Sasaki, M.\ 2004}, \apj, 600, 296 
\bibitem[Jouve et  al.(2015)]{2015A&A...575A.106J}\href{http://adsabs.harvard.edu/abs/2015A\%26A...575A.106J}{Jouve, L., Gastine, T., \& Ligni{\`e}res, F.\ 2015}, \aap, 575, A106 
\bibitem[Kaminker et al.(2006)]{Kaminker_06}\href{http://adsabs.harvard.edu/abs/2006MNRAS.371..477K}{Kaminker, A.~D., Yakovlev, D.~G., Potekhin, A.~Y., et al., 2006}, \mnras, 371, 477 
\bibitem[Kippenhahn et al.(2013)]{Kippenhahn_12}\href{http://adsabs.harvard.edu/abs/2012sse..book.....K}{Kippenhahn, R., Weigert, A., \& Weiss, A., 2012}, Stellar Structure and Evolution. Springer-Verlag Berlin Heidelberg
\bibitem[Kochukhov et al.(2002)]{Kochukhov_02}\href{http://adsabs.harvard.edu/abs/2002A\%26A...389..420K}{Kochukhov, O., Piskunov, N., Ilyin, I., Ilyina, S., \& Tuominen, I., 2002}, \aap, 389, 420
\bibitem[Kochukhov et al.(2004)]{Kochukhov_04}\href{http://adsabs.harvard.edu/abs/2004A\%26A...414..613K}{Kochukhov, O., Bagnulo, S., Wade, G.~A., et al., 2004}, \aap, 414, 613 
\bibitem[Kochukhov \& Wade(2010)]{KochukhovW10}\href{http://adsabs.harvard.edu/abs/2010A\%26A...513A..13K}{Kochukhov, O., \& Wade, G.~A., 2010}, \aap, 513, A13 
\bibitem[Kochukhov et al.(2011a)]{Kochukhov11a}\href{http://adsabs.harvard.edu/abs/2011ApJ...726...24K}{Kochukhov, O., Lundin, A., Romanyuk, I., \& Kudryavtsev, D., 2011a}, \apj, 726, 24 
\bibitem[Kochukhov et al.(2011b)]{Kochukhov11b}\href{http://adsabs.harvard.edu/abs/2011A\%26A...534L..13K}{Kochukhov, O., Makaganiuk, V., Piskunov, N., et al.\ 2011b}, \aap, 534, L13 
\bibitem[Kochukhov \& Sudnik(2013)]{KochukhovS13}\href{http://adsabs.harvard.edu/abs/2013A\%26A...554A..93K}{Kochukhov, O., \& Sudnik, N., 2013}, \aap, 554, A93 
\bibitem[Konstantinova-Antova et al.(2010)]{Konstantinova-Antova_10}\href{http://adsabs.harvard.edu/abs/2010A\%26A...524A..57K}{Konstantinova-Antova, R., Auri{\`e}re, M., Charbonnel, C., et al., 2010}, \aap, 524, A57
\bibitem[Kulsrud(2005)]{Kulsrud05}\href{http://adsabs.harvard.edu/abs/2005ppfa.book.....K}{Kulsrud, R.~M.\ 2005}, {\it Plasma physics for astrophysics}, Princeton University Press  
\bibitem[Landstreet(1991)]{Landstreet91} \href{http://adsabs.harvard.edu/abs/1991LNP...380..342L}{Landstreet, J.~D., 1991}, IAU Colloq.~130: The Sun and Cool Stars.~Activity, Magnetism, Dynamos, 380, 342
\bibitem[Landstreet(2011)]{Landstreet11} \href{http://adsabs.harvard.edu/abs/2011ASPC..449..249L}{Landstreet, J.~D., 2011}, Astronomical Society of the Pacific Conference Series, 449, 249 
\bibitem[Landstreet et al.(2009a)]{Landstreet_09a} \href{http://adsabs.harvard.edu/abs/2009ASPC..405..505L}{Landstreet, J.~D., Bagnulo, S., Andretta, V., et al., 2009a}, Solar Polarization 5: In Honor of Jan Stenflo, 405, 505
\bibitem[Landstreet et al.(2009b)]{Landstreet_09b} \href{http://adsabs.harvard.edu/abs/2009A\%26A...503..973L}{Landstreet, J.~D., Kupka, F., Ford, H.~A., et al., 2009b}, \aap, 503, 973
\bibitem[Landstreet \& Mathys(2000)]{LandstreetM00} \href{http://adsabs.harvard.edu/abs/2000A\%26A...359..213L}{Landstreet, J.~D., \& Mathys, G., 2000}, \aap, 359, 213
\bibitem[Levy \& Sonett(1978)]{LevyS78} \href{http://adsabs.harvard.edu/abs/1978prpl.conf..516L}{Levy, E.~H., \& Sonett, C.~P., 1978}, IAU Colloq.~52: Protostars and Planets, 516 
\bibitem[Ligni{\`e}res et al.(2009)]{Lignieres_09} \href{http://adsabs.harvard.edu/abs/2009A\%26A...500L..41L}{Ligni{\`e}res, F., Petit, P., B{\"o}hm, T., \& Auri{\`e}re, M., 2009}, \aap, 500, L41
\bibitem[Lyutikov(2010)]{Lyutikov10} \href{http://adsabs.harvard.edu/abs/2010MNRAS.402..345L}{Lyutikov, M., 2010}, \mnras, 402, 345
\bibitem[MacDonald \& Mullan(2004)]{MacDonaldM04} \href{http://adsabs.harvard.edu/abs/2004MNRAS.348..702M}{MacDonald, J., \& Mullan, D.~J., 2004}, \mnras, 348, 702
\bibitem[MacGregor \& Cassinelli(2003)]{MacGregorC03} \href{http://adsabs.harvard.edu/abs/2003ApJ...586..480M}{MacGregor, K.~B., \& Cassinelli, J.~P., 2003}, \apj, 586, 480 
\bibitem[Machida et al.(2008)]{Machida_08} \href{http://adsabs.harvard.edu/abs/2008ApJ...677..327M}{Machida, M.~N., Tomisaka, K., Matsumoto, T., \& Inutsuka, S.-i., 2008}, \apj, 677, 327 
\bibitem[Maitzen et al.(2008)]{Maitzen_08} \href{http://adsabs.harvard.edu/abs/2008CoSka..38..385M}{Maitzen, H.~M., Paunzen, E., \& Netopil, M., 2008}, Contributions of the Astronomical Observatory Skalnate Pleso, 38, 385 
\bibitem[Marchant et al.(2011)]{Marchant_11} \href{http://adsabs.harvard.edu/abs/2011MNRAS.415.2426M}{Marchant, P., Reisenegger, A., \& Akg{\"u}n, T., 2011}, \mnras, 415, 2426
\bibitem[Markey \& Tayler(1973)]{MarkeyT73} \href{http://adsabs.harvard.edu/abs/1973MNRAS.163...77M}{Markey, P., \& Tayler, R.~J., 1973}, \mnras, 163, 77 
\bibitem[Markey \& Tayler(1974)]{MarkeyT74} \href{http://adsabs.harvard.edu/abs/1974MNRAS.168..505M}{Markey, P., \& Tayler, R.~J., 1974}, \mnras, 168, 505 
\bibitem[Marsden et al.(2014)]{2014MNRAS.444.3517M} \href{http://adsabs.harvard.edu/abs/2014MNRAS.444.3517M}{Marsden, S.~C., Petit, P., Jeffers, S.~V., et al.\ 2014}, \mnras, 444, 3517 
\bibitem[Mathis et al.(2008)]{Mathis_08} \href{http://adsabs.harvard.edu/abs/2008SoPh..251..101M}{Mathis, S., Talon, S., Pantillon, F.-P., \& Zahn, J.-P., 2008}, \solphys, 251, 101 
\bibitem[Mathys(2008)]{Mathys08} \href{http://adsabs.harvard.edu/abs/2008CoSka..38..217M}{Mathys, G., 2008}, Contributions of the Astronomical Observatory Skalnate Pleso, 38, 217 
\bibitem[Mathys(2012)]{Mathys12} \href{http://adsabs.harvard.edu/abs/2012ASPC..462..295M}{Mathys, G., 2012}, Progress in Solar/Stellar Physics with Helio- and Asteroseismology, 462, 295
\bibitem[Mathys(2016)]{Mathys16} \href{http://adsabs.harvard.edu/abs/2016arXiv161203632M}{Mathys, G., 2016}, 	eprint arXiv:1612.03632, accepted for publication in A\&A
\bibitem[Mereghetti(2008)]{Mereghetti08} \href{http://adsabs.harvard.edu/abs/2008A\%26ARv..15..225M}{Mereghetti, S., 2008}, \aapr, 15, 225 
\bibitem[Mestel(1953)]{Mestel53}\href{http://adsabs.harvard.edu/abs/1953MNRAS.113..716M}{Mestel, L., 1953}, \mnras, 113, 716
\bibitem[Mestel(1961)]{Mestel61} \href{http://adsabs.harvard.edu/abs/1961MNRAS.122..473M}{Mestel, L., 1961}, \mnras, 122, 473 
\bibitem[Mestel(1970)]{Mestel70} \href{http://adsabs.harvard.edu/abs/1970MSRSL..19..167M}{Mestel, L., 1970}, Memoires of the Societe Royale des Sciences de Liege, 19, 167
\bibitem[Michaud(1970)]{Michaud70} \href{http://adsabs.harvard.edu/abs/1970ApJ...160..641M}{Michaud, G., 1970}, \apj, 160, 641 
\bibitem[Michaud et al.(2013)]{Michaud13}\href{http://adsabs.harvard.edu/abs/2013AN....334..114M}{Michaud, G., Richer, 
J., \& Richard, O.\ 2013}, Astronomische Nachrichten, 334, 114 
\bibitem[Mikul{\'a}{\v s}ek et al.(2011a)]{Mikulasek_11a} \href{http://adsabs.harvard.edu/abs/2011A\%26A...534L...5M}{Mikul{\'a}{\v s}ek, Z., Krti{\v c}ka, J., Henry, G.~W., et al., 2011a}, \aap, 534, L5 
\bibitem[Mikul{\'a}{\v s}ek et al.(2011b)]{Mikulasek_11b} \href{http://adsabs.harvard.edu/abs/2011mast.conf...52M}{Mikul{\'a}{\v s}ek, Z., Krti{\v c}ka, J., Jan{\'{\i}}k, J., et al., 2011b}, Magnetic Stars, eds I. I. Romanyuk \& D. O. Kudryavtsev, p. 52 
\bibitem[Mikul{\'a}{\v s}ek et al.(2013)]{Mikulasek_13} \href{http://adsabs.harvard.edu/abs/2014psce.conf..270M}{Mikul{\'a}{\v s}ek, Z., Krti{\v c}ka, J., Jan{\'{\i}}k, J., et al., 2013}, Putting A Stars into Context, Proceedings June 3-7, 2013 at  Lomonosov State University, Eds.: G. Mathys, O. Kochukhov et al., p. 270-278
\bibitem[Mitchell et al.(2014)]{Mitchell14} \href{http://adsabs.harvard.edu/abs/2014IAUS..302..441M}{Mitchell, J.~P., Braithwaite, J., Langer, N., Reisenegger, A., \& Spruit, H.\ 2014}, IAU Symposium 302, 441 
\bibitem[Mitchell et al.(2015)]{2015MNRAS.447.1213M} \href{http://adsabs.harvard.edu/abs/2015MNRAS.447.1213M}{Mitchell, J.~P., Braithwaite, J., Reisenegger, A., et al.\ 2015}, \mnras, 447, 1213 
\bibitem[Morin et al.(2008a)]{Morin_08a} \href{http://adsabs.harvard.edu/abs/2008MNRAS.384...77M}{Morin, J., Donati, J.-F., Forveille, T., et al., 2008a}, \mnras, 384, 77
\bibitem[Morin et al.(2008b)]{Morin_08b} \href{http://adsabs.harvard.edu/abs/2008MNRAS.390..567M}{Morin, J., Donati, J.-F., Petit, P., et al., 2008b}, \mnras, 390, 567 
\bibitem[Morin et al.(2010)]{Morin_10} \href{http://adsabs.harvard.edu/abs/2010MNRAS.407.2269M}{Morin, J., Donati, J.-F., Petit, P., et al., 2010}, \mnras, 407, 2269 
\bibitem[Moss(1989)]{Moss89} \href{http://adsabs.harvard.edu/abs/1989MNRAS.236..629M}{Moss, D., 1989}, \mnras, 236, 629 
\bibitem[Moss \& Tayler(1969)]{MossT69} \href{http://adsabs.harvard.edu/abs/1969MNRAS.145..217M}{Moss, D.~L., \& Tayler, R.~J., 1969}, \mnras, 145, 217
\bibitem[Mosser et al.(2012)]{Mosser_12} \href{http://adsabs.harvard.edu/abs/2012A\%26A...548A..10M}{Mosser, B., Goupil, M.~J., Belkacem, K., et al., 2012}, \aap, 548, A10
\bibitem[Newcomb(1961)]{Newcomb61} \href{http://adsabs.harvard.edu/abs/1961PhFl....4..391N}{Newcomb, W.A., 1961}, Phys. Fluids 4, 391
\bibitem[Ogilvie \& Lin(2007)]{OgilvieL07} \href{http://adsabs.harvard.edu/abs/2007ApJ...661.1180O}{Ogilvie, G.~I., \& Lin, D.~N.~C., 2007}, \apj, 661, 1180 
\bibitem[Oskinova et al.(2012)]{Oskinova_12} \href{http://adsabs.harvard.edu/abs/2012ASPC..465..172O}{Oskinova, L., Hamann, W.-R., Todt, H., \& Sander, A., 2012}, Proceedings of a Scientific Meeting in Honor of Anthony F.~J.~Moffat, 465, 172
\bibitem[Palla \& Stahler(1993)]{1993ApJ...418..414P} \href{http://adsabs.harvard.edu/abs/1993ApJ...418..414P}{Palla, F., \& Stahler, S.~W.\ 1993}, \apj, 418, 414 
\bibitem[Parker(1963)]{Parker63} \href{http://adsabs.harvard.edu/abs/1963ApJ...138..552P}{Parker, E.~N., 1963}, \apj, 138, 552
\bibitem[Parker(1966)]{Parker66} \href{http://adsabs.harvard.edu/abs/1966ApJ...145..811P}{Parker, E.~N., 1966}, \apj, 145, 811 
\bibitem[Parker(1972)]{Parker72} \href{http://adsabs.harvard.edu/abs/1972ApJ...174..499P}{Parker, E.~N., 1972}, \apj, 174, 499 
\bibitem[Parker(1979a)]{Parker79a} \href{http://adsabs.harvard.edu/abs/1979ApJ...234..333P}{Parker, E.~N., 1979a}, \apj, 234, 333
\bibitem[Parker(1979b)]{Parker79b} \href{http://adsabs.harvard.edu/abs/1979cmft.book.....P}{Parker, E.~N. 1979b}, {\it Cosmical magnetic fields}, Clarendon, Oxford
\bibitem[Parker(1979c)]{Parker79c} \href{http://adsabs.harvard.edu/abs/1979Ap\%26SS..62..135P}{Parker, E.~N.\ 1979c}, \apss, 62, 135
\bibitem[Parker(2012)]{Parker12} \href{http://adsabs.harvard.edu/abs/2012ASSP...33....3P}{Parker, E.~N., 2012}, Astrophysics and Space Science Proceedings, 33, ISBN 978-3-642-30441-5, Springer, p. 3 
\bibitem[Pease et al.(2006)]{Pease_06} \href{http://adsabs.harvard.edu/abs/2006ApJ...636..426P}{Pease, D.~O., Drake, J.~J., \& Kashyap, V.~L., 2006}, \apj, 636, 426 
\bibitem[Pedlosky(1982)]{Pedlosky82} \href{http://adsabs.harvard.edu/abs/1982bsv..book.....P}{Pedlosky, J., 1982}, {\it Geophysical Fluid Dynamics}, New York and Berlin, Springer-Verlag, 1982.  
\bibitem[Petit et al.(2004)]{Petit_04} \href{http://adsabs.harvard.edu/abs/2004MNRAS.348.1175P}{Petit, P., Donati, J.-F., Wade, G.~A., et al., 2004}, \mnras, 348, 1175
\bibitem[Petit et al.(2010)]{Petit_10} \href{http://adsabs.harvard.edu/abs/2010A\%26A...523A..41P}{Petit, P., Ligni{\`e}res, F., Wade, G.~A., et al., 2010}, \aap, 523, A41 
\bibitem[Petit et al.(2011)]{Petit_11} \href{http://adsabs.harvard.edu/abs/2011A\%26A...532L..13P}{Petit, P., Ligni{\`e}res, F., Auri{\`e}re, M., et al., 2011}, \aap, 532, L13
\bibitem[Petit et al.(2013)]{Petit_13} \href{http://adsabs.harvard.edu/abs/2013MNRAS.429..398P}{Petit, V., Owocki, S.~P., Wade, G.~A., et al., 2013}, \mnras, 429, 398
\bibitem[Petrie(2012)]{Petrie12} \href{http://adsabs.harvard.edu/abs/2012SoPh..281..577P}{Petrie, G.~J.~D., 2012}, \solphys, 281, 577
\bibitem[Phan-Bao et al.(2009)]{Phan-Bao_09} \href{http://adsabs.harvard.edu/abs/2009ApJ...704.1721P}{Phan-Bao, N., Lim, J., Donati, J.-F., Johns-Krull, C.~M., \& Mart{\'{\i}}n, E.~L., 2009}, \apj, 704, 1721 
\bibitem[Piskunov \& Kochukhov(2003)]{PiskunovK03} \href{http://adsabs.harvard.edu/abs/2003ASPC..305...83P}{Piskunov, N.~E., \& Kochukhov, O., 2003}, Magnetic Fields in O, B and A Stars: Origin and Connection to Pulsation, Rotation and Mass Loss, 305, 83
\bibitem[Pitts \& Tayler(1985)]{PittsT85} \href{http://adsabs.harvard.edu/abs/1985MNRAS.216..139P}{Pitts, E., \& Tayler, R.~J., 1985}, \mnras, 216, 139  
\bibitem[Pizzolato et al.(2003)]{2003A&A...397..147P} \href{http://adsabs.harvard.edu/abs/2003A\%26A...397..147P}{Pizzolato, N., Maggio, A., Micela, G., Sciortino, S., \& Ventura, P.\ 2003}, \aap, 397, 147 
\bibitem[Power et al.(2007)]{Power_07} \href{http://adsabs.harvard.edu/abs/2007pms..conf...89P}{Power, J., Wade, G.~A., Hanes, D.~A., Aurier, M., \& Silvester, J., 2007}, Physics of Magnetic Stars, eds I. I. Romanyuk and D. O. Kudryavtsev, p. 89
\bibitem[Prendergast(1956)]{Prendergast56} \href{http://adsabs.harvard.edu/abs/1956ApJ...123..498P}{Prendergast, K.~H., 1956}, \apj, 123, 498 
\bibitem[Putney(1999)]{Putney99} \href{http://adsabs.harvard.edu/abs/1999ASPC..169..195P}{Putney, A., 1999}, 11th European Workshop on White Dwarfs, 169, 195
\bibitem[Pyper et al.(1998)]{Pyper_98} \href{http://adsabs.harvard.edu/abs/1998A\%26A...339..822P}{Pyper, D.~M., Ryabchikova, T., Malanushenko, V., et al., 1998}, \aap, 339, 822
\bibitem[Pyper et al.(2013)]{Pyper_13} \href{http://adsabs.harvard.edu/abs/2013MNRAS.431.2106P}{Pyper, D.~M., Stevens, I.~R., \& Adelman, S.~J., 2013}, \mnras, 431, 2106
\bibitem[R\"adler(1980)]{Raedler80} \href{http://adsabs.harvard.edu/abs/1980AN....301..101R}{R\"adler, K.-H., 1980}, Astron., Nachr. 301, 101
\bibitem[Reisenegger(2009)]{Reisenegger09} \href{http://adsabs.harvard.edu/abs/2009A\%26A...499..557R}{Reisenegger, A., 2009}, \aap, 499, 557
\bibitem[Rempel(2011)]{Rempel11} \href{http://adsabs.harvard.edu/abs/2011ApJ...740...15R}{Rempel, M., 2011}, \apj, 740, 15
\bibitem[Rempel(2014)]{2014ApJ...789..132R} \href{http://adsabs.harvard.edu/abs/2014ApJ...789..132R}{Rempel, M.\ 2014}, \apj, 789, 132 
\bibitem[Roberts(1967)]{Roberts67} \href{http://adsabs.harvard.edu/abs/1967imhd.book.....R}{Roberts, P.H., 1967}, {\it Magnetohydrodynamics}, Longmans, London.
\bibitem[Rogers et al.(2013)]{Rogers_13} \href{http://adsabs.harvard.edu/abs/2013ApJ...772...21R}{Rogers, T.~M., Lin, D.~N.~C., McElwaine, J.~N., \& Lau, H.~H.~B., 2013}, \apj, 772, 21 
\bibitem[Roxburgh(1966)]{Roxburgh66} \href{http://adsabs.harvard.edu/abs/1966MNRAS.132..347R}{Roxburgh, I.~W., 1966}, \mnras, 132, 347
\bibitem[Schmidt(2001)]{Schmidt01} \href{http://adsabs.harvard.edu/abs/2001ASPC..248..443S}{Schmidt G.D., 2001}, in Magnetic fields across the HR diagram, ASPC 248, eds. G.Mathys, S.K. Solanki \& D.T.Wickramasinghe. (San Francisco: ASP), 443 
\bibitem[Sch\"ussler \& V\"ogler(2006)]{SchuesslerV06} \href{http://adsabs.harvard.edu/abs/2006ApJ...641L..73S}{Sch\"ussler, M., \& V\"ogler, A. 2006}, ApJ, 641, L73
\bibitem[Schwarzschild(1970)]{schwarzschild70}\href{http://adsabs.harvard.edu/abs/1970QJRAS..11...12S}{Schwarzschild, M. 1970}, \qjras, 11, 12 
\bibitem[Sennhauser \& Berdyugina(2012)]{SennhauserB12} \href{http://adsabs.harvard.edu/abs/2012AIPC.1429...75S}{Sennhauser, C., \& Berdyugina, S.~V., 2012}, American Institute of Physics Conferences, 1429, 75 
\bibitem[Sorathia et al.(2012)]{Sorathia_12} \href{http://adsabs.harvard.edu/abs/2012ApJ...749..189S}{Sorathia, K.~A., Reynolds, C.~S., Stone, J.~M., \& Beckwith, K., 2012}, \apj, 749, 189 
\bibitem[Spruit(1991)]{Spruit91}\href{http://adsabs.harvard.edu/abs/1991suti.conf..118S}{Spruit, H.~C.\ 1991}, The Sun in Time, ed. Sonett, Tucson, AZ, University of Arizona Press, p.118 
\bibitem[Spruit(1998)]{Spruit98} \href{http://adsabs.harvard.edu/abs/1998A\%26A...333..603S}{Spruit, H.~C., 1998}, \aap, 333, 603 
\bibitem[Spruit(1999)]{Spruit99} \href{http://adsabs.harvard.edu/abs/1999A\%26A...349..189S}{Spruit, H.~C., 1999}, \aap, 349, 189
\bibitem[Spruit(2002)]{Spruit02} \href{http://adsabs.harvard.edu/abs/2002A\%26A...381..923S}{Spruit, H.~C., 2002}, \aap, 381, 923
\bibitem[Spruit(2008)]{Spruit08}\href{http://adsabs.harvard.edu/abs/2008AIPC..983..391S}{Spruit, H.~C.\ 2008}, 40 Years of Pulsars: Millisecond Pulsars, Magnetars and More, AIP Conference Proceedings 983, 391 
\bibitem[Spruit(2011)]{Spruit11} \href{http://adsabs.harvard.edu/abs/2011sswh.book...39S}{Spruit, H.~C., 2011}, in The Sun, the Solar Wind, and the Heliosphere, eds. M.P.\ Miralles and J.~S\'anchez Almeida, IAGA Special Sopron Book Series, Vol. 4. Berlin: Springer, p39 \url{http://arxiv.org/abs/1004.4545}
\bibitem[Spruit(2012)]{Spruit12} \href{http://adsabs.harvard.edu/abs/2012PThPS.195..185S}{Spruit, H.~C., 2012}, Progress of Theoretical Physics Supplement, 195, 185
\bibitem[Spruit(2013)]{Spruit13} \href{http://adsabs.harvard.edu/abs/2013arXiv1301.5572S}{Spruit, H.~C., 2013}, {\it Essential magnetohydrodynamics for astrophysics},  arXiv:1301.5572 , sect. 1.5.4, \url{http://www.mpa-garching.mpg.de/~henk/mhd12.zip} 
\bibitem[Spruit \& Weiss(1986)]{SpruitW86}\href{http://adsabs.harvard.edu/abs/1986A\%26A...166..167S} Spruit, H.~C., \& Weiss, A.\ 1986, \aap, 166, 167 
\bibitem[Spruit \& Phinney(1998)]{SpruitP98} \href{http://adsabs.harvard.edu/abs/1998Natur.393..139S}{Spruit, H.~C., \& Phinney, E.~S., 1998}, \nat, 393, 139 
\bibitem[Spruit \& Uzdensky(2005)]{SpruitU05} \href{http://adsabs.harvard.edu/abs/2005ApJ...629..960S}{Spruit, H.~C., \& Uzdensky, D.~A., 2005}, \apj, 629, 960 
\bibitem[Spruit \& Scharmer(2006)]{SpruitS06} \href{http://adsabs.harvard.edu/abs/2006A\%26A...447..343S}{Spruit, H.~C., \& Scharmer, G.~B., 2006}, \aap, 447, 343
\bibitem[Stahler \& Palla(2005)]{2005fost.book.....S} \href{http://adsabs.harvard.edu/abs/2005fost.book.....S}{Stahler, S.~W., \& Palla, F.\ 2005}, The Formation of Stars, ~ISBN 3-527-40559-3.~Wiley-VCH 
\bibitem[Stella et al.(2005)]{Stella05}\href{http://adsabs.harvard.edu/abs/2005ApJ...634L.165S
}{Stella, L., Dall'Osso, S., Israel, G.~L., \& Vecchio, A.\ 2005}, \apjl, 634, L165 
\bibitem[St\k{e}pie\'n(1998)]{Stepien98} \href{http://adsabs.harvard.edu/abs/1998A\%26A...337..754S}{St\k{e}pie\'n, K., 1998}, \aap, 337, 754
\bibitem[Stift \& Leone(2008)]{StiftL08} \href{http://adsabs.harvard.edu/abs/2008CoSka..38..185S}{Stift, M.~J., \& Leone, F., 2008}, Contributions of the Astronomical Observatory Skalnate Pleso, 38, 185
\bibitem[Strugarek et al.(2011)]{Strugarek11} \href{http://adsabs.harvard.edu/abs/2011A\%26A...532A..34S}{Strugarek, A., Brun, A.~S., \& Zahn, J.-P.\ 2011}, \aap, 532, A34 
\bibitem[Suijs et al.(2008)]{Suijs_08} \href{http://adsabs.harvard.edu/abs/2008A\%26A...481L..87S}{Suijs, M. P. L., Langer, N., Poelarends, A.-J., et al. 2008}, A\&A, 481, L87
\bibitem[Sundqvist et al.(2013)]{Sundqvist_13} \href{http://adsabs.harvard.edu/abs/2013MNRAS.433.2497S}{Sundqvist, J.~O., Petit, V., Owocki, S.~P., et al., 2013}, \mnras, 433, 2497 
\bibitem[Suwa(2013)]{Suwa13} \href{http://adsabs.harvard.edu/abs/2014PASJ...66L...1S}{Suwa, Y., 2013}, \pasj, 66, L1 
\bibitem[Tayler(1973)]{Tayler73} \href{http://adsabs.harvard.edu/abs/1973MNRAS.161..365T}{Tayler, R.~J., 1973}, \mnras, 161, 365 
\bibitem[Tchekhovskoy et al.(2011)]{Tchekhovskoy_11} \href{http://adsabs.harvard.edu/abs/2011MNRAS.418L..79T}{Tchekhovskoy, A., Narayan, R., \& McKinney, J.~C., 2011}, \mnras, 418, L79 
\bibitem[Thompson \& Duncan(1993)]{ThompsonD93} \href{http://adsabs.harvard.edu/abs/1993ApJ...408..194T}{Thompson, C., \& Duncan, R.~C., 1993}, \apj, 408, 194
\bibitem[Turcotte(2003)]{Turcotte03} \href{http://adsabs.harvard.edu/abs/2003ASPC..305..199T}{Turcotte, S., 2003}, Magnetic Fields in O, B and A Stars: Origin and Connection to Pulsation, Rotation and Mass Loss, ASP Conference 305, 199 
\bibitem[Urpin \& Yakovlev(1980)]{Urp80}\href{http://adsabs.harvard.edu/abs/1980SvA....24..425U} {Urpin, V.~A., \& Yakovlev, D.~G.\ 1980}, \sovast, 24, 425 
\bibitem[van Ballegooijen(1989)]{vanBallegooijen89} \href{http://adsabs.harvard.edu/abs/1989ASSL..156...99V}{van Ballegooijen, A.~A., 1989}, in Accretion Disks and Magnetic Fields in Astrophysics, Kluwer Academic Publishers, (ASSL vol., 156),  p99 
\bibitem[Vidotto et al.(2014)]{2014MNRAS.441.2361V} \href{http://adsabs.harvard.edu/abs/2014MNRAS.441.2361V}{Vidotto, A.~A., Gregory, S.~G., Jardine, M., et al.\ 2014}, \mnras, 441, 2361 
\bibitem[Vieira et al.(2003)]{Vieira_03} \href{http://adsabs.harvard.edu/abs/2003AJ....126.2971V}{Vieira, S.~L.~A., Corradi, W.~J.~B., Alencar, S.~H.~P., et al., 2003}, \aj, 126, 2971
\bibitem[Vink \& Kuiper(2006)]{VinkK06} \href{http://adsabs.harvard.edu/abs/2006MNRAS.370L..14V}{Vink, J., \& Kuiper, L., 2006}, \mnras, 370, L14
\bibitem[Vlemmings et al.(2010)]{Vlemmings_10} \href{http://adsabs.harvard.edu/abs/2010MNRAS.404..134V}{Vlemmings, W.~H.~T., Surcis, G., Torstensson, K.~J.~E., \& van Langevelde, H.~J., 2010}, \mnras, 404, 134 
\bibitem[Wade et al.(2005)]{Wade_05} \href{http://adsabs.harvard.edu/abs/2005A\%26A...442L..31W}{Wade, G.~A., Drouin, D., Bagnulo, S., et al., 2005}, \aap, 442, L31
\bibitem[Wade et al.(2013)]{Wade_13} \href{http://adsabs.harvard.edu/abs/2014IAUS..302..265W}{Wade, G.~A., Grunhut, J., Alecian, E., et al., 2013}, IAU Symposium, 302, 265 
\bibitem[Weiss(1966)]{Weiss66} \href{http://adsabs.harvard.edu/abs/1966RSPSA.293..310W}{Weiss, N.~O., 1966}, Royal Society of London Proceedings Series A, 293, 310
\bibitem[Woltjer(1958)]{Woltjer58} \href{http://adsabs.harvard.edu/abs/1958PNAS...44..833W}{Woltjer, L., 1958}, Proceedings of the National Academy of Science, 44, 833 
\bibitem[Wongwathanarat et al.(2013)]{Wongwathanarat_13} \href{http://adsabs.harvard.edu/abs/2013A\%26A...552A.126W}{Wongwathanarat, A., Janka, H.-T., M{\"ul}ller, E., 2013}, \aap, 552, A126 
\bibitem[Wright(1973)]{Wright73} \href{http://adsabs.harvard.edu/abs/1973MNRAS.162..339W}{Wright, G.~A.~E., 1973}, \mnras, 162, 339
\bibitem[Wright et al.(2011)]{2011ApJ...743...48W} \href{http://adsabs.harvard.edu/abs/2011ApJ...743...48W}{Wright, N.~J., Drake, J.~J., Mamajek, E.~E., \& Henry, G.~W.\ 2011}, \apj, 743, 48 
\bibitem[Yang \& Johns-Krull(2011)]{YangJ-K11} \href{http://adsabs.harvard.edu/abs/2011ApJ...729...83Y}{Yang, H., \& Johns-Krull, C.~M., 2011}, \apj, 729, 83 
\bibitem[Yoshida et al.(2006)]{Yoshida_06} \href{http://adsabs.harvard.edu/abs/2006ApJ...651..462Y}{Yoshida, S., Yoshida, S., \& Eriguchi, Y., 2006}, \apj, 651, 462
\bibitem[Zahn(1992)]{Zahn92} \href{http://adsabs.harvard.edu/abs/1992A\%26A...265..115Z}{Zahn, J.-P., 1992}, \aap, 265, 115 
\bibitem[Zahn et al.(2007)]{2007A&A...474..145Z} \href{http://adsabs.harvard.edu/abs/2007A\%26A...474..145Z}{Zahn, J.-P., Brun, A.~S., \& Mathis, S.\ 2007}, \aap, 474, 145 
\bibitem[Zeldovich(1956)]{Zeldovich56}{Zeldovich, Ya. B. 1956}, JETP 31, 154 [Sov. Phys. JETP 4, 460 (1957)] 
\bibitem[Zhang \& Low(2003)]{ZhangL03} \href{http://adsabs.harvard.edu/abs/2003ApJ...584..479Z}{Zhang, M., \& Low, B.~C., 2003}, \apj, 584, 479 
\bibitem[Zhdankin et al.(2013)]{Zhdankin13}\href{http://adsabs.harvard.edu/abs/2007ARA\%26A..45..481Z}{Zhdankin, V., Uzdensky, D.~A., Perez, J.~C., \& Boldyrev, S.\ 2013}, \apj, 771, 124 
\bibitem[Zinnecker \& Yorke(2007)]{ZinneckerY07} \href{http://adsabs.harvard.edu/abs/2007ARA\%26A..45..481Z}{Zinnecker, H., \& Yorke, H.~W., 2007}, \araa, 45, 481 

 \end{thebibliography}

\end{document}